\documentclass[a4paper,12pt]{article}
\pdfoutput=1 
\usepackage{a4wide}
\usepackage[bookmarks=false,colorlinks]{hyperref}
\usepackage{rotating,amsmath,bm}
\usepackage{tabularx}
\usepackage{multicol,multirow}
\usepackage{latexsym}
\usepackage{relsize}
\usepackage[center]{subfigure}
\usepackage{float}
\usepackage{caption}
\usepackage{slashed}
\usepackage{lineno}
\usepackage{cite,mcite}
\usepackage{appendix}
\usepackage{url}
\hypersetup{urlcolor=blue, citecolor=blue}

\date{\today}
 
\normalsize

\def\Bbar{\overline{B}}

\def\cbar{\overline{c}}

\def\ellbar{{\overline{\ell}}}
\def\nubar{{\overline{\nu}}}
\def\btod{{\Bbar\to D\ell\nubar_\ell}}
\def\btodstar{{\Bbar\to D^\ast\ell\nubar_\ell}}
\def\btodscal{{\Bbar\to D_0^\ast\ell\nubar_\ell}}
\def\Heff{\mathcal{H}_{\rm eff}}
\def\M{\mathcal{M}}

\def\Re{\mathcal{R}e}
\def\Im{\mathcal{I}m}

\newcommand{\bea}{\begin{eqnarray}}
\newcommand{\eea}{\end{eqnarray}}
\newcommand{\beq}{\begin{equation}}
\newcommand{\eeq}{\end{equation}}
\newcommand{\ec}{\end{center}}
\newcommand{\bc}{\begin{center}}
\newcommand{\gev}{{\rm GeV}}
\newcommand{\mev}{{\rm MeV}}
\newcommand{\kev}{{\rm keV}}
\newcommand{\nn}{\nonumber}
\def\Dst{{D^\ast}}
\def\DOst{{D_0^\ast}}
\def\p{{\hat{\bm p}}}
\def\q{{\bm q}}

\def\thl{{\theta_\ell}}
\def\thD{{\theta_D}}

\begin{document}
%\leftline\today
\thispagestyle{empty} 
\begin{flushright}
\begin{tabular}{l}
{\footnotesize \tt LPT 15-77}\\
{\footnotesize \tt DO-TH 16/04}\\
\end{tabular}
\end{flushright}
\begin{center}
\vskip 2.5cm\par
{\par\centering \textbf{\LARGE  
\Large \bf Angular distributions of $\Bbar\to D^{(\ast)}\ell\nubar_\ell$ decays }}\\
\vskip .25cm\par
{\par\centering \textbf{\LARGE  
\Large  and search of New Physics} }\\
\vskip 1.25cm\par
{\scalebox{.82}{\par\centering \large  
\sc Damir Be\v cirevi\'c$^a$, Svjetlana Fajfer$^b$, Ivan Ni\v{s}and\v{z}i\'c$^c$, Andrey Tayduganov$^d$ }}
{\par\centering \vskip 0.3 cm\par}
{\sl \small
$^a$~Laboratoire de Physique Th\'eorique,  B\^at.~210 (UMR 8627)\\
Universit\'e Paris Sud, Universit\'e Paris-Saclay, 91405 Orsay cedex, France}\\
{\par\centering \vskip 0.3 cm\par}
{\sl \small
$^b$~J. Stefan Institute, Jamova 39, P. O. Box 3000, 1001 Ljubljana, Slovenia , and \\
Department of Physics, University of Ljubljana, Jadranska 19, 1000 Ljubljana, Slovenia }\\
{\par\centering \vskip 0.3 cm\par}
{\sl \small
$^c$~Institut f\"ur Physik, Technische Universit\"at Dortmund, D-44221 Dortmund, Germany }\\
{\par\centering \vskip 0.3 cm\par}
{\sl \small
$^d$~Centre de Physique des Particules de Marseille, \\ 
Aix-Marseille Universit\'e, CNRS/IN2P3,  3288 Marseille, France}\\

{\vskip 1.cm\par}
\end{center}

\vskip 0.55cm
\begin{abstract}
We derive the expressions for the full angular distributions of $\Bbar\to D\ell\nubar_\ell$ and $\Bbar\to D^{\ast }\ell\nubar_\ell$ decays and discuss the spectra on each angle separately. The coefficient functions, depending on helicity amplitudes, can then be combined in an ensemble of observables which can then be used to check for the presence of New Physics. We examine the sensitivity of each of these observables on the presence of non-Standard Model interaction terms at low energies. The expressions presented here are general, and can be used for studying any other semileptonic pseudoscalar to pseudoscalar/vector meson decay.  We also examine the problem of pollution of the $\Bbar\to D^{\ast}(\to D\pi)_S\ell\nubar_\ell$ decay sample by the $\Bbar\to D_0^{\ast}(\to D\pi)\ell\nubar_\ell$ events, and point out that a measurement of two particular quantities could clarify whether or not the $(D\pi)_{S-\rm wave}$ in the vicinity of $D^\ast$-peak is (approximately) described by the Breit-Wigner formula.
\end{abstract}
%\vskip 2.2 cm 
\newpage
\setcounter{page}{1}
\setcounter{footnote}{0}
\setcounter{equation}{0} 
%%%%%%%%%%%%%%%%%%%%%%%%%%%%%%%%%%%%%%%%
\noindent
\allowdisplaybreaks[1]
\renewcommand{\thefootnote}{\arabic{footnote}}

\setcounter{footnote}{0}
%%%%%%%%%%%  Section 1
\section{Introduction}
For many years the main motivation to study the leptonic and semileptonic meson decays was to extract the Cabibbo-Kobayashi-Maskawa (CKM) couplings through a comparison of  theoretical expressions with the experimentally measured branching fractions. Although not directly accessible, the CKM couplings that involve the top quark are indirectly obtained from the low energy processes driven by the flavor changing neutral currents (FCNC). Concerning the FCNC processes a more appealing exercise is to assume the CKM matrix to be unitary, which is a sufficient condition to fix the values of $\vert V_{td,ts,tb}\vert$, and then check for discrepancies between the measured and predicted rates that could be interpreted as signals of non-Standard Model heavy particles propagating in the loops. Such a strategy to search for the effects of New Physics (NP) has been extensively explored in the past couple of decades but no significant discrepancy with respect to the Standard Model (SM) expectations has been found so far. As a striking example one can 
quote the results obtained by the CKM-fitter and the UT-fit, showing that, to a present-day accuracy, the unitarity triangle reconstructed by using the tree-level decays does not differ from the one obtained by relying on the loop-induced processes~\cite{CKMology}. Therefore the effects of NP are either absent or small.

Looking for the small departures of measured branching fractions from their SM predictions is extremely difficult as it requires a precision determination of hadronic matrix elements which have to be computed non-perturbatively from the first theory principles of QCD. Only for a very limited number of quantities such a percent precision accuracy, based on numerical simulations of QCD on the lattice, has been achieved so far~\cite{FLAG}. In such a situation the angular analysis of $B\to K^{(\ast)}\ell^+\ell^-$ decays proved to be particularly interesting as it allowed to define a number of observables that are accessible to the modern day experiments and are highly sensitive to the effects of physics beyond the SM (BSM). More interestingly, a subset of these observables appeared to be mildly sensitive to the hadronic uncertainties. In this paper we show that a similar strategy can be adopted to study the tree level processes and  check for the effects of NP through a comparison of the SM predictions of the angular distribution of $\bar B\to D\ell\bar \nu_\ell$ and  $\bar B\to D^\ast\ell\bar \nu_\ell$ decay modes with experiment.  $\bar B\to D^{(\ast)}\ell\bar \nu_\ell$ have been studied at the $B$-factories (BaBar and Belle) to a very good accuracy~\cite{Aubert:2008yv,Dungel:2010uk,Amhis:2014hma}.

The samples of these decay modes will be much larger at Belle~II and a detailed precision study of their angular distribution will become feasible. Although we focus onto the $\bar B\to D\ell\bar \nu_\ell$ and  $\bar B\to D^\ast\ell\bar \nu_\ell$ modes, the discussion we make in this paper is equally applicable to all the other semileptonic decays in which a pseudoscalar meson decays to another pseudoscalar or a vector meson, namely $D/B\to \pi\ell\bar \nu$, $D/B\to \rho \ell\bar \nu$,  $D_{(s)}\to K^{(\ast)}\ell\bar \nu$, $B_s\to K^{(\ast)} \ell\bar \nu$, $D_s\to \phi \ell\bar \nu$, $K\to \pi \ell \nu_\ell$, $B_c\to J/\psi \ell\nu_\ell$, $B_c\to \eta_c \ell\nu_\ell$, $B_c\to B_{d,s} \ell\nu_\ell$, or the semileptonic $B_s$-meson decays.~\footnote{Results of one such a study have been recently presented in Ref.~\cite{Feldmann:2015xsa} where the authors focused onto the phenomenologically appealing $B_s\to K^{(\ast )}\ell \nu_\ell$ decay mode. 
} Our choice to focus on $\bar B\to D\ell\bar \nu_\ell$ and  $\bar B\to D^\ast\ell\bar \nu_\ell$ is related to the fact that a small but intriguing disagreement between experiment and theory has been recently reported in the case of $R_{D^{(\ast )}}= {\cal B}(\bar B\to D^\ast\tau\bar \nu_\tau )/{\cal B}(\bar B\to D^\ast\mu\bar \nu_\mu )$~\cite{RDst_exp,RDst_th,distrBD}.

We will first derive the expressions for the full angular distribution of these decays and then address the following questions:
\begin{itemize}
\item Which observables can be extracted from the angular distributions that are sensitive to the effects of physics BSM?
\item In the case of a heavy lepton in the final state, a significant fraction of the $\bar B\to D^\ast (\to D\pi) \tau \bar \nu_\tau$ events involves $D\pi$ pairs in the $S$-wave. Could these events be polluted by the $D\pi$ pairs emerging from the scalar $D^\ast_0$ state, i.e. $\bar B\to D_0^\ast (\to D\pi) \tau \bar \nu_\tau$?
\end{itemize}
To our knowledge, several quantities have been proposed to study so far in refs.~\cite{RDst_th,distrBD}. Here we consider the full ensemble of observables that can be derived from the angular distribution. Like in the seminal paper of Ref.~\cite{Korner:1989qb} our expressions for the angular distribution coefficients are given in terms of helicity amplitudes, and as such they are completely general. We will adopt a particular effective Hamiltonian to examine the effect of non-SM interactions, and therefore only after we express the helicity amplitudes in terms of kinematic variables, the hadronic form factors and the NP couplings, our expressions will become (slightly) model dependent.~\footnote{Model dependence in this case means the assumptions concerning the possible extensions of the SM at high energy which at low energies are manifested by a handful of additional operators. More important model dependence comes with the choice of the hadronic form factors for which the uncertainties are still not at the percent level, at least not in the full range of available $q^2$'s.}

Concerning the second of the above questions the answer is negative if one adopts a simple Breit-Wigner (BW) function and the width of $D_0^{\ast}$ state reported in PDG~\cite{PDG}. If the deviations from the BW form occur, similar to those present in e.g. the tail of $K_0^\ast$, then this problem can be experimentally important to address. We found quantities that can be studied through the angular distribution and which are nonzero only if there is interference between the $D\pi$-pairs in $S$-wave coming from $\bar B\to D^\ast (\to D\pi) \tau \bar \nu_\tau$ with those coming from some other source, $\bar B\to D_{\rm scal} (\to D\pi) \tau \bar \nu_\tau$, similar to the situation of $K_0^\ast \to K\pi$ and $\kappa\to K\pi$ in the $B\to K^\ast \ell^+\ell^-$ decay~\cite{s-wave-kpi}. 
This problem is much less relevant in the case of $\bar B_s\to D_s^\ast\tau\bar \nu_\tau$ because the scalar state $D_{s0}^\ast$, in the corresponding $\bar B_s\to D_{s0}^\ast\tau\bar \nu_\tau$ is extremely narrow~\cite{PDG}.

In Sec.~\ref{sec:2} of what follows we provide the explicit expressions for the angular distribution of semileptonic decays, define the full set of observables and discuss the terms that are nonzero only if there is interference between semileptonic decays to a vector and to a scalar meson. In Sec.~\ref{sec:3} we check the sensitivity of the observables defined in Sec.~\ref{sec:2} with respect to the NP quark operators at low energies. We summarize our findings in Sec.~\ref{sec:4}.

\section{Full distributions of $\Bbar\to D\ell\nubar_\ell$  and  $\Bbar \to D^\ast\ell\bar \nu_\ell$ decays\label{sec:2}}

In this section we sketch the derivation of expressions for the full two-fold and five-fold distribution of the $\Bbar\to D\ell\nubar_\ell$ and $\Bbar\to D^\ast\ell\nubar_\ell$, respectively. We will keep the non-zero mass of the lepton in all our formulas. The SM expressions that we derive coincide with those presented in Ref.~\cite{Korner:1989qb}. Since our aim is to study the possible NP effects, we will go a step beyond Ref.~\cite{Korner:1989qb} and include the terms that are absent in the SM but can be non-zero in a generic NP scenario. We then consider an effective Hamiltonian in which the NP effects could affect only the quark sector, while leaving the lepton sector universal, in its SM form. Other possibilities for the NP effective Hamiltonian can, of course, be envisaged.

\subsection{Effective Hamiltonian}
At the level of an effective theory we consider~\cite{Dassinger:2008as}~\footnote{We use the definition $\sigma_{\mu\nu}=(i/2) [\gamma_\mu,\gamma_\nu]$.} 
\begin{equation}
	\begin{split}
		\Heff = {G_F \over \sqrt2}V_{cb} &\ H_\mu L^\mu + {\rm h.c}\cr
		 ={G_F \over \sqrt2}V_{cb} & \biggl[(1+g_V)\cbar\gamma_\mu b + (-1+g_A)\cbar\gamma_\mu\gamma_5 b + g_S\ i\partial_\mu(\cbar b) + g_P\ i\partial_\mu(\cbar\gamma_5 b) \biggr. \\
		      & \biggl.\quad  + g_T\ i\partial^\nu(\cbar i\sigma_{\mu\nu} b) + g_{T5}\  i\partial^\nu(\cbar i\sigma_{\mu\nu}\gamma_5 b) \biggr]\  \ellbar\gamma^\mu(1-\gamma_5)\nu_\ell + {\rm h.c} \,,
	\end{split}
	\label{eq:Heff}
\end{equation}
which is the most general if the coupling to leptons is of the $V-A$ form, like in the SM.  While $g_{V,A}$ are dimensionless, the couplings $g_{S,P,T,T5}$ are dimensionfull as to compensate for the fact that that the corresponding quark operators have mass dimension equal to four. Furthermore the couplings $g_{S,P,T,T_5}\equiv g_{S,P,T,T_5}(\mu)$ carry the QCD anomalous dimension which is the inverse of the anomalous dimension of the bilinear quark operator they multiply as to leave $\Heff$ scale independent.~\footnote{ $g_T$ and $g_{T_5}$ are obviously not independent. We define them separately for computational commodity, but when doing phenomenology we take into account the fact that $\sigma_{\mu\nu}\gamma_5 =(i/2)\epsilon_{\mu\nu\alpha\beta}\sigma^{\alpha\beta}$.
} Finally, quite obviously, by setting $g_{S,P,V,A,T,T5}=0$ in eq.~(\ref{eq:Heff}) one retrieves the usual SM effective Hamiltonian.

\subsection{$\btod$ decay\label{sec:dlnu}}

We begin with the expression for the full spectrum of $\btod$ decay which has a very simple form, 
\begin{equation}
	{d^2\Gamma \over dq^2d\cos\thl} = {1 \over 32(2\pi)^3m_B^2}|\q|\left(1-{m_\ell^2\over q^2}\right)|\M(\Bbar\to D\ell\nubar_\ell)|^2 \,,
	\label{eq:dG2}
\end{equation}
where $\q$ stands for the three-momentum of the $\ell\nubar_\ell$ pair in the $B$-meson rest frame, and  $\thl$ is the angle between the direction of flight of $D$ and $\ell$ in the center of mass frame  of $\ell\nubar_\ell$~\cite{Korner:1989qb}. To write the amplitude $\M(\btod)$ explicitly we decompose the non-vanishing hadronic matrix elements of the quark operators in eq.~(\ref{eq:Heff}) in terms of the Lorentz invariant hadronic form factors, 
\bea\label{eq:B-D_FF}
	\langle D(k)|\cbar\gamma_\mu b|\Bbar(p)\rangle &=& \left[(p+k)_\mu-{m_B^2-m_D^2 \over q^2}q_\mu\right]f_+(q^2)+q_\mu{m_B^2-m_D^2 \over q^2}f_0(q^2) \,,\nn
	\\
	\hfill\nn\\
	\langle D(k)|[\cbar b](\mu)|\Bbar(p)\rangle &=& {1 \over m_b(\mu) -m_c(\mu)}q^\mu\langle D(k)|\cbar\gamma_\mu b|\Bbar(p)\rangle = {m_B^2-m_D^2 \over m_b(\mu) -m_c(\mu)}f_0(q^2) \,,\nn
	\\
	\hfill\nn\\
	\langle D(k)|[\cbar\sigma_{\mu\nu} b](\mu)|\Bbar(p)\rangle  &=& -i\left( p_\mu k_\nu - k_\mu p_\nu\right)\ {2\ f_T(q^2,\mu) \over m_B+m_D} \,,
\eea
where the form factors $f_{+,0,T}(q^2)$ are functions of $q^2=(p-k)^2$. As mentioned above, the scalar and tensor densities in QCD, at short distances, each acquire anomalous dimension. Their respective scale dependence is indicated in the argument  of the operators on the left hand side (l.h.s.). The $\mu$-dependence of the form factor $f_T(q^2,\mu)$ and of the quark mass difference $m_b(\mu) -m_c(\mu)$ cancel against the $\mu$-dependence of $g_T(\mu)$ and $g_S(\mu)$, respectively. In what follows the $\mu$-dependence will be implicit and the value $\mu=m_b$ will be assumed.  

With the above definitions in hands and with $\widetilde\varepsilon_{0,t}^{\mu}$, polarization vectors of the virtual vector boson $V^\ast$ specified  in Appendix~\ref{app:polarization}, we can now write the helicity amplitudes for $\overline B\to V^\ast D$ decay as
\begin{equation}\label{eq:ha1}
   h_{0,t}(q^2) = \widetilde\varepsilon_{0,t}^{\mu \ast }\ \langle D| H_\mu|\Bbar\rangle \,,
\end{equation}
or explicitly,
\begin{equation}
        \begin{split}
          h_0(q^2) =& \biggl[1+g_V-g_T{q^2 \over m_B+m_D}{f_T(q^2) \over f_+(q^2)}\biggr]{\sqrt{\lambda(m_B^2,m_D^2,q^2)} \over \sqrt{q^2}}f_+(q^2) \,, \\
          h_t(q^2) =& \left[1+g_V+g_S{q^2 \over m_b-m_c}\right]{m_B^2-m_D^2 \over \sqrt{q^2}}f_0(q^2) \,,
        \end{split}
\end{equation}
where $\lambda(x^2,y^2,z^2)=[x^2-(y-z)^2] [x^2-(y+z)^2]$. The full two-fold decay distribution~(\ref{eq:dG2}) then reads: 
\begin{equation}\label{eq:10}
	{d^2\Gamma \over dq^2d\cos\thl} = a_\thl (q^2)+b_\thl(q^2) \cos\thl+c_\thl(q^2)\cos^2\thl \,,
\end{equation}
where the $q^2$-dependent coefficient functions are given by
\begin{subequations}\label{eq:11}
	\begin{align}
		\begin{split}
			& a_\thl(q^2) = {G_F^2|V_{cb}|^2 \over 256\pi^3m_B^3}\lambda^{1/2}(m_B^2,m_D^2,q^2)\ q^2 \left(1-{m_\ell^2 \over q^2}\right)^2 \ \biggl[ 
			\left|h_0(q^2) \right|^2 + {m_\ell^2 \over q^2} \left|h_t(q^2) \right|^2\biggr] \,,
		\end{split}\\
		& \nonumber \\
		\begin{split}
			& b_\thl(q^2) = -{G_F^2|V_{cb}|^2 \over 128\pi^3m_B^3} \lambda^{1/2}(m_B^2,m_D^2,q^2)\ q^2 \left(1-{m_\ell^2 \over q^2}\right)^2 \ 
			{m_\ell^2 \over q^2} \   \Re\left[ h_0(q^2) h_t^\ast(q^2)\right] \,,
		\end{split}\\
		& \nonumber \\
		\begin{split}
			& c_\thl(q^2) = -{G_F^2|V_{cb}|^2 \over 256\pi^3m_B^3}\lambda^{1/2}(m_B^2,m_D^2,q^2)\ q^2  \left(1-{m_\ell^2 \over q^2}\right)^3 \ \left|h_0(q^2) \right|^2  \,.
		\end{split}
	\end{align}
\end{subequations}
Out of three functions, $a_\thl(q^2)$, $b_\thl(q^2)$, $c_\thl(q^2)$, one can derive at most three independent observables.  The first of those is the differential decay rate which is simply obtained from 
\bea\label{eq:bd1}
{d\Gamma \over dq^2} &=& \int_{-1}^1  {d^2\Gamma \over dq^2d\cos\thl} d\cos\thl	= 2 \left[ a_\thl(q^2) +\frac{1}{ 3} c_\thl(q^2)\right]\nn\\
&&\hfill \nn\\
 &=& {G_F^2|V_{cb}|^2 \over 192\pi^3m_B^3} \lambda^{1/2}(m_B^2,m_D^2,q^2)\left(1-{m_\ell^2 \over q^2}\right)^2 |f_+(q^2)|^2 \times \nn\\
			&&\quad\quad\quad\quad\times  \biggl\{\biggr.   \left|1+g_V-g_T{q^2 \over m_B+m_D}{f_T(q^2) \over f_+(q^2)}\right|^2\lambda(m_B^2,m_D^2,q^2) \left(1+{m_\ell^2 \over 2 q^2}\right) \nn \\
			&& \quad\quad\quad\quad\biggl.~~+\left|1+g_V+g_S{q^2 \over m_b-m_c}\right|^2{3m_\ell^2 \over 2 q^2}(m_B^2-m_D^2)^2\left| {f_0(q^2)\over f_+(q^2)} \right|^2\biggr\} \,,
\eea
which for $g_{S,V,T}=0$ gives the familiar SM expression. 
The full decay width of $\btod$ is then obtained after integrating in $q^2$,
\bea
\Gamma(\btod)= \int_{m_\ell^2}^{q^2_{\rm max}} {d\Gamma \over dq^2}  \ dq^2 \,,
\eea
with $q^2_{\rm max} = (m_B-m_D)^2$, and assuming neutrinos to be massless. 
\subsubsection{Two more observables}

Apart from the differential decay rate~(\ref{eq:bd1}) we can construct two more observables that are experimentally accessible via the angular distribution of $\btod$: the forward-backward asymmetry and the lepton polarization asymmetry. They are both sensitive to the lepton mass and are therefore interesting to study when the $\tau$-lepton is in the final state. 

As it can be seen from eq.~(\ref{eq:10}), the linear dependence on $\cos\thl$ in $d\Gamma/dq^2$ is lost after integration in $\thl$,  but it can be retrieved when considering   
the forward-backward asymmetry, 
\bea
A_{FB}^D(q^2)& =&{\displaystyle{  \int_{0}^1 {d^2\Gamma \over dq^2d\cos\thl} d\cos\thl  - \int_{-1}^0  {d^2\Gamma \over dq^2d\cos\thl} d\cos\thl} \over\displaystyle{d\Gamma\over dq^2} }\nn\\
&=&  { b_\thl(q^2)  \over {d\Gamma/dq^2} }=
- {3 \over 2}{m_\ell^2 \over q^2}{\Re[h_0(q^2)h_t^*(q^2)] \over {|h_0(q^2)|^2\left(1+\displaystyle{m_\ell^2 \over 2q^2}\right)+\displaystyle{{3 \over 2}{m_\ell^2 \over q^2}}|h_t(q^2)|^2}}\,,
 \eea
where the normalization is conventionally made to  ${d\Gamma/dq^2}$. One can also compute its integrated characteristics
\bea
\langle A_{FB}^D\rangle = {1\over \Gamma} \int_{m_\ell^2}^{q^2_{\rm max}}  b_\thl(q^2)  dq^2\,.
\eea

Another quantity that can be interesting in studying the NP effects is the lepton polarization asymmetry. It is defined from the differential decay rates with definite lepton helicity, $\lambda_\ell = \pm 1/2$:
   \begin{align}
      & {d\Gamma^+ \over dq^2}\equiv  \left.{d\Gamma \over dq^2}\right|_{\lambda_\ell=+1/2} \!\!\!\!= {G_F^2|V_{cb}|^2 q^2 \over 192\pi^3m_B^3} \lambda^{1/2}(m_B^2,m_D^2,q^2)\left(1-{m_\ell^2 \over q^2}\right)^2 {m_\ell^2 \over 2q^2}\biggl[ |h_0(q^2)|^2 +3|h_t(q^2)|^2 \biggr] \,,\nn
      & \nonumber \\
      &{d\Gamma^- \over dq^2}\equiv  \left.{d\Gamma \over dq^2}\right|_{\lambda_\ell=-1/2} \!\!\!\! = {G_F^2|V_{cb}|^2 q^2 \over 192\pi^3m_B^3} \lambda^{1/2}(m_B^2,m_D^2,q^2)\left(1-{m_\ell^2 \over q^2}\right)^2 |h_0(q^2)|^2 \,,  
   \end{align}
which obviously verify $\Gamma = \Gamma^+ + \Gamma^-$. The lepton polarization asymmetry $A_{\lambda_\ell}(q^2)$ then reads
\begin{equation}
   \begin{split}
      A_{\lambda_\ell}^D(q^2)  =& \ 1 -\ 2\  {  d\Gamma^+/dq^2   \over d\Gamma/dq^2} \\
      =& 1- {m_\ell^2 \over q^2} { |h_0(q^2)|^2+3|h_t(q^2)|^2  \over {|h_0(q^2)|^2\left(1+\displaystyle{m_\ell^2 \over 2q^2}\right)+\displaystyle{{3 \over 2}{m_\ell^2 \over q^2}}|h_t(q^2)|^2}}\,.
   \end{split}
\end{equation}
It can be convenient to compute its value integrated over available $q^2$'s,
\bea
\langle 1- A_{\lambda_\ell}^D \rangle =  {2 \over \Gamma} \int_{m_\ell^2}^{q^2_{\rm max}}   {d\Gamma^+ \over dq^2} \  dq^2\,.
\eea
Being proportional to the squared lepton mass, $A_{FB}^D(q^2)$ and $1-A_{\lambda_\ell}^D(q^2)$ are sensibly different from zero in the SM only in the case of the $\tau$-lepton in the final state.  These quantities with $e$ or $\mu$ in the final state could be used as null-tests of the SM, the non-zero value of which would suggest the presence of operators that lift the helicity suppression. Note also that, like all the observables we consider here (apart from differential decay rates), the above asymmetries do not depend on the CKM parameter and that they are functions of the ratios of form factors, $f_0(q^2)/f_+(q^2)$ and $f_T(q^2)/f_+(q^2)$, for which the hadronic uncertainties are generally smaller than those for the absolute values of form factors.

\subsection{$\btodstar$ decay}

We now proceed along the lines discussed in Sec.~\ref{sec:dlnu} and write the five-fold differential decay rate of the $\Bbar\to D\pi\ell\nubar_\ell$ decay as follows:
\begin{equation}\label{eq:dG5}
	{d^5\Gamma \over dq^2dm_{D\pi}^2d\cos\thD d\cos\thl d\chi} = {1 \over 128(2\pi)^6 m_B^2}|\q|\left(1-{m_\ell^2\over q^2}\right){|\p_D| \over m_{D\pi}}|\M(\Bbar\to D\pi\ell\nubar_\ell)|^2 \,,
\end{equation}
where $\p_D$ is the three-momentum of $D$ in the rest frame of $D\pi$, and the three angles are specified in the Appendix~\ref{app:polarization} of the present paper. 
We focus onto the first two $D\pi$-resonances and write the decay amplitude as
\begin{equation}\label{eq:dG6}
	\M(\Bbar\to D\pi\ell\nubar_\ell) = \sum_{D_{\rm res}=\Dst,\,\DOst} \langle D \pi |D_{\rm res}\rangle \langle D_{\rm res}| H_\mu |\Bbar\rangle L^\mu \,\widetilde{BW}_{D_{\rm res}} \,,
\end{equation}
where the propagation of the intermediate resonant state is parametrized by
\begin{equation}\label{eq:bw}
	\begin{split}
		\widetilde{BW}_{D_{\rm res}}(m_{D\pi}^2) &= {1 \over m_{D\pi}^2-m_{D_{\rm res}}^2+im_{D_{\rm res}}\Gamma_{D_{\rm res}}} \,, \\
			    BW_{D_{\rm res}}(m_{D\pi}^2) &= {\sqrt{m_{D_{\rm res}}\Gamma_{D_{\rm res}}/\pi} \over m_{D\pi}^2-m_{D_{\rm res}}^2+im_{D_{\rm res}}\Gamma_{D_{\rm res} }} \,.
	\end{split}
\end{equation}
The first and second line in the above equation respectively correspond to the non-normalized and normalized BW function. 

Like in the previous section, we need to specify the decomposition of hadronic matrix elements in terms of the Lorentz invariant form factors.  
Concerning the $B\to D^\ast$ transition we write~\footnote{We use the convention with $ \epsilon_{0123}=1$.}
\begin{equation}
	\begin{split}
		\langle\Dst(k,\varepsilon)|\cbar\gamma_\mu b|\Bbar(p)\rangle =& \epsilon_{\mu\nu\rho\sigma}\varepsilon^{\nu*}p^\rho k^\sigma{2V(q^2) \over m_B+m_\Dst} \,, \\
	\langle\Dst(k,\varepsilon)|\cbar\gamma_\mu\gamma_5 b|\Bbar(p)\rangle =& i\varepsilon_{\mu}^\ast (m_B+m_\Dst)A_1(q^2) - i(p+k)_\mu(\varepsilon^*q){A_2(q^2) \over m_B+m_\Dst} \\
									      & -iq_\mu(\varepsilon^*q){2m_\Dst \over q^2}\Bigl[ A_3(q^2)-A_0(q^2)\Bigr] \,,
	\end{split}
	\label{eq:B-Dst_FF_VA}
\end{equation}
with
\begin{equation}\label{eq:B-Dst_FF_VAbis}
	A_3(q^2) = {m_B+m_\Dst \over 2m_\Dst}A_1(q^2)-{m_B-m_\Dst \over 2m_\Dst}A_2(q^2) \,,
\end{equation}
satisfying the condition $A_3(0)=A_0(0)$. The matrix element of the pseudoscalar density is related to the one of the axial current via the axial Ward identity, i.e., 
\begin{equation}
	\begin{split}
		\langle\Dst(k,\varepsilon)|[\cbar\gamma_5 b](\mu)|\Bbar(p)\rangle =& -{1 \over m_b(\mu) +m_c(\mu)}q_\nu\langle\Dst(k,\varepsilon)|\cbar\gamma^\nu\gamma_5 b|\Bbar(p)\rangle \\
										 =& -i(\varepsilon^\ast q){2m_\Dst \over m_b(\mu)+m_c (\mu)}A_0(q^2) \,,
	\end{split}
	\label{eq:B-Dst_FF_P}
\end{equation}
where the $\mu$-dependence of the operator is carried by the quark mass on the right hand side (r.h.s).
When considering $g_{T,T5}(\mu)\neq 0$ we will also need, 
\begin{equation}
	\begin{split}
		\langle\Dst(k,\varepsilon)|[\cbar\sigma_{\mu\nu}q^\nu b](\mu)|\Bbar(p)\rangle =& i\epsilon_{\mu\nu\rho\sigma}\varepsilon^{\nu\ast}p^\rho k^\sigma \ 2T_1(q^2,\mu) \,, \\
	\langle\Dst(k,\varepsilon)|[\cbar\sigma_{\mu\nu}\gamma_5q^\nu b](\mu)|\Bbar(p)\rangle =& \Bigl[(m_B^2-m_\Dst^2)\varepsilon^{\mu\ast} - (\varepsilon^\ast q)(p+k)_\mu\Bigr] T_2(q^2,\mu) \\
											& +(\varepsilon^\ast q)\left[q_\mu-{q^2 \over m_B^2-m_\Dst^2}(p+k)_\mu\right]T_3(q^2,\mu) \,,
	\end{split}
	\label{eq:B-Dst_FF_TT5}
\end{equation}
where the $\mu$-dependence of the operator is then carried by the form factors and it will cancel against those in the couplings $g_{T,T5}(\mu)$.

We will also consider the decay to the scalar meson, $\overline B \to D_0^\ast (\to D\pi) \ell \bar \nu_\ell$, for which the relevant hadronic matrix element is decomposed as:
\begin{equation}
	\begin{split}
		\langle\DOst(k)|\cbar\gamma_\mu\gamma_5 b|\Bbar(p)\rangle =& -i\left[(p+k)_\mu-{m_B^2-m_\DOst^2 \over q^2}q_\mu\right]F_1(q^2) \\
									   & \quad -iq_\mu{m_B^2-m_\DOst^2 \over q^2} F_0(q^2) \,,
	\end{split}
	\label{eq:B-D0st_FF_A}
\end{equation}
where we opted for the conventions of Ref.~\cite{cheng}. By  means of the axial Ward identity we have,
\begin{equation}
	\begin{split}
		\langle\DOst(k)|[\cbar\gamma_5 b](\mu)|\Bbar(p)\rangle =& -{q_\nu \over m_b(\mu) +m_c(\mu)}  \langle\DOst(k)|\cbar\gamma^\nu\gamma_5 b|\Bbar(p)\rangle \\
								      =& i{m_B^2-m_{\DOst}^2 \over m_b(\mu)+m_c(\mu)}F_0(q^2) \,.
	\end{split}
	\label{eq:B-D0st_FF_P}
\end{equation}
Finally, in eq.~(\ref{eq:dG6}) we also need, 
\bea
\langle D\pi\vert D^\ast\rangle  &=& g_{\Dst D\pi} p_D^\mu\varepsilon_\mu \,, \nn\\
	\langle D\pi|\DOst\rangle &=& g_{\DOst D\pi} \,,
\eea 
where the coupling $g_{\Dst D\pi}$ parameterizes the physical $D^\ast \to D\pi$ decay and can be extracted from the numerical simulations of QCD on the lattice~\cite{notre-g} and its value agrees with the result extracted from the width of the charged $D^\ast$, $\Gamma(D^{\ast +})$, recently measured at BaBar~\cite{babar-g}. Similarly, $g_{\DOst D\pi}$ describes the decay $D_0^\ast \to D\pi$ which has been computed on the lattice in the static limit~\cite{hs} and in the case of propagating charm quark in~\cite{hc}. In terms of the above couplings we have
\bea\label{eq:width}
\Gamma(D^{\ast}\to D\pi) = {C \over 24 \pi m_{D^{\ast }}^2}\  g_{D^\ast D\pi}^2  |\p_D|^3,\qquad \Gamma(D_0^{\ast}\to D\pi) = {C \over 8 \pi m_{D_0^{\ast }}^2}\  g_{D_0^\ast D\pi}^2 |\p_D^\prime|  \,, 
\eea
where $C=1$ if the outgoing pion is charged, and $C=1/2$ if it is neutral, and  
\bea
|\p_D^{(\prime )}| = {\sqrt{\lambda(m_{D_{(0)}^\ast}^2,m_D^2,m_\pi^2)} \over 2 m_{D_{(0)}^\ast} } \,.
\eea
We should stress again that our $g_{\Dst D\pi}$ and $g_{\DOst D\pi}$ are  $m_{D\pi}^2$-independent, and the entire dependence of the amplitude (\ref{eq:dG6}) on $m_{D\pi}^2$ is assumed to be described by the corresponding BW functions.

Similarly to the previous section [cf. around eq.~(\ref{eq:ha1})] we define the helicity amplitudes of the $\overline B\to V^\ast D^{\ast}$ decay as~\footnote{Please note that the polarization vector of $D^\ast$ is denoted by $\varepsilon_\mu$ while the one of the virtual $V^\ast$ is labelled by $\widetilde\varepsilon_\mu$. They are specified in Appendix~\ref{app:polarization}.}
\begin{subequations}
	\begin{align}
		 H_{\pm,0}& = \widetilde\varepsilon_{\pm,0}^{\mu\ast}\langle\Dst(\varepsilon_{\pm,0})| H_\mu |\Bbar\rangle \,,\\
		 &\nn\\
		 H_t &= \widetilde\varepsilon_t^{\mu*}\langle\Dst(\varepsilon_0)|H_\mu |\Bbar\rangle \,, \\
		 &\nn\\
		 H_0^\prime &= \widetilde\varepsilon_0^{\mu*}\langle\DOst| H_\mu |\Bbar\rangle \,,\\
		 &\nn\\
		H_t^\prime &= \widetilde\varepsilon_t^{\mu*}\langle\DOst| H_\mu |\Bbar\rangle \,,
	\end{align}
\end{subequations}
where $H_\mu$ stands for the hadronic part of the effective Hamiltonian [$\bar c \gamma_\mu(1-\gamma_5)b$ in the SM]. After using eq.~(\ref{eq:Heff}) and 
the definitions~(\ref{eq:B-Dst_FF_VA},\ref{eq:B-Dst_FF_VAbis},\ref{eq:B-Dst_FF_P},\ref{eq:B-Dst_FF_TT5}), the explicit expressions of our helicity amplitudes read: 
\begin{equation}\label{eq:eq00}
	\begin{split}
		H_\pm(q^2) =& i\biggl\{\mp\left[1+g_V-g_T(m_B+m_\Dst){T_1(q^2) \over V(q^2)}\right]{\sqrt{\lambda(m_B^2,m_\Dst^2,q^2)} \over m_B+m_\Dst}V(q^2)\biggr. \\
			    & \biggl.\quad-\left[1-g_A-g_{T5}(m_B-m_\Dst){T_2(q^2) \over A_1(q^2)}\right](m_B+m_\Dst)A_1(q^2)\biggr\} \,, \\
		 &\\
    H_0(q^2) =& -{i \over 2m_\Dst\sqrt{q^2}}\biggl\{\left[1-g_A-g_{T5}(m_B-m_\Dst){T_2(q^2) \over A_1(q^2)}\right] \\
			    & \quad\times (m_B+m_\Dst)(m_B^2-m_\Dst^2-q^2)A_1(q^2) \biggr. \\
			    & \biggl.-\left[1-g_A-g_{T5}\left((m_B+m_\Dst){T_2(q^2) \over A_2(q^2)} + {q^2 \over m_B-m_\Dst}{T_3(q^2) \over A_2(q^2)}\right)\right] \\
			    & \quad\times {\lambda(m_B^2,m_\Dst^2,q^2) \over m_B+m_\Dst}A_2(q^2) \biggr\} \,, \\
		 &\\
		  H_t(q^2) =& -i\left[1-g_A+g_P{q^2 \over m_b+m_c}\right]{\sqrt{\lambda(m_B^2,m_\Dst^2,q^2)} \over \sqrt{q^2}}A_0(q^2) \,,
	\end{split}
\end{equation}
and in a way analogous to eq.~(\ref{eq:ha1}) the helicity amplitudes parametrizing $\overline B \to D_0^\ast V^\ast$ decay are, 
\begin{equation}\label{eq:HA2}
	\begin{split}
		H_0^\prime(q^2) =& -i(1-g_A){\sqrt{\lambda(m_B^2,m_\DOst^2,q^2)} \over \sqrt{q^2}}F_1(q^2) \,, \\
		H_t^\prime(q^2) =& -i\left[1-g_A+g_P{q^2 \over m_b+m_c}\right]{m_B^2-m_\DOst^2 \over \sqrt{q^2}}F_0(q^2) \,. 
	\end{split}
\end{equation}
We stress once again that the $\mu$-dependence of $g_{S,P,T,T5}$, $m_{c,b}$ and $T_{1,2,3}(q^2)$ is implicit, and that the helicity amplitudes are, of course, scale independent.

\subsubsection{Partially integrated decay distributions}

Using the above definitions one can write the full five-fold distribution that not only includes the $B\to D^\ast$ transition but also the $B\to D^\ast_0$ one. The complete expression is provided in Appendix~\ref{app:2}. Here we will first focus on the separate distributions  on each of the three angles separately. Before spelling out these expressions, we first integrate over all angles to get
		\begin{equation}\label{eq:diffdecaystar}
			\begin{split}
				& {d^2\Gamma \over dq^2dm_{D\pi}^2} = {d^2\Gamma_0 \over dq^2dm_{D\pi}^2} + {d^2\Gamma_S \over dq^2dm_{D\pi}^2} = {G_F^2|V_{cb}|^2|\q|q^2 \over 96\pi^3m_B^2}\left(1-{m_\ell^2 \over q^2}\right)^2 \times \biggl\{\biggr. \\
				& \quad\quad \left[(|H_+|^2+|H_-|^2+|H_0|^2)\left(1+{m_\ell^2 \over 2q^2}\right) + {3 \over 2}{m_\ell^2 \over q^2}|H_t|^2 \right]|BW_\Dst(m_{D\pi}^2)|^2 \\
				& \quad\quad +\biggl.\left[|H_0^\prime|^2\left(1+{m_\ell^2 \over 2q^2}\right)+{3 \over 2}{m_\ell^2 \over q^2}|H_t^\prime|^2 \right]|BW_\DOst(m_{D\pi}^2)|^2\biggr\} \,,
			\end{split}
		\end{equation}
		where $\Gamma_0$ denotes the pure vector meson ($D^*$) contribution, while $\Gamma_S$ denotes the pure scalar meson ($D_0^*$) part. Obviously, after setting $H_{0,t}^\prime =0$ one retrieves the familiar differential decay rate for the pseudoscalar to vector meson semileptonic decay.

\begin{itemize}
	\item $\thl$ distribution: After integrating over $\chi$ and $\theta_D$ we get 
		\begin{equation}\label{eq:thetaellST}
			{d^3\Gamma \over dq^2dm_{D\pi}^2 d\cos\thl} = a_\thl+b_\thl\cos\thl+c_\thl\cos^2\thl \,,
		\end{equation}
where the coefficient functions $a_\thl$,  $b_\thl$, $c_\thl$ depend on $q^2$ and on $m_{D\pi}^2$, and they read
		\begin{subequations}
			\begin{align}
				\begin{split}
					& a_\thl(q^2,\,m_{D\pi}^2) = {G_F^2|V_{cb}|^2|\q|q^2 \over 256\pi^3m_B^2}\left(1-{m_\ell^2 \over q^2}\right)^2 \times \biggl\{\biggr. \\
					& \quad\quad \left[(|H_+|^2+|H_-|^2)\left(1+{m_\ell^2 \over q^2}\right) + 2\left(|H_0|^2+{m_\ell^2 \over q^2}|H_t|^2\right)\right]|BW_\Dst(m_{D\pi}^2)|^2 \\
					& \quad\quad \biggl.+2\left[|H_0^\prime|^2+{m_\ell^2 \over q^2}|H_t^\prime|^2\right]|BW_\DOst(m_{D\pi}^2)|^2 \biggr\} \,,
				\end{split}   \\
				& \nonumber \\
				\begin{split}
					& b_\thl(q^2,\,m_{D\pi}^2) = {G_F^2|V_{cb}|^2|\q|q^2 \over 128\pi^3m_B^2}\left(1-{m_\ell^2 \over q^2}\right)^2 \times \biggl\{\biggr. \\
					& \quad\quad \left[|H_+|^2-|H_-|^2 + 2{m_\ell^2 \over q^2}\Re[H_0H_t^*]\right]|BW_\Dst(m_{D\pi}^2)|^2 \\
					& \quad\quad \biggl. +2{m_\ell^2 \over q^2}\Re[H_0^\prime H_t^{\prime*}]|BW_\DOst(m_{D\pi}^2)|^2 \biggr\} \,,
				\end{split}  
							\end{align}
			\begin{align}
				\begin{split}
					& c_\thl(q^2,\,m_{D\pi}^2) = {G_F^2|V_{cb}|^2|\q|q^2 \over 256\pi^3m_B^2}\left(1-{m_\ell^2 \over q^2}\right)^3 \times \biggl\{\biggr. \\
					& \quad\quad \left[|H_+|^2+|H_-|^2-2|H_0|^2\right]|BW_\Dst(m_{D\pi}^2)|^2 \biggl.-2|H_0^\prime|^2|BW_\DOst(m_{D\pi}^2)|^2 \biggr\} \,.
				\end{split}
			\end{align}
		\end{subequations}

	\item $\thD$ distribution: If, instead, we integrate eq.~(\ref{eq:dG5}) in $\chi$ and in $\theta_\ell$ we get 
		\begin{equation}\label{eq:thetaDST}
			{d^3\Gamma \over dq^2dm_{D\pi}^2 d\cos\thD} = a_\thD+b_\thD\cos\thD+c_\thD\cos^2\thD \,,
		\end{equation}
		where
		\begin{subequations}
			\begin{align}
				\begin{split}
					& a_\thD(q^2,\,m_{D\pi}^2) = {G_F^2|V_{cb}|^2|\q|q^2 \over 128\pi^3m_B^2}\left(1-{m_\ell^2 \over q^2}\right)^2 \times \biggl\{\biggr. \\
					& \quad\quad \left[(|H_+|^2+|H_-|^2)\left(1+{m_\ell^2 \over 2q^2}\right)\right]|BW_\Dst(m_{D\pi}^2)|^2 \\
					& \quad\quad \biggl.+{2 \over 3}\left[|H_0^\prime|^2\left(1+{m_\ell^2 \over 2q^2}\right)+{3 \over 2}{m_\ell^2 \over q^2}|H_t^\prime|^2\right]|BW_\DOst(m_{D\pi}^2)|^2 \biggr\} \,,
				\end{split} \\
				& \nonumber \\
				\begin{split}
					& b_\thD(q^2,\,m_{D\pi}^2) = {\sqrt3 G_F^2|V_{cb}|^2|\q|q^2 \over 96\pi^3m_B^2}\left(1-{m_\ell^2 \over q^2}\right)^2 \times \\
					& \quad\quad \Re\left[\left(H_0 H_0^{\prime*}\left(1+{m_\ell^2 \over 2q^2}\right) + {3 \over 2}{m_\ell^2 \over q^2}H_t H_t^{\prime*}\right)BW_\Dst(m_{D\pi}^2)BW_\DOst^*(m_{D\pi}^2)\right] \,,
				\end{split} \\
				& \nonumber \\
				\begin{split}
					& c_\thD(q^2,\,m_{D\pi}^2) = -{G_F^2|V_{cb}|^2|\q|q^2 \over 128\pi^3m_B^2}\left(1-{m_\ell^2 \over q^2}\right)^2 \times \\
					& \quad\quad \left[(|H_+|^2+|H_-|^2-2|H_0|^2)\left(1+{m_\ell^2 \over 2q^2}\right) - 3{m_\ell^2 \over q^2}|H_t|^2\right]|BW_\Dst(m_{D\pi}^2)|^2 \,.
				\end{split}
			\end{align}
		\end{subequations}

		\item $\chi$ distribution: Finally, integration over $\theta_D$ and $\theta_\ell$ results in,
		\begin{equation}\label{eq:chiST}
			{d^3\Gamma \over dq^2dm_{D\pi}^2 d\chi} = a_\chi+b_\chi^c\cos\chi+b_\chi^s\sin\chi+c_\chi^c\cos2\chi+c_\chi^s\sin2\chi \,,
		\end{equation}
		with the coefficient functions, 
		\begin{subequations}
			\begin{align}
				\begin{split}
					& a_\chi(q^2,\,m_{D\pi}^2) = {G_F^2|V_{cb}|^2|\q|q^2 \over 192\pi^4m_B^2}\left(1-{m_\ell^2 \over q^2}\right)^2 \times \biggl\{\biggr. \\
					& \quad\quad \left[(|H_+|^2+|H_-|^2+|H_0|^2)\left(1+{m_\ell^2 \over 2q^2}\right) + {3 \over 2}{m_\ell^2 \over q^2}|H_t|^2 \right]|BW_\Dst(m_{D\pi}^2)|^2 \\
				& \quad\quad +\biggl.\left[|H_0^\prime|^2\left(1+{m_\ell^2 \over 2q^2}\right)+{3 \over 2}{m_\ell^2 \over q^2}|H_t^\prime|^2 \right]|BW_\DOst(m_{D\pi}^2)|^2\biggr\} \,,
				\end{split} 
			\end{align}
			\begin{align}
				\begin{split}
					& b_\chi^c(q^2,\,m_{D\pi}^2) = -{\sqrt3 G_F^2|V_{cb}|^2|\q|q^2 \over 2048\pi^2m_B^2}\left(1-{m_\ell^2 \over q^2}\right)^2 \times \\
					& \quad\quad \Re\left[\left( (H_+ -H_- )H_0^{\prime*}-{m_\ell^2 \over q^2}(H_+ +H_- ) H_t^{\prime*})\right)BW_\Dst(m_{D\pi}^2)BW_\DOst^*(m_{D\pi}^2)\right] \,,
				\end{split} \\
				& \nonumber \\
				\begin{split}
					& b_\chi^s(q^2,\,m_{D\pi}^2) = -{\sqrt3 G_F^2|V_{cb}|^2|\q|q^2 \over 2048\pi^2m_B^2}\left(1-{m_\ell^2 \over q^2}\right)^2 \times \\
					& \quad\quad \Im\left[\left( (H_+ +H_- )H_0^{\prime*}-{m_\ell^2 \over q^2}(H_+ -H_- )H_t^{\prime*})\right)BW_\Dst(m_{D\pi}^2)BW_\DOst^*(m_{D\pi}^2)\right] \,,
				\end{split} \\
				& \nonumber \\
				\begin{split}
					& c_\chi^c(q^2,\,m_{D\pi}^2) = -{G_F^2|V_{cb}|^2|\q|q^2 \over 192\pi^4m_B^2}\left(1-{m_\ell^2 \over q^2}\right)^3 \times \Re\left[H_+H_-^*\right]|BW_\Dst(m_{D\pi}^2)|^2 \,,
				\end{split} \\
				& \nonumber \\
				\begin{split}
					& c_\chi^s(q^2,\,m_{D\pi}^2) = -{G_F^2|V_{cb}|^2|\q|q^2 \over 192\pi^4m_B^2}\left(1-{m_\ell^2 \over q^2}\right)^3 \times \Im\left[H_+H_-^*\right]|BW_\Dst(m_{D\pi}^2)|^2 \,.
				\end{split}
			\end{align}
		\end{subequations}
\end{itemize}
We emphasize once again that all of the above distributions are written in terms of helicity amplitudes and therefore are completely general. Only after choosing a specific NP scenario the helicity amplitudes are expressed in terms of form factors, kinematic variables and the NP couplings, which is where the model dependence enters the discussion. 

From any of the above angular distributions~(\ref{eq:thetaellST},\ref{eq:thetaDST},\ref{eq:chiST}) one can easily reproduce the differential decay rate~(\ref{eq:diffdecaystar}).
Helicity amplitudes, for our specific choice of $\Heff$ in eq.~(\ref{eq:Heff}), are explicitly given in eqs.~(\ref{eq:eq00},\ref{eq:HA2}). 

\subsection{Comment on the pollution of $\btodstar$ by  $\btodscal$} 

\begin{figure}[t!]
\begin{center}
{\resizebox{8.4cm}{!}{\includegraphics{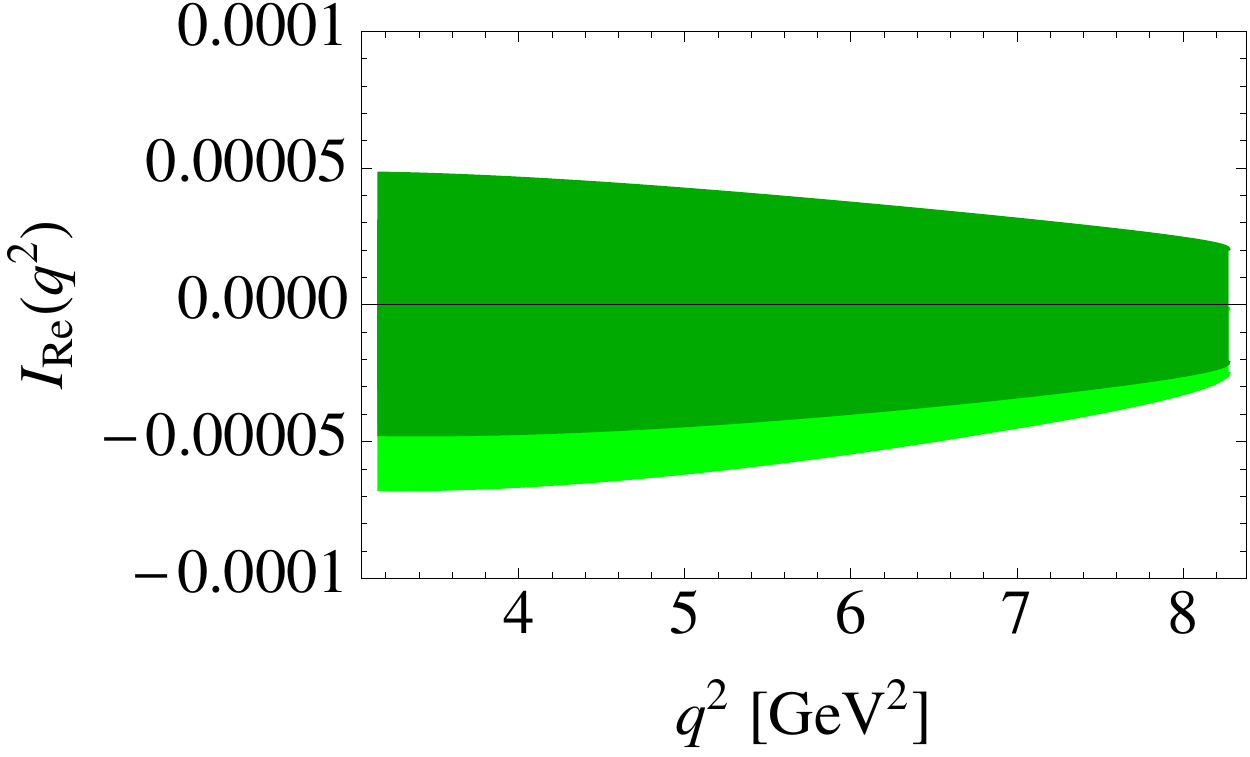}}}~{\resizebox{8cm}{!}{\includegraphics{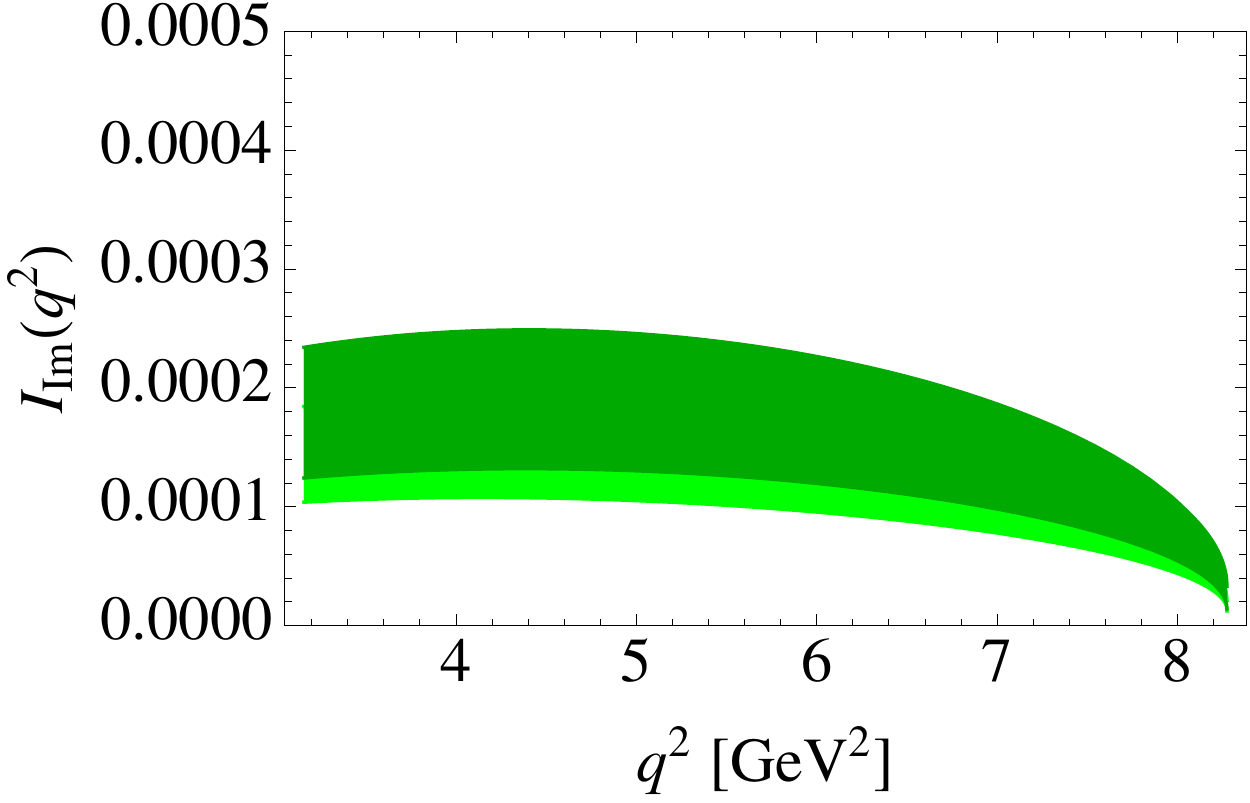}}} 
\caption{\label{fig:REIM}\footnotesize{\sl 
The quantities  $I_{\rm Re}(q^2)$ and $I_{\rm Im}(q^2)$ defined in eq.~(\ref{eq:Ireim})] are plotted for two values of $\delta$: brighter regions for $\delta=100$~MeV and darker regions for $\delta = 300$~MeV.   }} 
\end{center}
\end{figure}

As it can be seen from the above expressions, the $D\pi$ pair emerging from the $\btodscal$ can be mistakenly identified as an $S$-wave contribution to the $\btodstar$ 
if the range around the $D^\ast$-resonance, $m_{D\pi}^2 \in [(m_\Dst -\delta)^2 , (m_\Dst +\delta)^2]$, is relatively large with respect to the width of the $\DOst$ state. However, knowing that the experimentally measured~\cite{PDG}:
\begin{align}
m_{D^{\ast\pm}_0}&=2403(40)\ \mev\,,\qquad &\Gamma(D^{\ast\pm}_0)&=283(40)~\mev\,,\nn\\
  m_{D^{\ast 0}_0}&=2318(29)\ \mev\,,\qquad &\Gamma(D^{\ast 0}_0)&=267(40)~\mev\,,\nn \\
 m_{D^{\ast\pm}}&=2010\ \mev\,,\qquad &\Gamma(D^{\ast\pm})&=83(2)~\kev\,,\nn\\
  m_{D^{\ast 0}}&=2007\ \mev\,,\qquad &\Gamma(D^{\ast 0})&=60(3)~\kev\,,
\end{align}
we see that there is no interference between the two decays as long as $\delta$ is kept smaller than about $ \Gamma(\DOst)/2$, which is relatively easy to ensure in experiments since the width of $D^\ast$ is very small. 
This reasoning relies on the assumption that the shape of the scalar state $\DOst$ can be described by the BW formula~(\ref{eq:bw}), which is not {\sl a priori} clear for a pair of hadrons in their $S$-wave. For example, large deviations from the BW shape in the case of $(K\pi)_{S-\rm wave}$  turned out to be very important in the analysis of $D^+\to \overline K^{\ast 0}(\to K^-\pi^+) \ell \nu_\ell$~\cite{Link:2002ev}. To check whether or not a similar phenomenon appears also in the case of $(D\pi)_{S-\rm wave}$, we find it informative to consider the angular distribution in $\chi$, cf. eq.~(\ref{eq:chiST}), because the terms proportional to $\cos\chi$ and $\sin\chi$ measure the real and imaginary part of the interference between the  $(D\pi)$ coming from $\btodstar$, coupling to non-transversely polarized virtual $W$, and those emerging from $\btodscal$. More specifically, we consider the quantities
\bea\label{eq:Ireim}
&&I_{\rm Re}(q^2) = {1\over d\Gamma/dq^2 } \int_{(m_\Dst -\delta)^2}^{(m_\Dst +\delta)^2} b_\chi^c(q^2,m_{D\pi}^2)dm_{D\pi}^2 \,,\nn\\
&& I_{\rm Im}(q^2) = {1\over d\Gamma/dq^2  } \int_{(m_\Dst -\delta)^2}^{(m_\Dst +\delta)^2} b_\chi^s(q^2,m_{D\pi}^2)dm_{D\pi}^2 \,,
\eea
where $\Gamma$ encapsulates the $\btodstar$ events and the leakage of $\btodscal$ that are included in the sample if $\delta$ is large enough. 
If we assume that both resonances can be described by the BW formula, we find that for either $\delta = 100$~MeV or $\delta = 300$~MeV, both above quantities remain negligibly small, cf. Fig.~\ref{fig:REIM}.
%%%%%%%%%%%%%%%%%%%%%%%%%%%%%%%%%%%%%%%%%%%%%%%%%%%%%%%%%%%
This conclusion remains as such not only in the SM [$g_{S,V,P,A,T,T5}=0$] but also in its extensions [$g_{S,V,P,A,T,T5}\approx 1$]. It is important to check whether or not this is indeed the case in realistic experimental studies 
because the nonzero values of $I_{\rm Re}(q^2)$ and/or $I_{\rm Im}(q^2)$ would suggest that the $(D\pi)_{S-\rm wave}$ amplitude contains contributions that are not captured by the BW formula. 
If it turns out to be zero, this would represent an important check of non-pollution of the sample of $\btodstar$, which is prerequisite for a precision determination of $\vert V_{cb}\vert$ and/or for distinguishing the effects of NP.  
Notice again that in producing the plot in Fig.~\ref{fig:REIM} we used the values of the form factors given in Refs.~\cite{melikhov,cheng}.~\footnote{We use Refs.~\cite{melikhov,cheng} because they contain the full list of form factors needed for our discussion. 
In this paper all the plots will be obtained by using the $B\to D^\ast$ transition form factors from Ref.~\cite{melikhov}, and $B\to D_0^\ast$ ones from Ref.~\cite{cheng}. }

Finally, from eq.~(\ref{eq:thetaDST}) we see that the forward-backward asymmetry in $\theta_D$, 
\begin{align}
A_{FB}^{\theta_D}(q^2)& ={\displaystyle{ \int_{-1}^0  {d^2\Gamma \over dq^2d\cos\thD} d\cos\thD  -  \int_{0}^1 {d^2\Gamma \over dq^2d\cos\thD} d\cos\thD} \over\displaystyle{d\Gamma\over dq^2} } = - { b_\thD (q^2)  \over {d\Gamma/dq^2} } \nn\\
& \propto \int_{(m_\Dst -\delta)^2}^{(m_\Dst +\delta)^2}  \Re\left[ \left(H_0 H_0^{\prime*}\left(1+{m_\ell^2 \over 2q^2}\right) + {3 \over 2}{m_\ell^2 \over q^2}H_t H_t^{\prime*}\right)BW_\Dst(m_{D\pi}^2)BW_\DOst^*(m_{D\pi}^2)\right],
 \end{align}
which is obviously only non-zero in the case of pollution of the $\btodstar$ sample by the $\btodscal$ events. If experimentally feasible, this quantity could be a good way to address this issue which is one of the major worries in assessing the systematic uncertainties of the experimental results. 
In what follows we will assume that the $(D\pi)$ emerging from the $\btodstar$ decay are not polluted by those coming from $\btodscal$.

\subsection{Eleven Observables in $\btodstar$}

Similarly to what we discussed in the case of $\btod$, where the three independent structures were probed by three different observables, we can now form $11$ different quantities 
that can be studied in the full angular analysis of $\btodstar$ decay, the expression of which is given in eq.~(\ref{eq:MDst2}).

\noindent 
{\underline{1. Differential decay rate}}:
\begin{align}
				\begin{split}
					 {d\Gamma\over dq^2}(\btodstar )= &{G_F^2|V_{cb}|^2|\q|q^2 \over 96\pi^3 m_B^2}\left(1-{m_\ell^2 \over q^2}\right)^2 \times   \\
					& \left[(|H_+|^2+|H_-|^2+|H_0|^2)\left(1+{m_\ell^2 \over 2q^2}\right) + {3 \over 2}{m_\ell^2 \over q^2}|H_t|^2 \right] \,.
				\end{split} 
\end{align}
{\underline{2. Forward-Backward asymmetry}}: 
\begin{align}
A_{FB}^{D^\ast}(q^2)& ={\displaystyle{ \int_{0}^1  {d^2\Gamma \over dq^2d\cos\thl} d\cos\thl  -  \int_{-1}^0  {d^2\Gamma \over dq^2d\cos\thl} d\cos\thl} \over\displaystyle{d\Gamma/dq^2} }=  { \displaystyle{b_\thl(q^2)}  \over \displaystyle{d\Gamma/dq^2} }\nn\\
&= {\displaystyle{G_F^2|V_{cb}|^2|\q|q^2} \over \displaystyle{128\pi^3m_B^2 (d\Gamma/dq^2)}}\left(1-{m_\ell^2 \over q^2}\right)^2 \times  \left[|H_+|^2-|H_-|^2 + 2{m_\ell^2 \over q^2}\Re[H_0H_t^*]\right].
 \end{align}
{\underline{3. Lepton-polarization asymmetry}}: We define the differential decay rates, $d\Gamma^\pm/dq^2$, with the spin of the charged lepton projected along the $z$-axis and with $\lambda_\ell=\pm1/2$. In other words,
\begin{align}
					 {d\Gamma^-\over dq^2}(\btodstar )= &{G_F^2|V_{cb}|^2|\q|q^2 \over 96\pi^3 m_B^2}\left(1-{m_\ell^2 \over q^2}\right)^2 \times   \left( |H_+|^2+|H_-|^2+|H_0|^2 \right) \,,\nn\\
					 {d\Gamma^+\over dq^2}(\btodstar )= &{G_F^2|V_{cb}|^2|\q|q^2 \over 96\pi^3 m_B^2}\left(1-{m_\ell^2 \over q^2}\right)^2 {m_\ell^2 \over 2 q^2} \times   \left( |H_+|^2+|H_-|^2+|H_0|^2 +3 |H_t|^2\right) \,,
\end{align}
and the lepton polarization asymmetry reads, 
\bea
A_{\lambda_\ell}^{D^\ast}(q^2) = {\displaystyle{d\Gamma^-/dq^2} - \displaystyle{d\Gamma^+/dq^2} \over \displaystyle{d\Gamma/dq^2} } =1 - 2 { \displaystyle{d\Gamma^+/dq^2} \over \displaystyle{d\Gamma/dq^2}}.
\eea
{\underline{4. Partial decay rate according to the polarization of $D^\ast$}}: Splitting the decay rate according to the polarization of the $D^\ast$-meson amounts to,
\bea
{d\Gamma_L \over dq^2} = \frac{2}{3} \left[ a_\thD (q^2) + c_\thD (q^2) \right], \quad {d\Gamma_T \over dq^2} = \frac{4}{3} a_\thD (q^2) ,
\eea
where the functions on the r.h.s. are given in eq.~(\ref{eq:thetaDST}). One of these components is independent, while the other can be obtained from $\Gamma = \Gamma_L+\Gamma_T$. To cancel the CKM and kinematic factors
we can define 
\bea
R_{L,T} = {\displaystyle{d\Gamma_L/ dq^2}\over \displaystyle{d\Gamma_T/ dq^2}} = {|H_0|^2 + 3 |H_t|^2 \left[ 1- 1/(1+m_\ell^2/2q^2)\right] \over |H_+|^2+|H_-|^2}.
\eea
{\underline{ 5. $A_5$}}: We see that three of the above observables involve the squares of the absolute values of four helicity amplitudes, $|H_{+,-,0,t}|^2$. We can build the fourth observable as follows. 
After integrating in $\chi$, we consider 
\bea \label{eq:phi}\Phi(q^2, \thD) = 
{\displaystyle{ \int_{-1}^0  {d^3\Gamma \over dq^2d\cos\thD d\cos\thl} d\cos\thl  -  \int_{0}^1 {d^3\Gamma \over dq^2d\cos\thD d\cos\thl} d\cos\thl}  },
\eea 
and then integrate in $\thD$ as,
\bea 
A_5(q^2) &=&  { \displaystyle{ \left[ 7 \int_{-1/2}^{1/2}-\int_{1/2}^{1} - \int_{-1}^{-1/2}\right] \Phi(q^2, \thD) \ d\cos\thD   } \over\displaystyle{d\Gamma/dq^2} }\nn\\
&=& -
{9 G_F^2|V_{cb}|^2|\q|q^2 \over 256\pi^3 m_B^2 (d\Gamma/dq^2)}\left(1-{m_\ell^2 \over q^2}\right)^2\biggl[ |H_+|^2 - |H_-|^2 \biggr] .
\eea 
{\underline{6. and 7. $C_\chi(q^2)$ and $S_\chi(q^2)$}}: From the distribution in $\chi$~(\ref{eq:chiST}), we see that $d\Gamma/dq^2=2\pi a_\chi (q^2)$, 
while from the terms proportional to $\sin2\chi$ and $\cos 2\chi$ one can get the additional information about the real and imaginary part of $H_+H_-^*$. To that end we define,   
\bea
C_\chi(q^2) &=& {c^c_\chi(q^2)\over a^c_\chi(q^2)} =-  { \left(1-{m_\ell^2 \over q^2}\right) \ \Re[H_+H_-^*]
\over (|H_+|^2+|H_-|^2+|H_0|^2)\left(1+{m_\ell^2 \over 2q^2}\right) + {3 \over 2}{m_\ell^2 \over q^2}|H_t|^2 
} \,, \nn\\
S_\chi(q^2) &=& {c^s_\chi(q^2)\over a^c_\chi(q^2)} =-  { \left(1-{m_\ell^2 \over q^2}\right) \ \Im[H_+H_-^*]
\over (|H_+|^2+|H_-|^2+|H_0|^2)\left(1+{m_\ell^2 \over 2q^2}\right) + {3 \over 2}{m_\ell^2 \over q^2}|H_t|^2 
}
\,, \eea
{\underline{8. and 9.  $A_8$ and $A_9$}}: We now first integrate the full distribution in $\thl$ and then define
\bea \label{eq:phi2}\widetilde \Phi(q^2, \chi ) = 
{\displaystyle{ \int_{-1}^0  {d^3\Gamma \over dq^2d\chi d\cos\thD} d\cos\thD  -  \int_{0}^1 {d^3\Gamma \over dq^2d\chi d\cos\thD} d\cos\thD}  },
\eea 
from which we can build the following two quantities
\begin{align}
A_8(q^2)&=  { \displaystyle{ \left[ \int_{0}^{\pi}-\int_{\pi}^{2\pi} \right] \widetilde \Phi(q^2, \chi) \ d\chi  } \over\displaystyle{d\Gamma/dq^2} }\nn\\
& = { G_F^2|V_{cb}|^2|\q|q^2 \over 128\pi^3 m_B^2}{\left(1-{m_\ell^2 /q^2}\right)^2\over d\Gamma/dq^2} \Im\left[ (H_+  + H_- )H_0^* - {m_\ell^2 \over q^2}\left(H_+  - H_- \right) H_t^* \right],\\
A_9(q^2)&=  - { \displaystyle{ \left[ \int_{\pi/2}^{3\pi/2}-\int_{0}^{\pi/2} -\int_{3\pi/2 }^{2\pi}  \right] \widetilde \Phi(q^2, \chi) \ d\chi  } \over\displaystyle{d\Gamma/dq^2} }\nn\\
& = { G_F^2|V_{cb}|^2|\q|q^2 \over 128\pi^3 m_B^2}{\left(1-{m_\ell^2 /q^2}\right)^2\over d\Gamma/dq^2} \Re\left[ (H_+  - H_- )H_0^* - {m_\ell^2 \over q^2}\left(H_+ + H_- \right) H_t^* \right].
\end{align}
{\underline{10. and 11.  $A_{10}$ and $A_{11}$}}: Finally, by forming the quantity
\bea \label{eq:phi3} \phi(q^2, \chi ,\thl ) &=& 
{\displaystyle{ \int_{-1}^0  {d^4\Gamma \over dq^2d\chi d\cos\thl d\cos\thD} d\cos\thD  -  \int_{0}^1 {d^4\Gamma \over dq^2d\chi d\cos\thl d\cos\thD} d\cos\thD}  },\nn\\
\widetilde \phi(q^2, \chi ) &=& 
{\displaystyle{ \int_{-1}^0  \phi(q^2, \chi ,\thl ) d\cos\thl  -  \int_{0}^1 \phi(q^2, \chi ,\thl ) d\cos\thl}  } \,
\eea
we can isolate the remaining two terms from the full angular distribution~(\ref{eq:MDst2}) as,
\begin{align}
A_{10}(q^2)&=  { \displaystyle{ \left[ \int_{0}^{\pi}-\int_{\pi}^{2\pi} \right] \widetilde \phi(q^2, \chi) \ d\chi  } \over\displaystyle{d\Gamma/dq^2} }\nn\\
& = -{ G_F^2|V_{cb}|^2|\q|q^2 \over 96\pi^4 m_B^2}{\left(1-{m_\ell^2 /q^2}\right)^3\over d\Gamma/dq^2} \Im[ (H_+ - H_- ) H_0^*], 
\end{align}
\begin{align}
A_{11}(q^2)&=  { \displaystyle{ \left[ \int_{\pi/2}^{3\pi/2}-\int_{0}^{\pi/2} -\int_{3\pi/2 }^{2\pi}  \right] \widetilde \phi(q^2, \chi) \ d\chi  } \over\displaystyle{d\Gamma/dq^2} }\nn\\
& = { G_F^2|V_{cb}|^2|\q|q^2 \over 96\pi^4 m_B^2}{\left(1-{m_\ell^2 /q^2}\right)^3\over d\Gamma/dq^2} \Re[ (H_+  + H_- ) H_0^*].
\end{align}

\begin{figure}[t!]
\begin{center}
\hspace*{-9mm}{\resizebox{5.4cm}{!}{\includegraphics{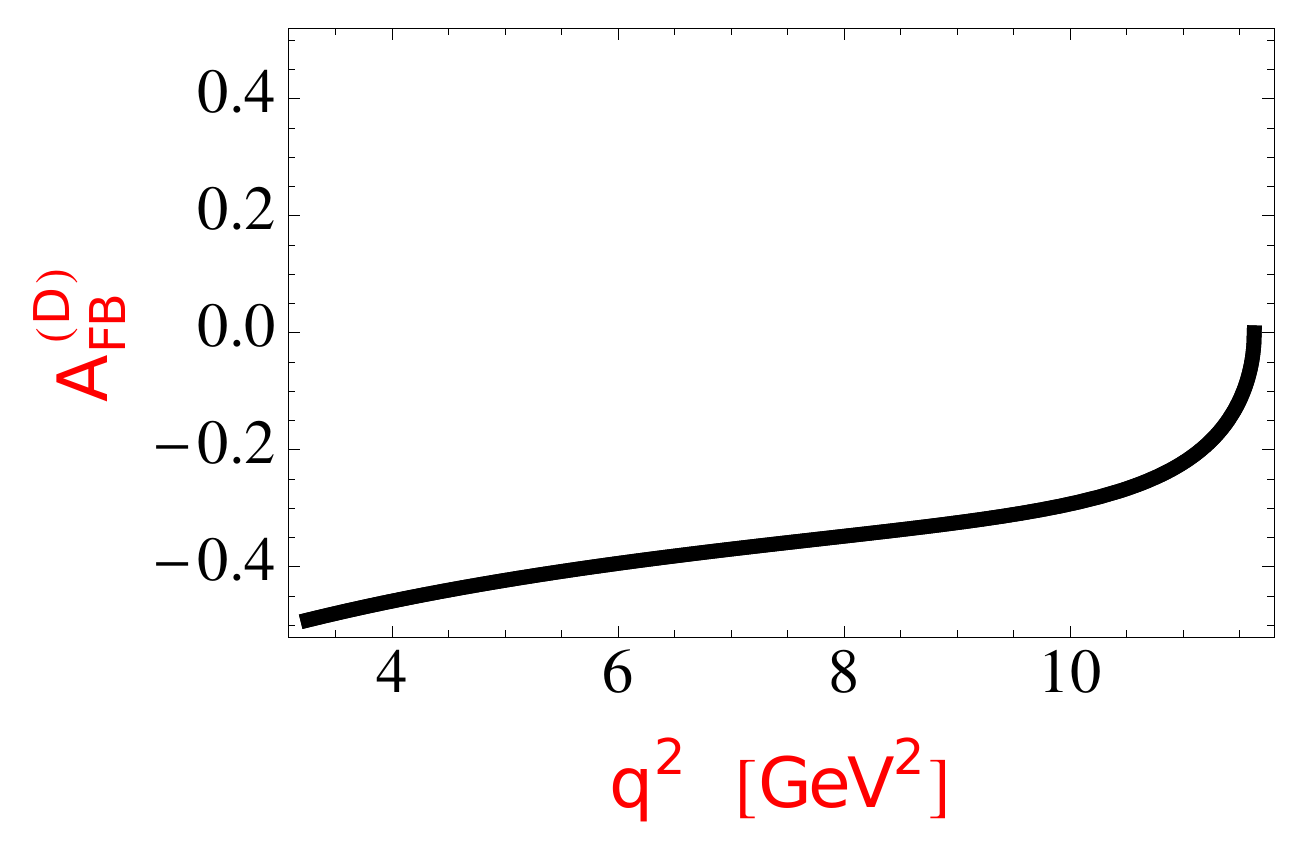}}}~{\resizebox{5.4cm}{!}{\includegraphics{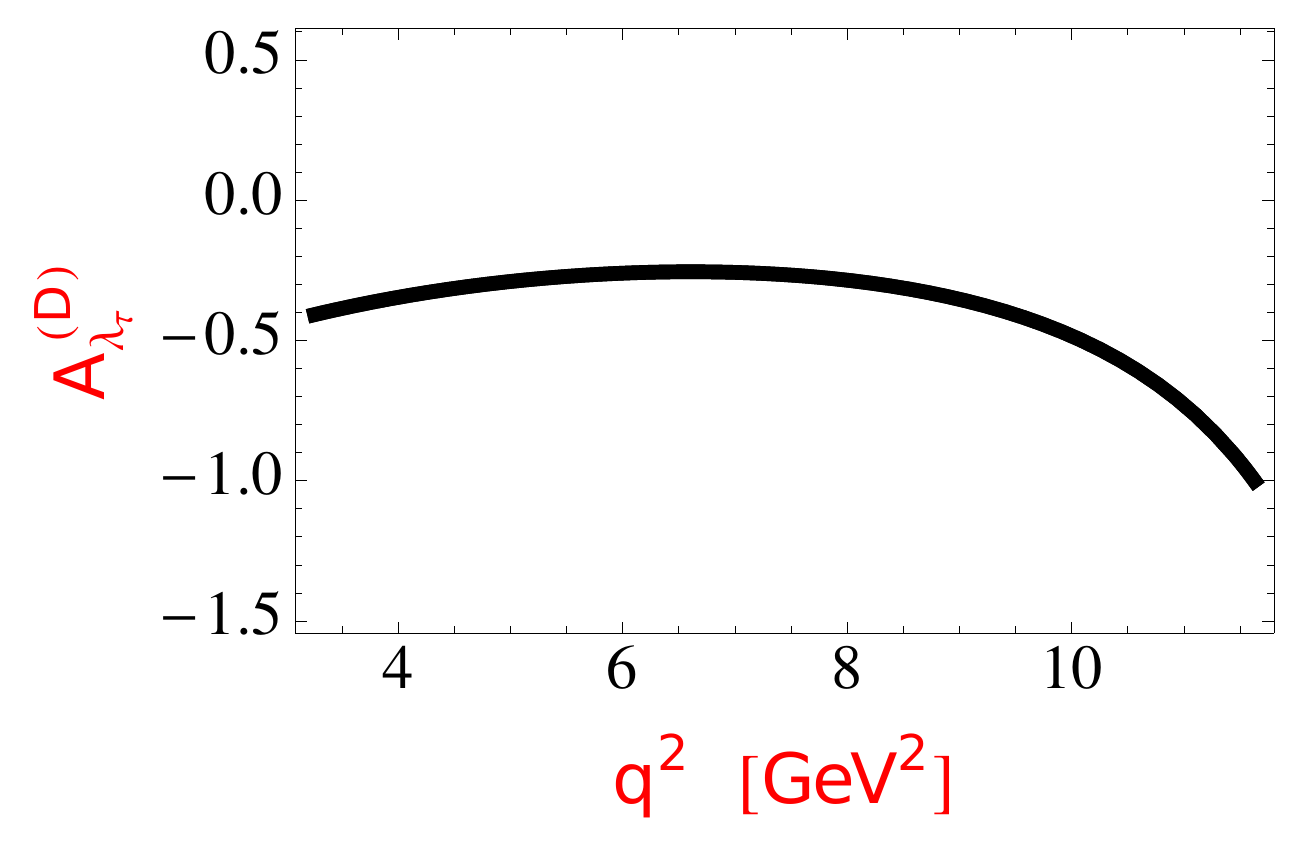}}}~{\resizebox{5.4cm}{!}{\includegraphics{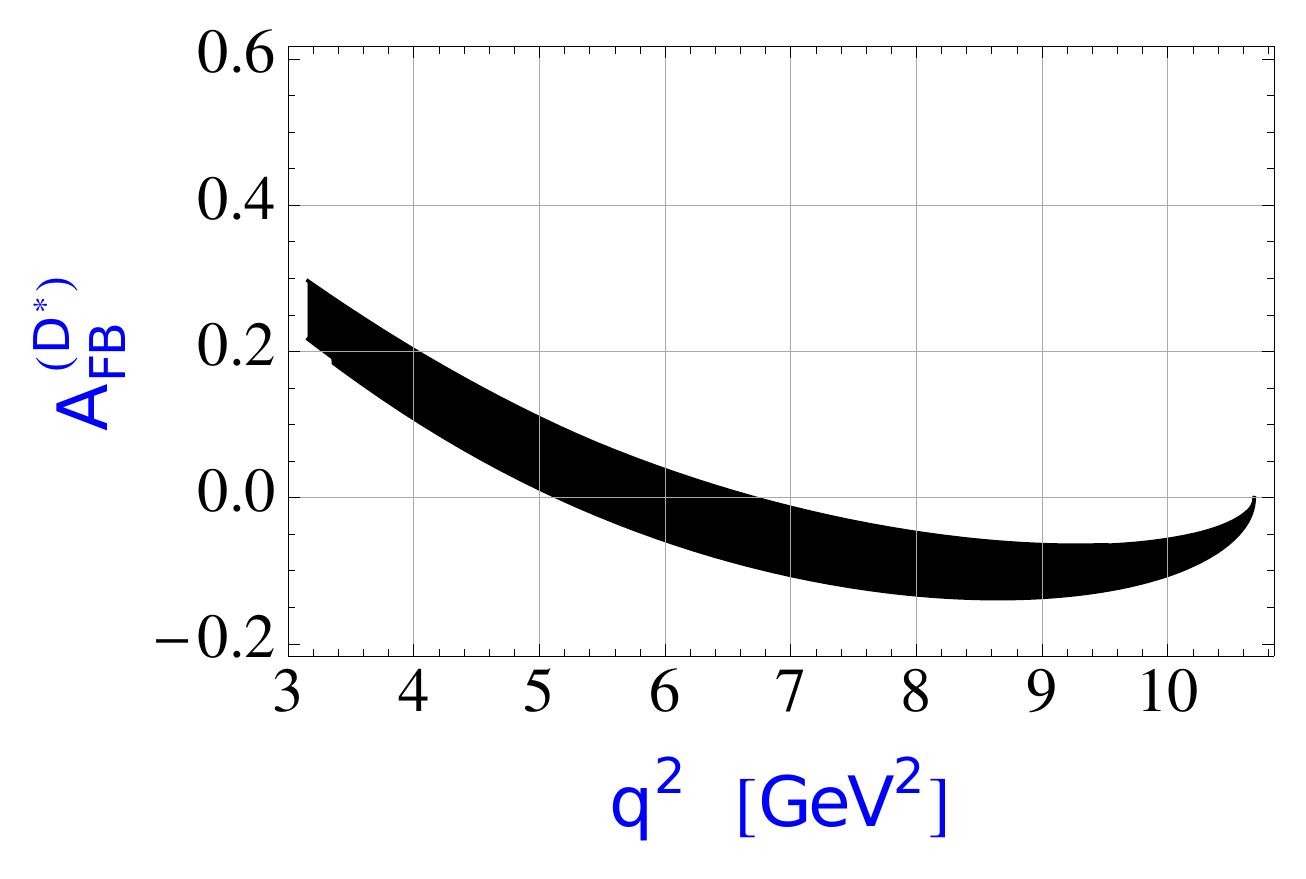}}} \\
\hspace*{-9mm}{\resizebox{5.5cm}{!}{\includegraphics{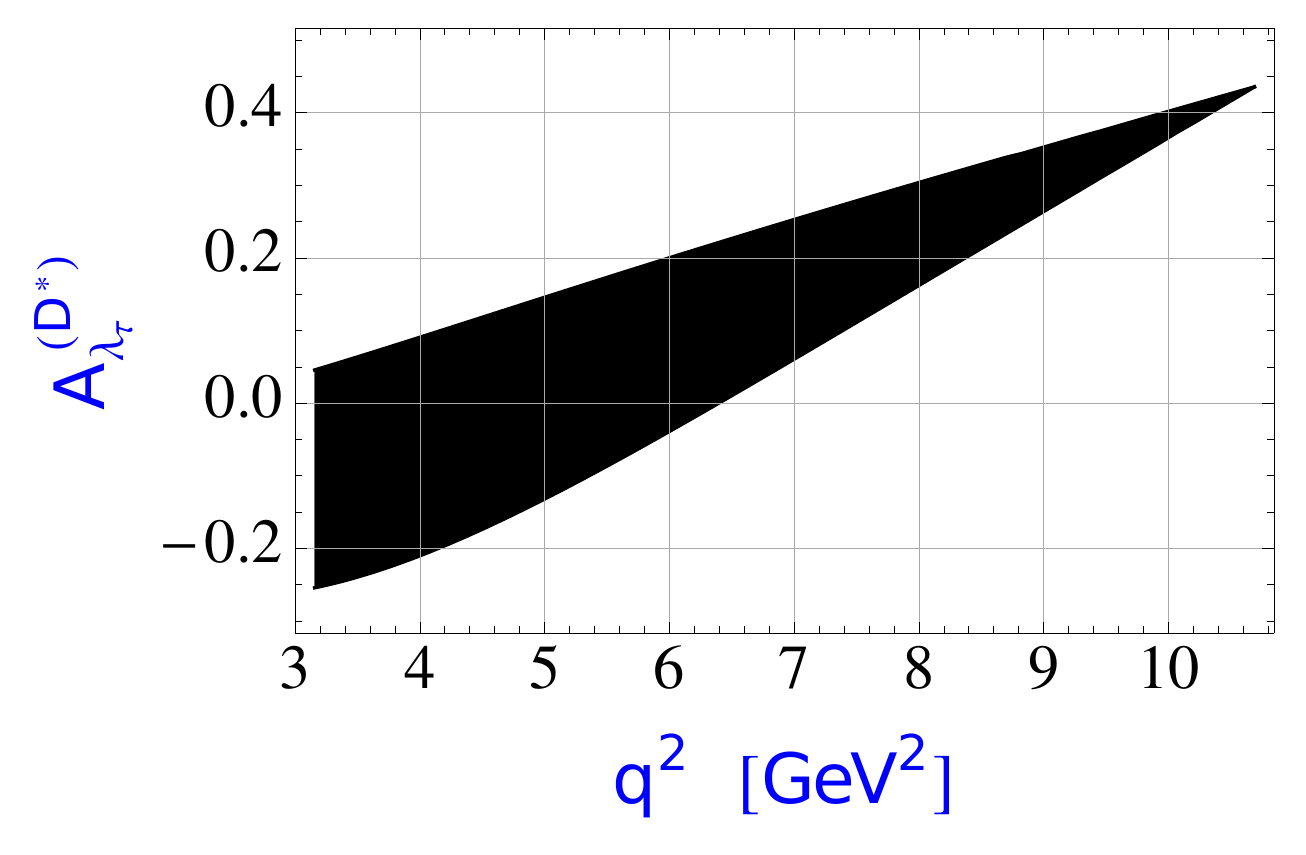}}}~{\resizebox{5.1cm}{!}{\includegraphics{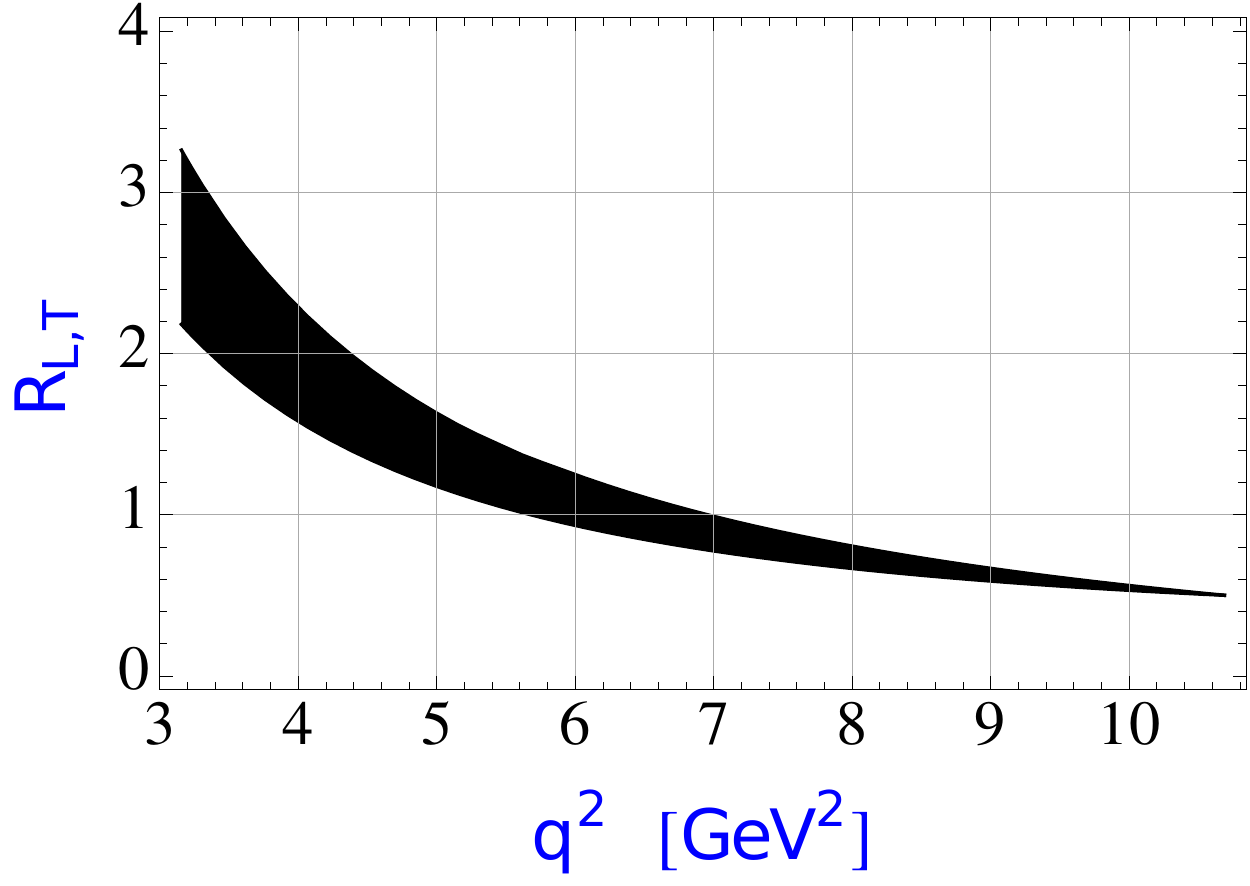}}}~{\resizebox{5.4cm}{!}{\includegraphics{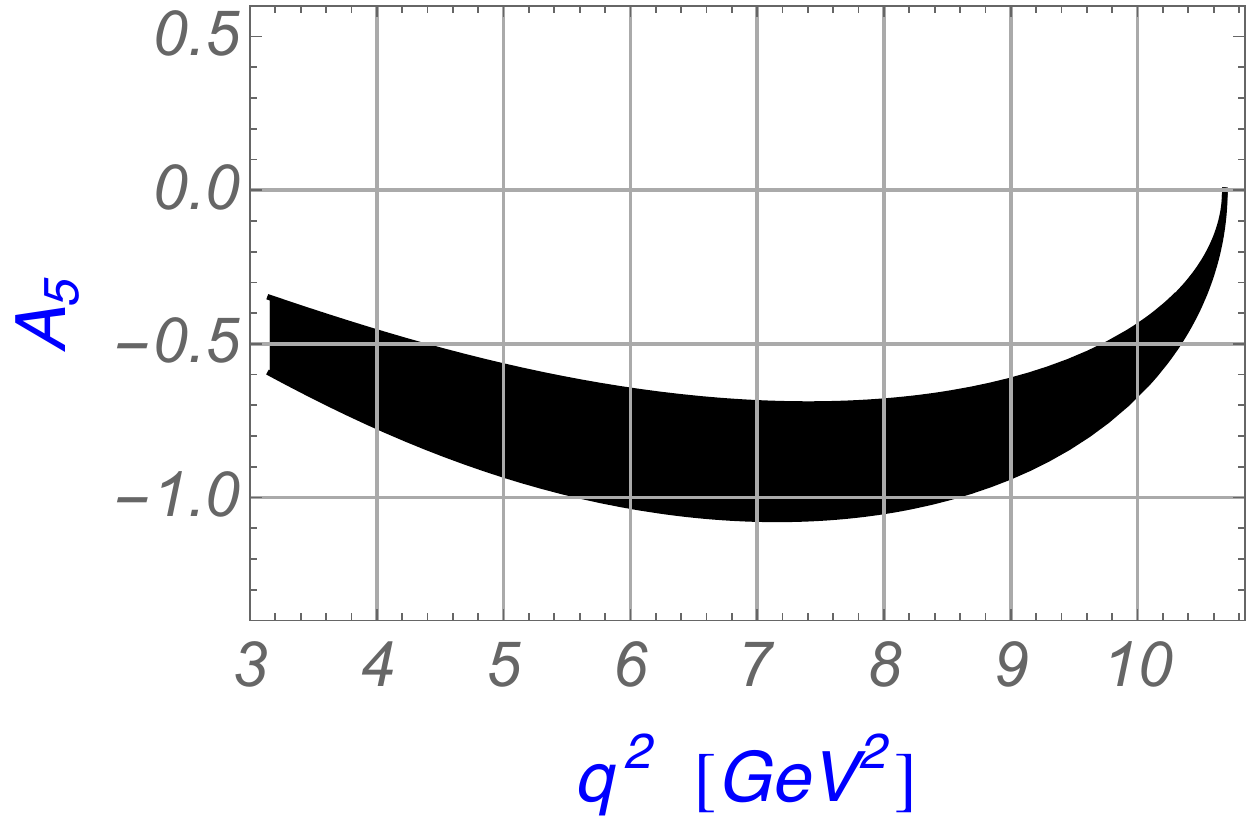}}} \\
\hspace*{-9mm}{\resizebox{5.4cm}{!}{\includegraphics{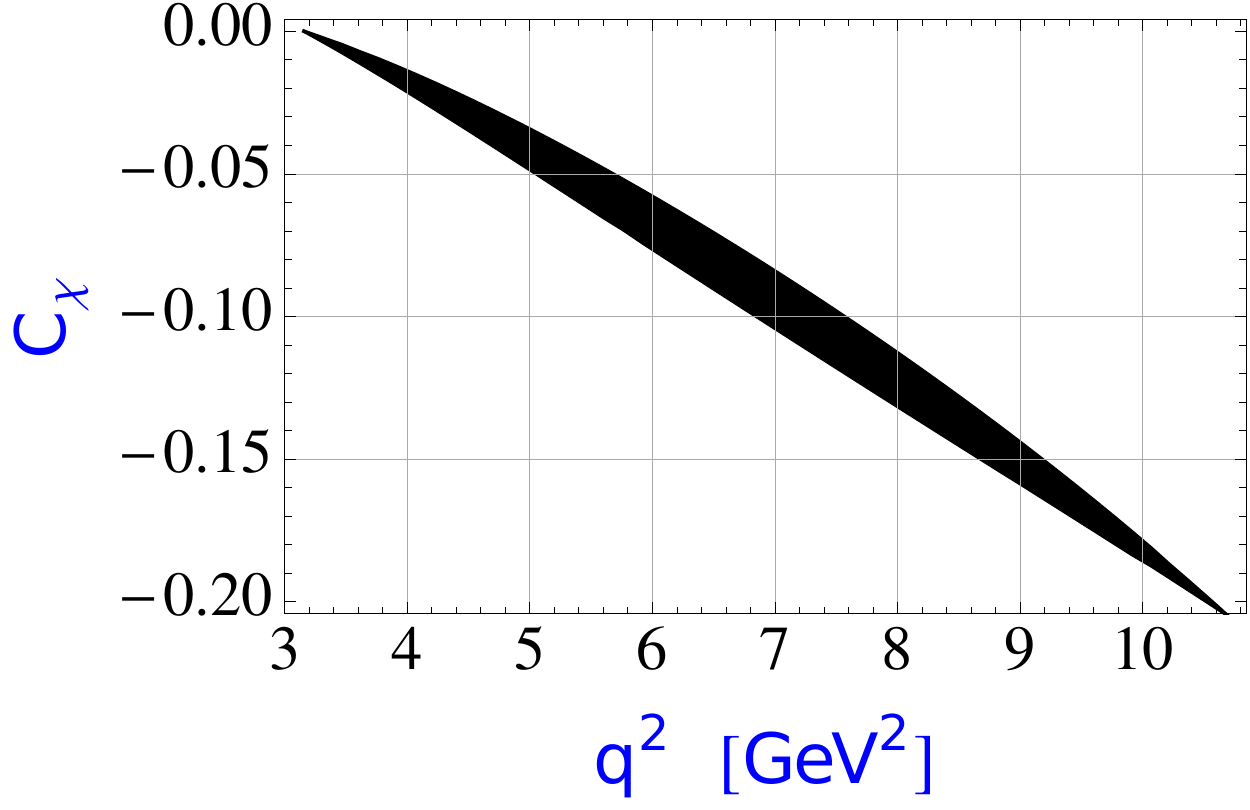}}}~{\resizebox{5.5cm}{!}{\includegraphics{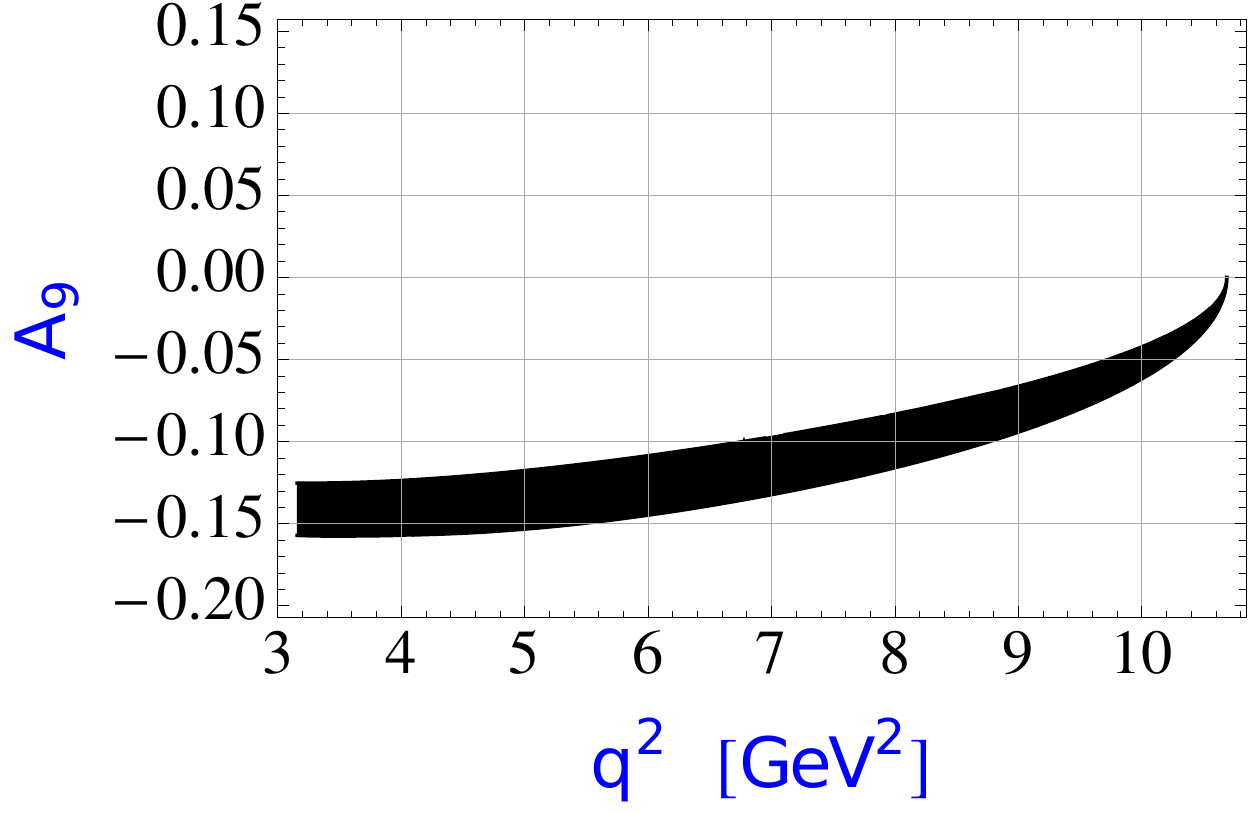}}}~{\resizebox{5.4cm}{!}{\includegraphics{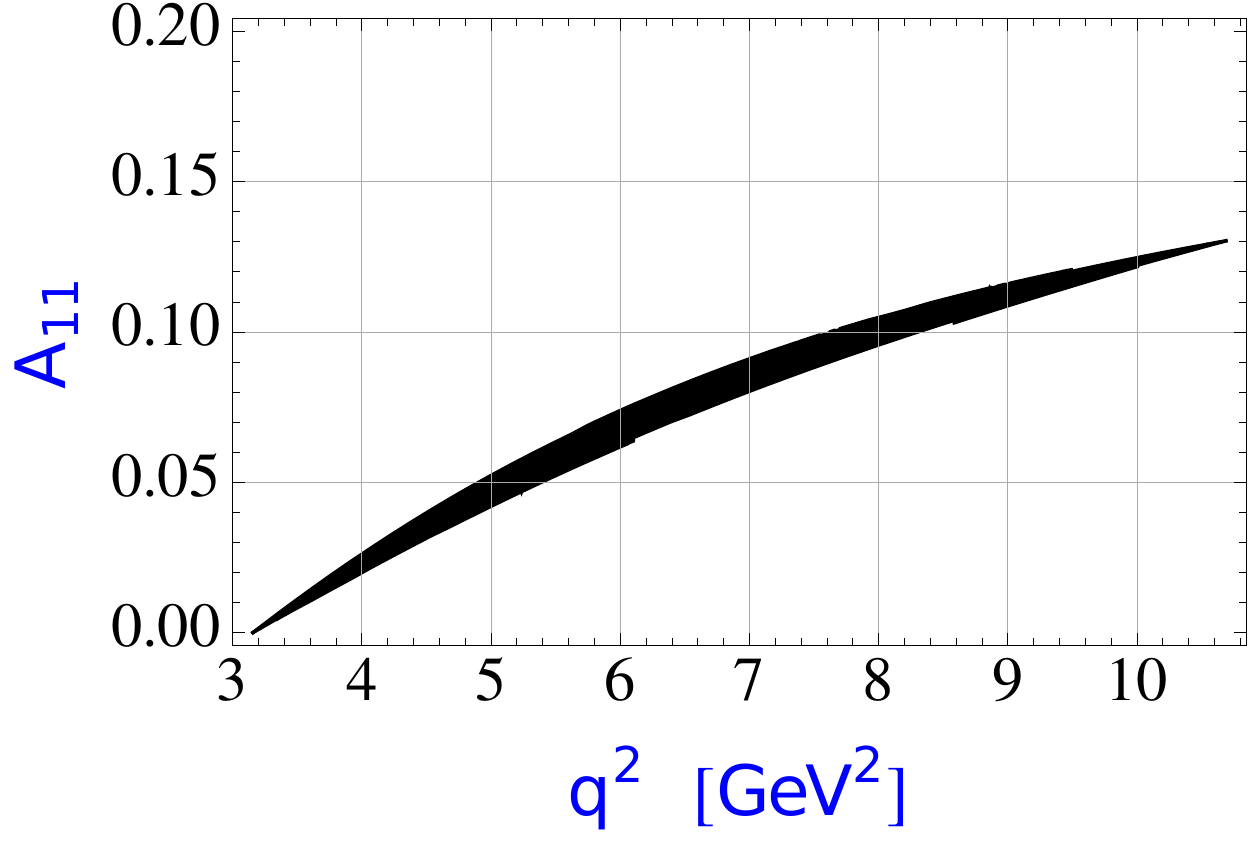}}} \\
\caption{\label{fig:0}\footnotesize{\sl Theory estimates of the observables constructed from the angular distribution of the semileptonic $\overline B\to D\tau \overline\nu_\tau$ (without grid-lines) and $\overline B\to D^\ast\tau \overline\nu_\tau$ (with grid-lines) decays in the Standard Model and by using the hadronic form factors from Ref.~\protect\cite{melikhov}.}} 
\end{center}
\end{figure}

Notice that the quantities $A_8$, $A_{10}$ and $S_\chi$ are non-zero only in the case of the non-zero NP phase. In other words, a nonzero measurement of these quantities would be a clear signal of NP. 
In Fig.~\ref{fig:0} we show the Standard Model shapes of the above quantities as functions of $q^2$ and by using the hadronic form factors to be those of Ref.~\cite{melikhov}.
%%%%%%%%%%%%%%%%%%%%%%%%%%%%%%%%%%%%%%%%%%%%%%%%%%%%%%%%%%%

\section{Illustration of numerical sensitivity to physics BSM in the quark sector\label{sec:3}}

In order to numerically illustrate the sensitivity of observables defined in the previous Section to the presence of physics BSM, we proceed as follows:
\begin{itemize}
\item[--] We use the effective Hamiltonian~(\ref{eq:Heff}), which amounts to replacing the helicity amplitudes by the explicit expressions given in eq.~(\ref{eq:eq00}).
\item[--] We use the experimental results for $R_D= {\cal B}(\Bbar\to D\tau\nubar_\tau)/{\cal B}(\Bbar\to D\mu\nubar_\mu)$ as obtained by BaBar and Belle, and $R_{D^\ast}={\cal B}(\Bbar\to D^\ast\tau\nubar_\tau)/{\cal B}(\Bbar\to D^\ast\mu\nubar_\mu)$ measured at BaBar, Belle and LHCb, and combine them with the form factors computed in Ref.~\cite{melikhov}. We use that latter reference because it contains the full list of form factors needed for this study.~\footnote{ Obviously, for a more viable theoretical description one should use the form factors obtained through numerical simulations of QCD on the lattice. However, since the full set of form factors obtained on the lattice is not available, and since the purpose of this work is to point out the usefulness of the above observables in searching for the effects of NP, we will satisfy ourselves by the form factors of Ref.~\cite{melikhov}.}
\item[--] After switching on the NP couplings, one at the time, we compare theory with experiment and find the range of allowed values for $g_i\equiv g_{V,A,S,P,T,T5}\neq 0$. Since we allow the couplings to be complex, we can choose them to be either fully real, or with a significant imaginary part, and then examine each of the $2$+$10$ observables discussed in this paper, to check on their sensitivity with respect to $g_i\neq 0$.~\footnote{Notice that the differential decay rates are used as input (through $R_{D^{(\ast)}}$), which is why instead of $3$+$11$ observables for $\btod$ and $\btodstar$, we consider the sensitivity of $2$+$10$ observables on $g_i\neq 0$. }  
\end{itemize}
\subsection{Allowed values of $g_{V,A,S,P,T}$}
\begin{figure}[t!]
%\psfrag{aa}{\color{blue}\Huge $q^2$} 
%\psfrag{Standard}{\color{blue}\huge \hspace*{-30mm}$ $} 
\begin{center}
{\resizebox{5.4cm}{!}{\includegraphics{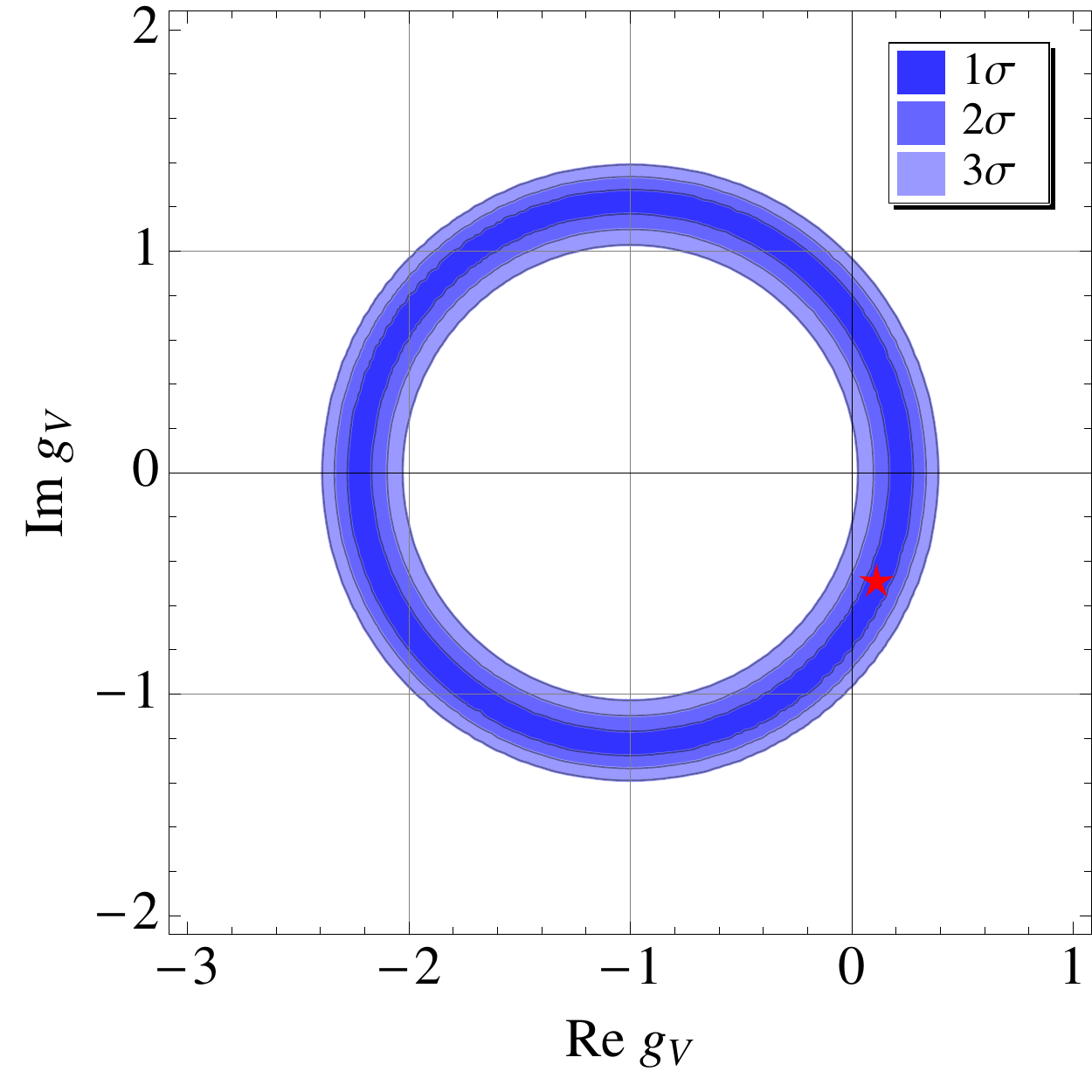}}}~{\resizebox{5.4cm}{!}{\includegraphics{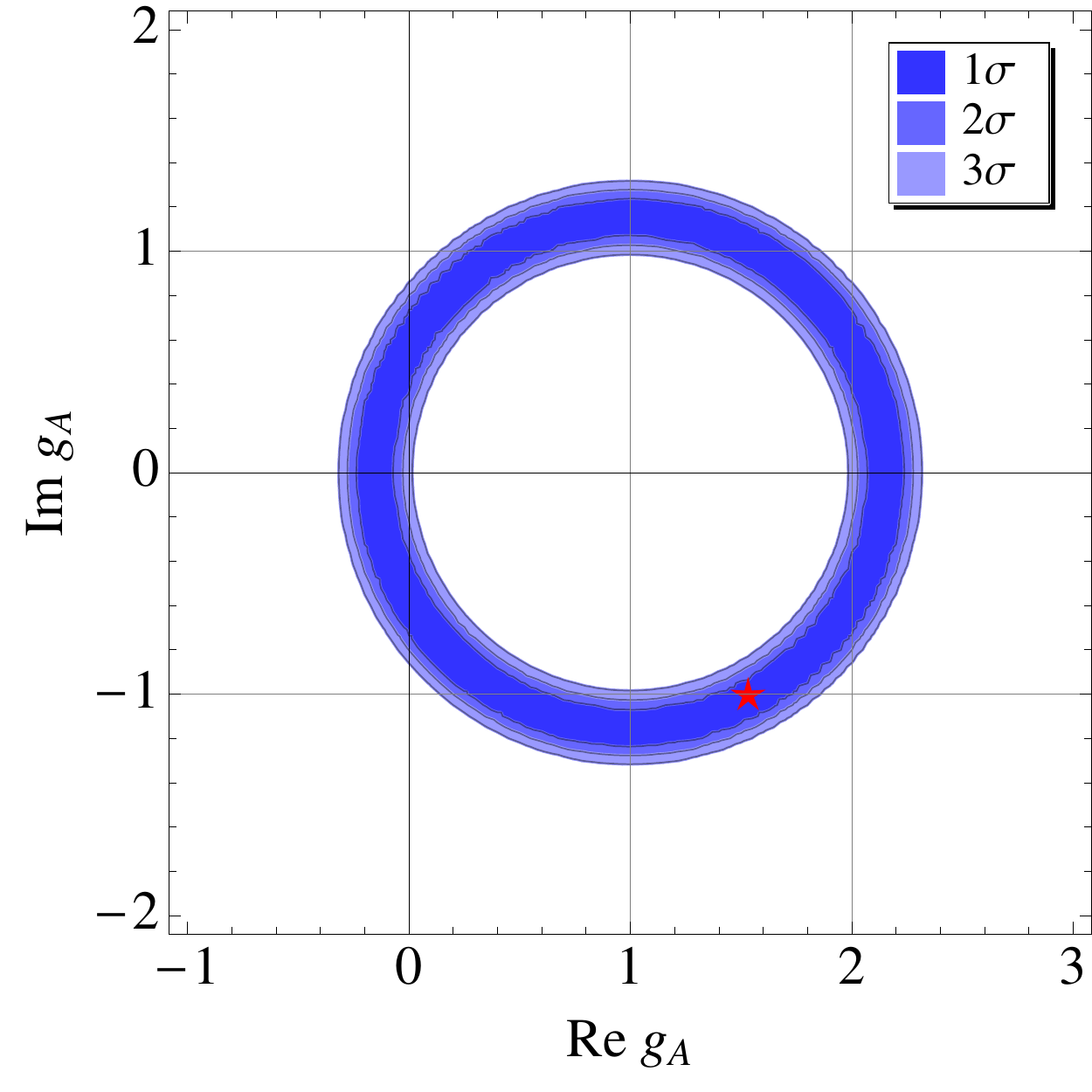}}} \\
{\resizebox{5.4cm}{!}{\includegraphics{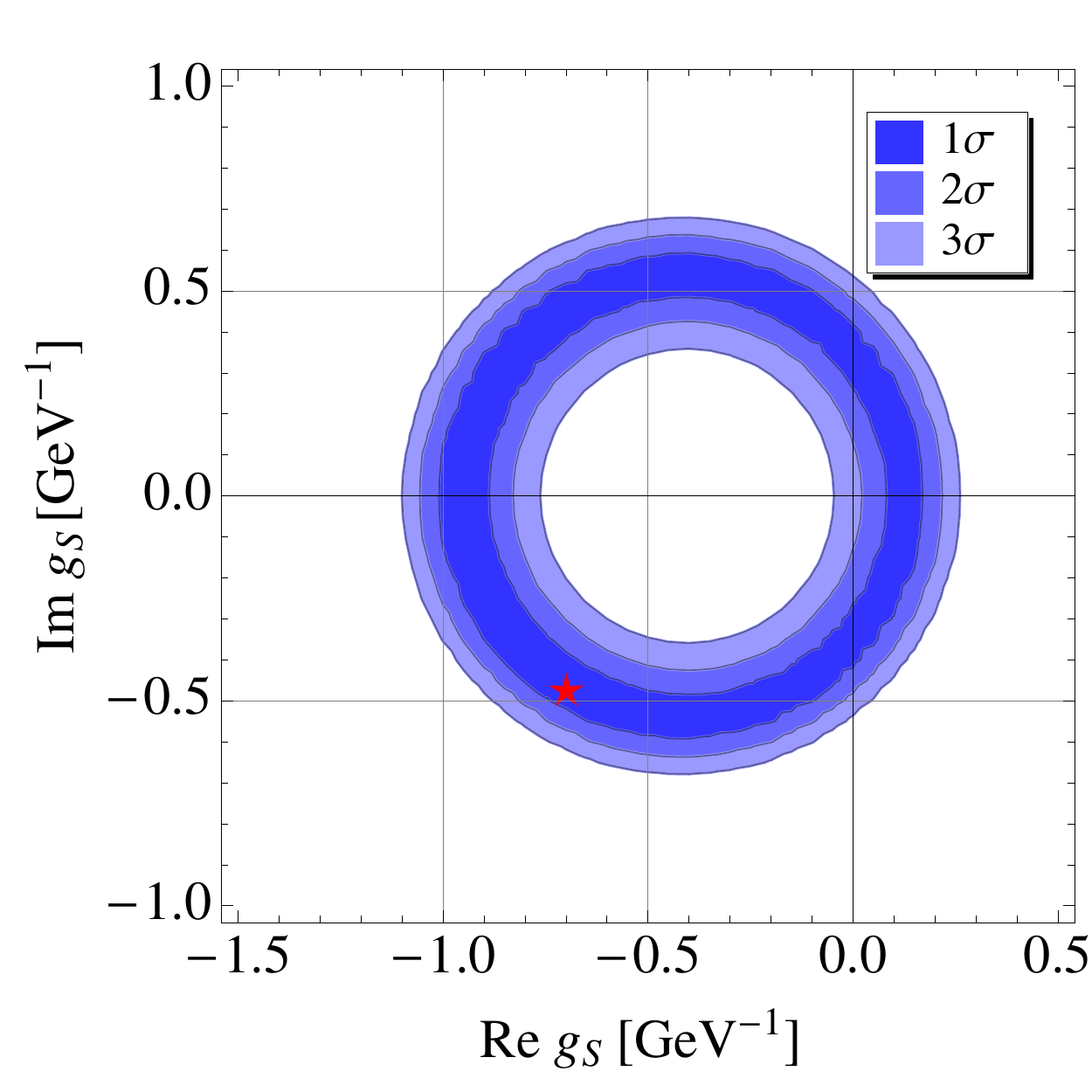}}}~{\resizebox{5.1cm}{!}{\includegraphics{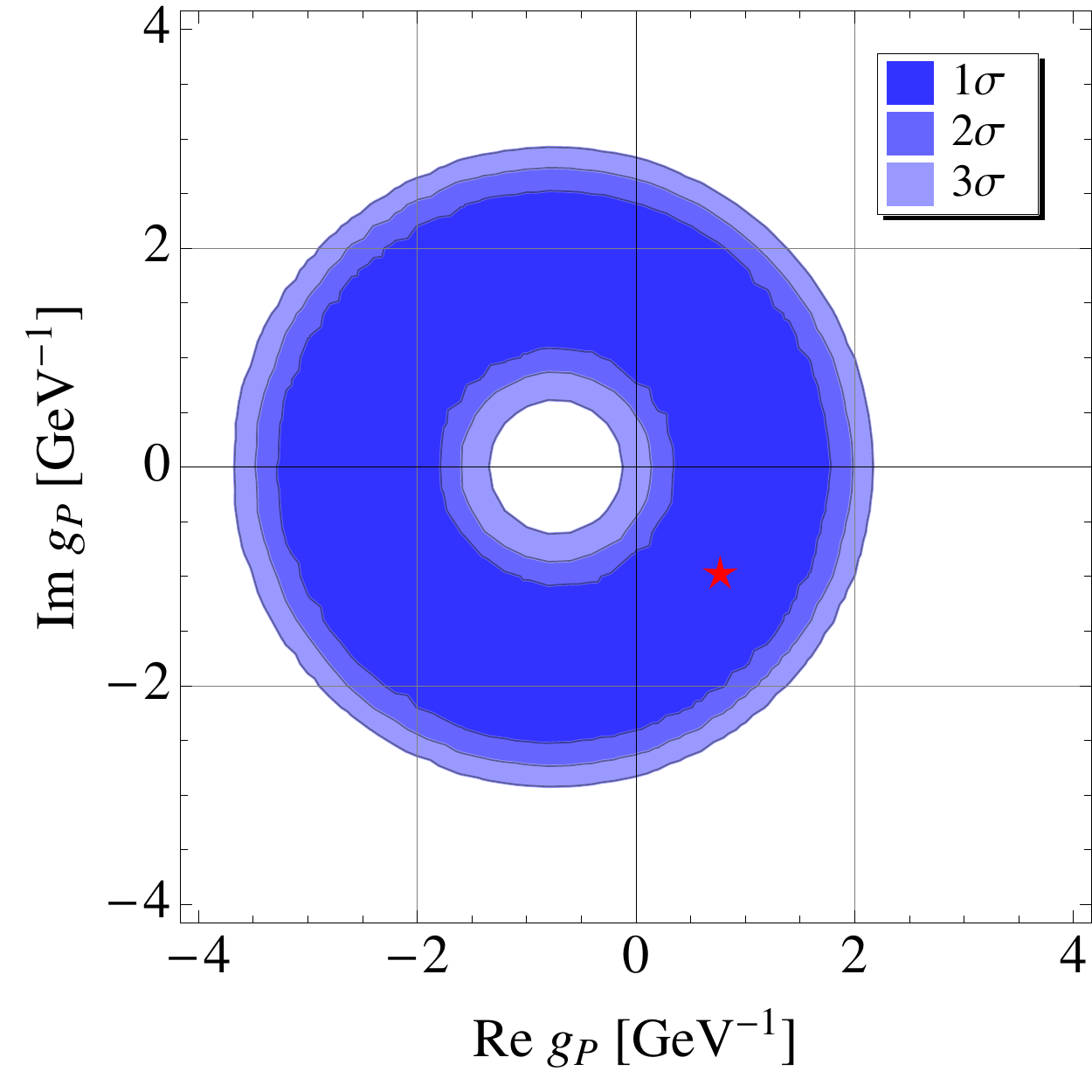}}}~{\resizebox{5.4cm}{!}{\includegraphics{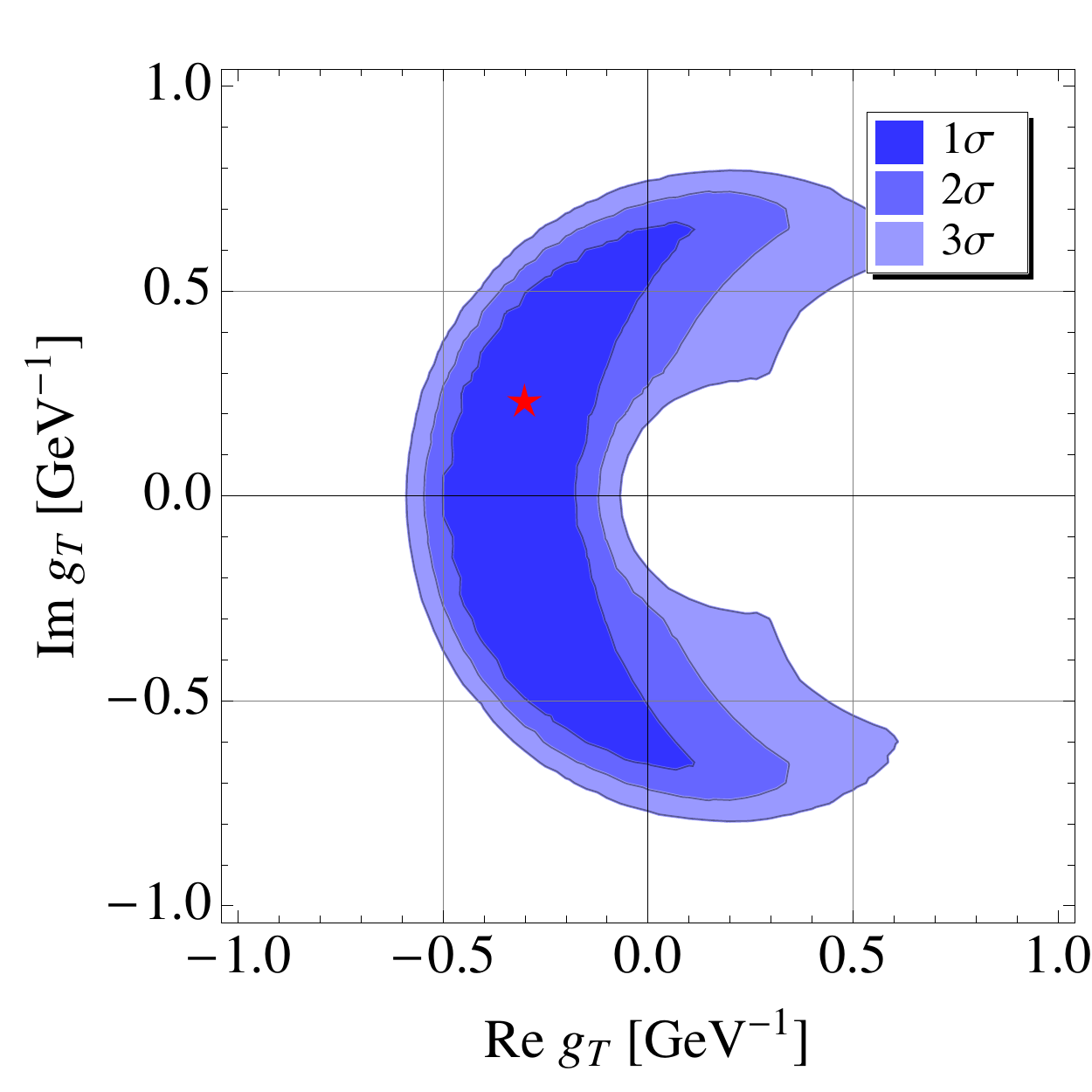}}} \\
\caption{\label{fig:2}\footnotesize{\sl The allowed values for the NP couplings $g_{V,A,S,P,T}$ as obtained from the comparison of experimental values for $R_D$ and $R_{D^\ast}$ with the theoretical estimates obtained by using the hadronic form factors from Ref.~\protect\cite{melikhov} and by switching on one coupling $g_i$ at the time. Stars denote the the best-fit values. Note also that $({\rm Re}[g_i],{\rm Im} [g_i])=(0,0)$ are those of the Standard Model.  Red star in each plot corresponds to the best fit value.}} 
\end{center}
\end{figure}

We now illustrate the allowed values of the NP couplings $g_{V,S,T}$ obtained from $R_{D}$, and $g_{V,A,P,T}$ from $R_{D^{\ast}}$. 
Furthermore we will assume that NP affects the $B\to D^{(\ast )}\tau \nu_\tau$ decay only. After switching on one coupling at the time we obtain the plots shown in Fig.~\ref{fig:2}.
%%%%%%%%%%%%%%%%%%%%%%%%%%%%%%%%%%%%%%%%%%%%%%%%%%%%%%%%%%%
The best fit values obtained in this way are:
\begin{align}\label{bfvalue}
&g_V= 0.21 - i\ 0.76,&&  g_A= -0.18 - i\ 0.05,&&& \nn\\
&g_S= -0.92 - i\ 0.38,&&  g_P= 0.91 + i\ 0.38,& g_T=-0.42+i\ 0.15,&&
\end{align}
and are labeled by red stars in Fig.~\ref{fig:2}. We reiterate that in the notation of eq.~(\ref{eq:Heff}) the couplings $g_{S,P,T}$ are dimensionful and are given in $\gev^{-1}$. 
To illustrate the effect of $g_i\neq 0$ on the observables discussed in the previous Section, we examine them in the case of $\Bbar\to D^{(\ast )}\tau\nubar_\tau$ 
for four different values of $g_i$:  the SM ones ($g_i =0$), the best fit values given above, and for the extreme case of $g_i\neq 0$ allowed from the fits, as shown in Fig.~\ref{fig:2}.  
The scale in $g_{S,P,T}(\mu)$ is implicit and is chosen to be $\mu=m_b$.

After examining each observable on $g_i\neq 0$, we make the following observations: 
\begin{itemize}
\item $A_{ FB}^{D}$ highly (weakly) depends on the value of $g_S$ ($g_T$) but is insensitive to its imaginary part, $\Im\ g_S$ ($\Im\ g_T$). Instead, it is completely insensitive to $g_V$; 
\item $A_{\lambda_\tau}^{D}$ behaves similarly to $A_{ FB}^{D}$ with respect to the variation of $g_{V,S,T}$, especially for the intermediate values of $q^2$. 
The above two observables are related to the decay to a pseudoscalar meson, $\bar B \to D\tau\bar \nu_\tau$, and their pronounced dependence on $g_{S,T}$ is shown in Fig.~\ref{fig:A};
 %%%%%%%%%%%%%%%%%%%%%%%%%%%%%%%%%%%%%%%%%%%%%%%%%%%%%%%%%%%
 \begin{figure}[h!!!]
\begin{center}
{\resizebox{5.6cm}{!}{\includegraphics{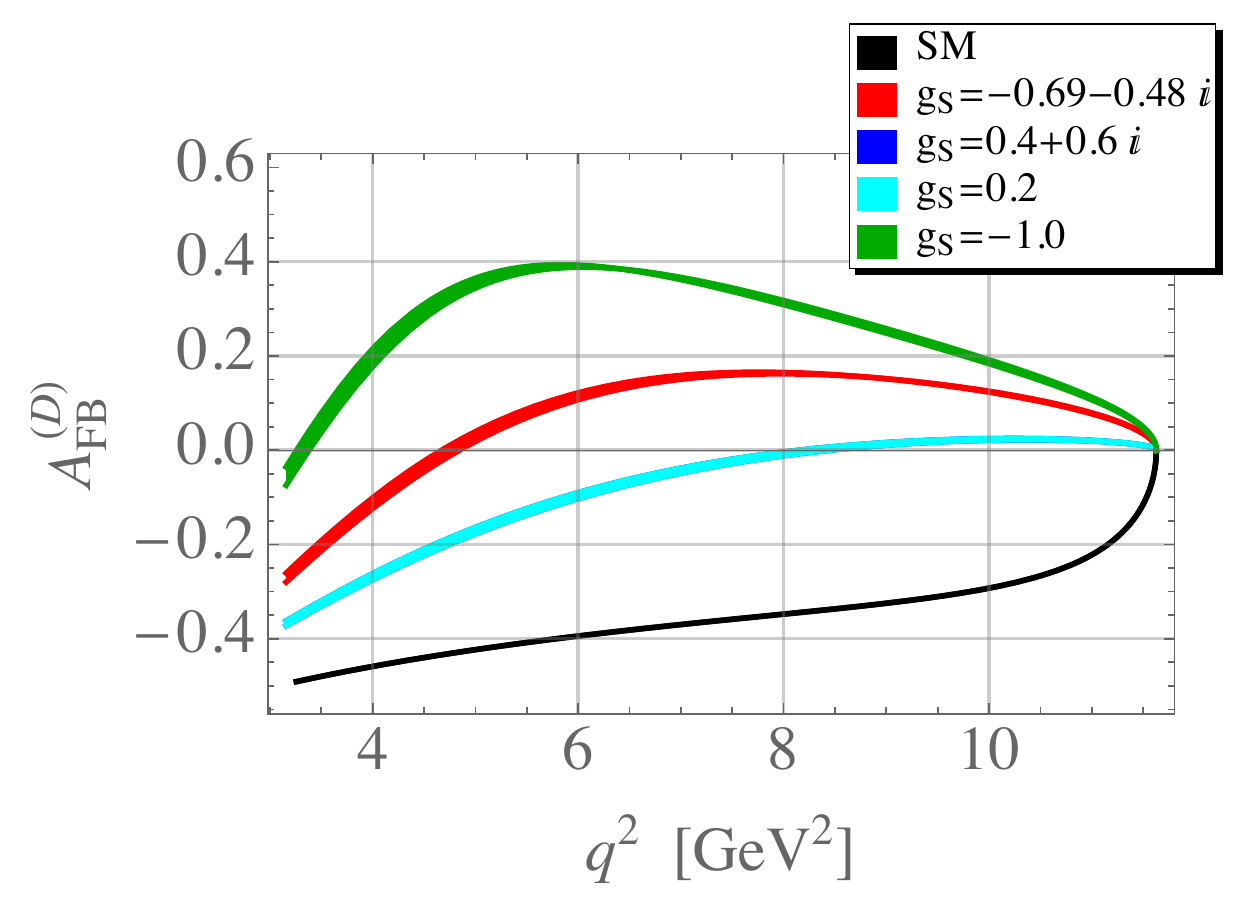}}}~{\resizebox{5.6cm}{!}{\includegraphics{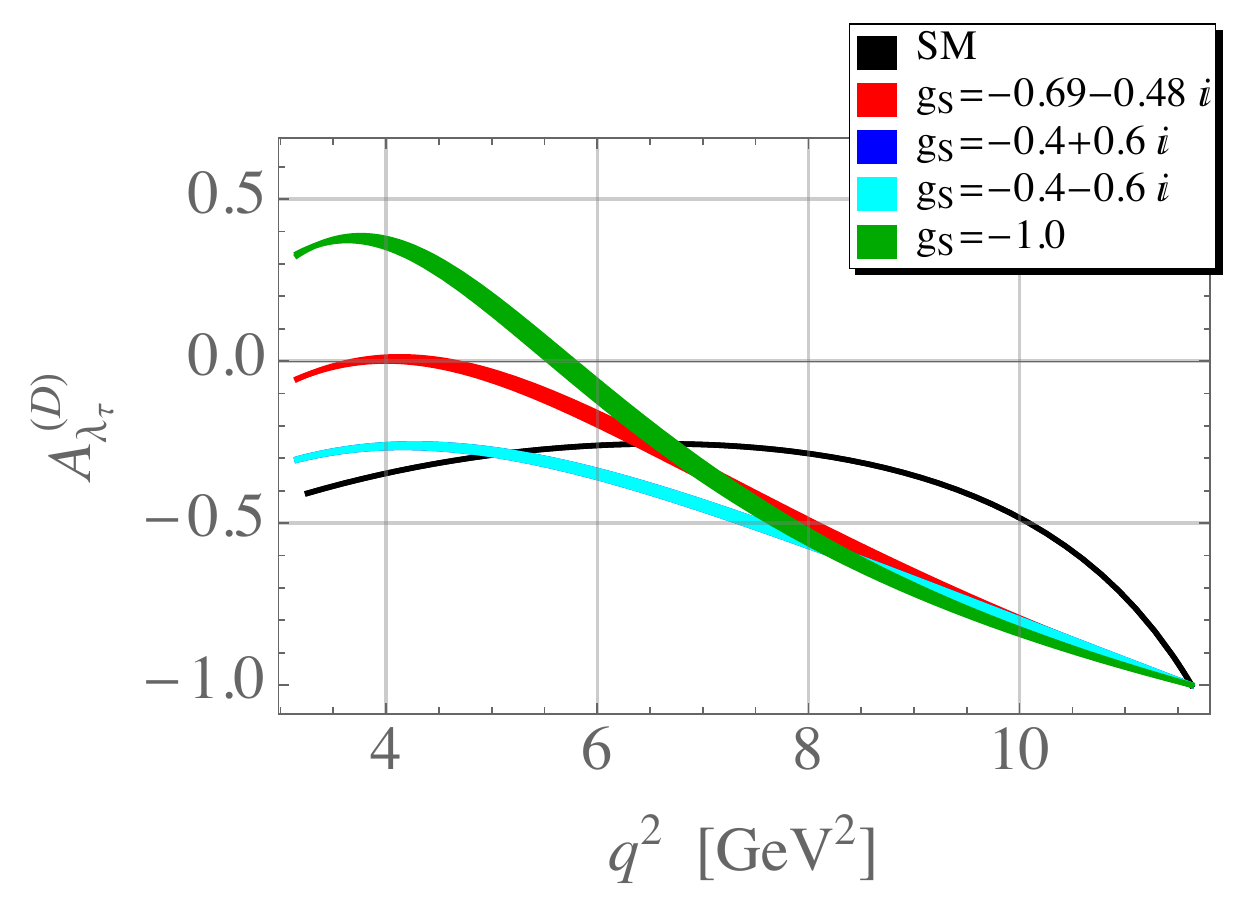}}}~{\resizebox{5.6cm}{!}{\includegraphics{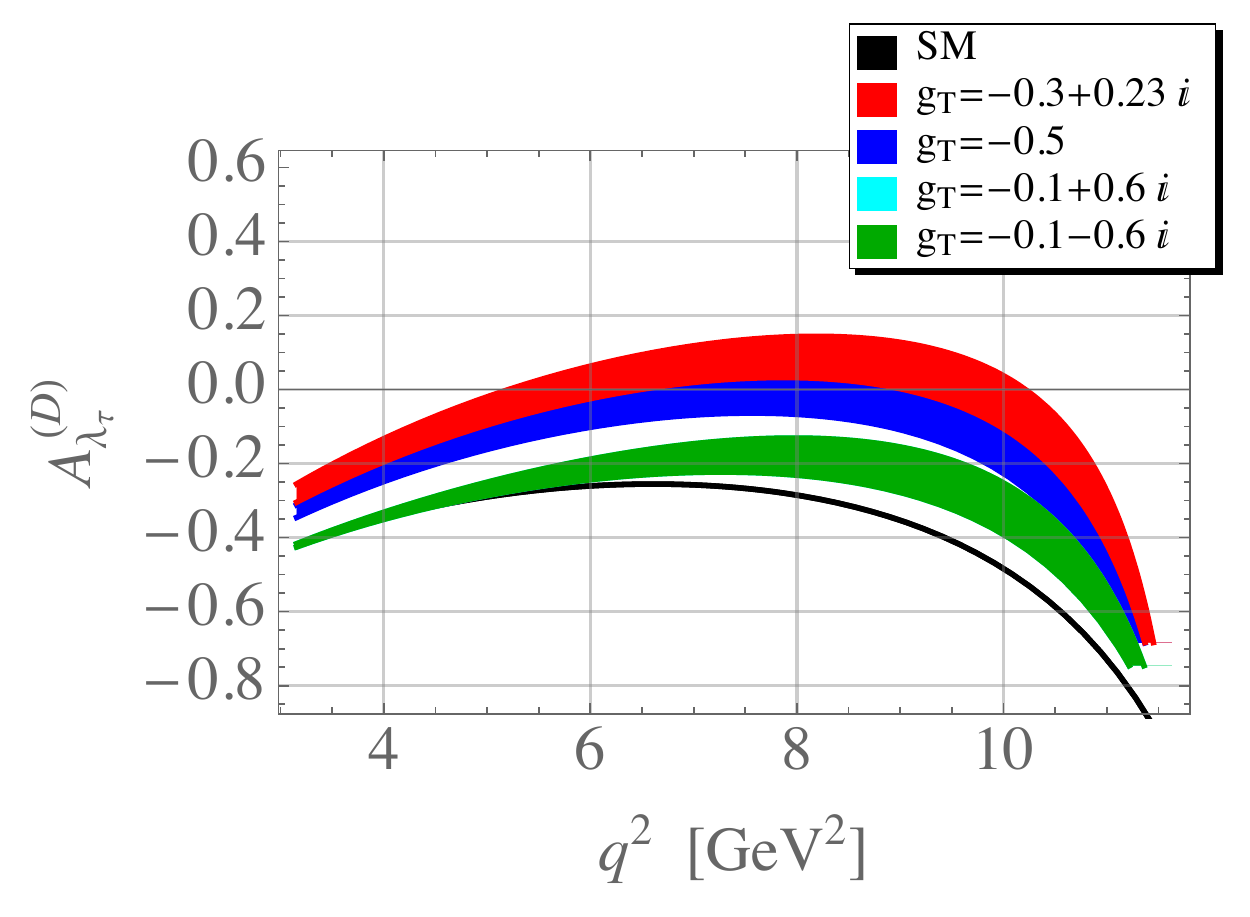}}} 
\caption{\label{fig:A}\footnotesize{\sl 
Forward-backward and the lepton polarization asymmetries in $\bar B \to D\tau\bar \nu_\tau$: sensitivity on the variation of $g_S$ and $g_T$. The values of $g_{S,T}$ are chosen: zero as in the SM, the best-fit values~(\ref{bfvalue}), and the other three values consistent with the results shown in Fig.~\ref{fig:2}.}} 
\end{center}
\end{figure}
%%%%%%%%%%%%%%%%%%%%%%%%%%%%%%%%%%%%%%%%%%%%%%%%%%%%%%%%%%%

\item $A_{ FB}^{D^\ast}$ depends on the sign of $\Re\ g_{V}$, but its deviation from the SM is more pronounced in the case of $g_{A,P,T}$, cf.~Fig.~\ref{fig:B}. However, and provided one observes a deviation with respect to the SM value, one cannot tell which $g_{A,P}\neq 0$ from this quantity alone. Variation of $g_T\neq 0$, instead, results in smaller departures of this quantity from the SM;
\item $A_{\lambda_\tau}^{D^\ast}$ does not depend on $g_{V,A}\neq 0$, but its shape can change in the case of $g_{P,T}\neq 0$. It does not depend on the size of the imaginary part in $g_i$; 
%%%%%%%%%%%%%%%%%%%%%%%%%%%%%%%%%%%%%%%%%%%%%%%%%%%%%%%%%%%
 \begin{figure}[h!!!]
\begin{center}
{\resizebox{5.4cm}{!}{\includegraphics{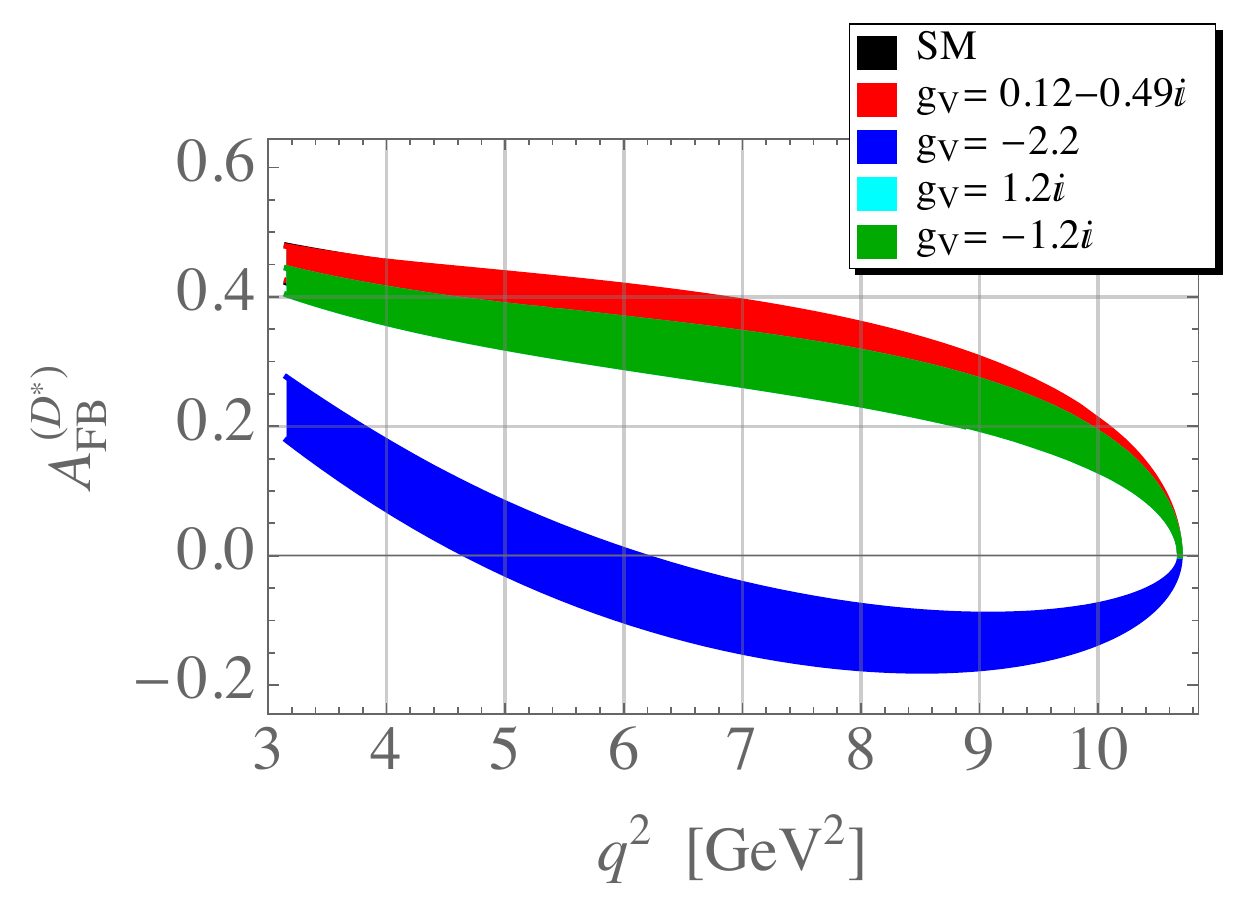}}}~{\resizebox{5.4cm}{!}{\includegraphics{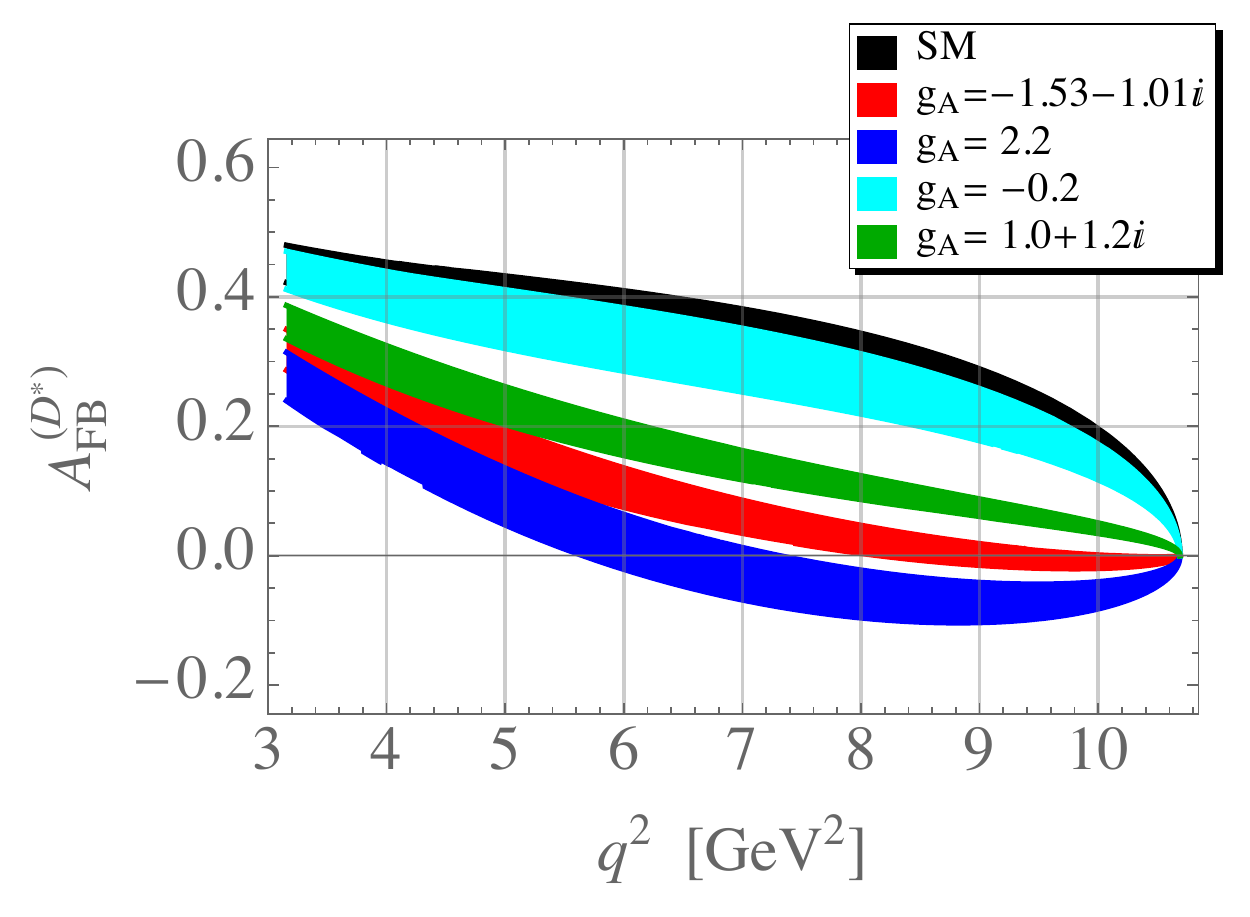}}}~{\resizebox{5.4cm}{!}{\includegraphics{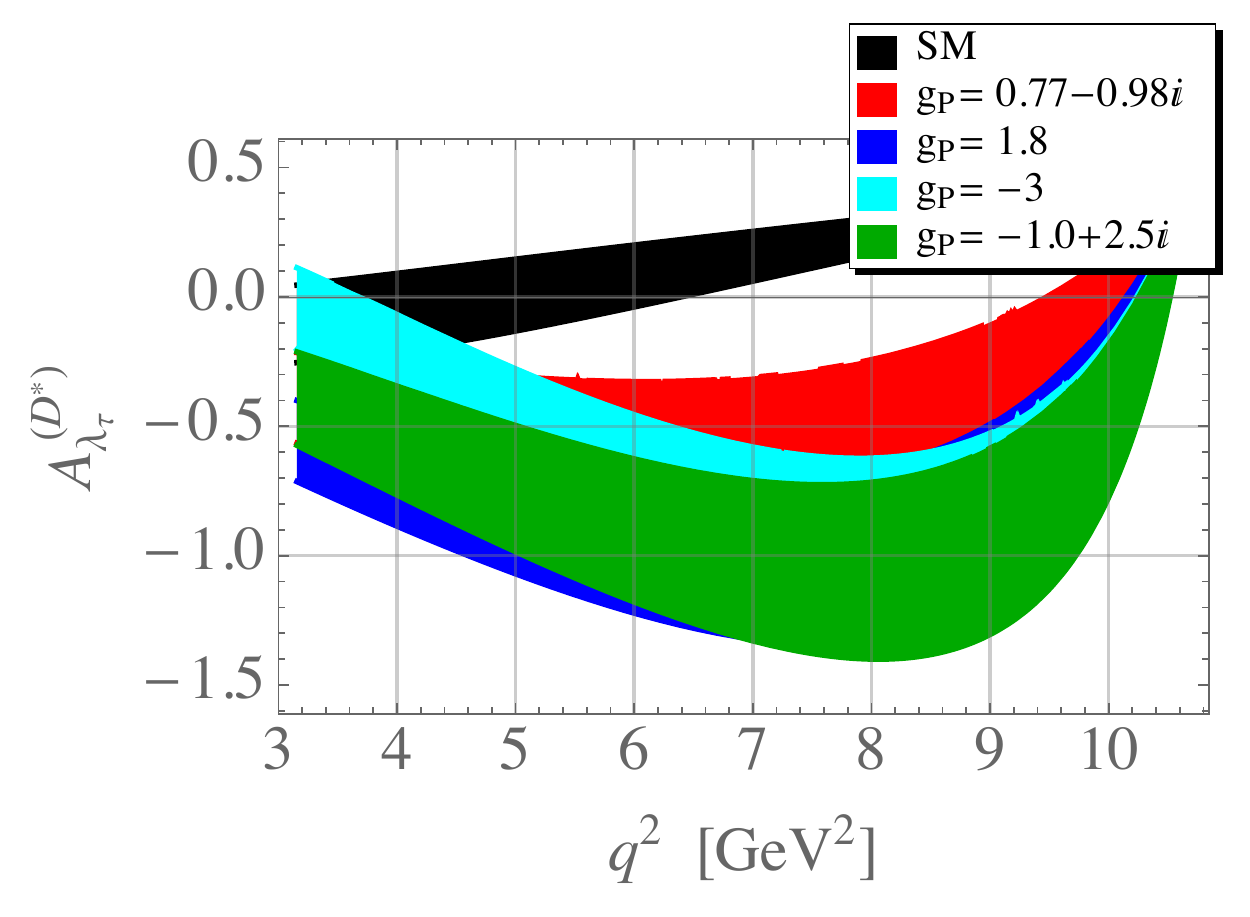}}} 
\caption{\label{fig:B}\footnotesize{\sl 
Forward-backward and the lepton polarization asymmetries in $\bar B \to D^\ast\tau\bar \nu_\tau$: sensitivity on the sign of $g_V$, on the variation of $g_{A}$ and, in the case of $A_{\lambda_\tau}^{B\to D^\ast}$ on the variation of $g_P$.}} 
\end{center}
\end{figure}
%%%%%%%%%%%%%%%%%%%%%%%%%%%%%%%%%%%%%%%%%%%%%%%%%%%%%%%%%%%
\item $R_{L,T}$ too depends on the variation of $\Re\ g_{P,T}\neq 0$, but is insensitive to $g_{V,A}\neq 0$. It is particularly sensitive to $g_T\neq 0$ so that its value falls from the SM ($\sim 3$) to about $\sim 1$ at $q^2 = m_\tau^2$, as shown in Fig.~\ref{fig:C};
\item $A_5$ depends on the sign of $\Re\ g_{V}$, it significantly changes with $\Re\ g_{A}$, only weakly depends on $\Re\ g_{P}$ and it is very sensitive to $\Re\ g_{T}$; 
\item $C_\chi$ only weakly depends on $g_{V}$, and is independent on $g_A$. Instead its linear dependence in the SM is modified to an arc-like behavior in $q^2$ for $g_{P,T}\neq 0$. In Fig.~\ref{fig:C} this is shown for the case of $g_T\neq 0$;
%%%%%%%%%%%%%%%%%%%%%%%%%%%%%%%%%%%%%%%%%%%%%%%%%%%%%%%%%%%
 \begin{figure}[h!!!]
\begin{center}
{\resizebox{5.4cm}{!}{\includegraphics{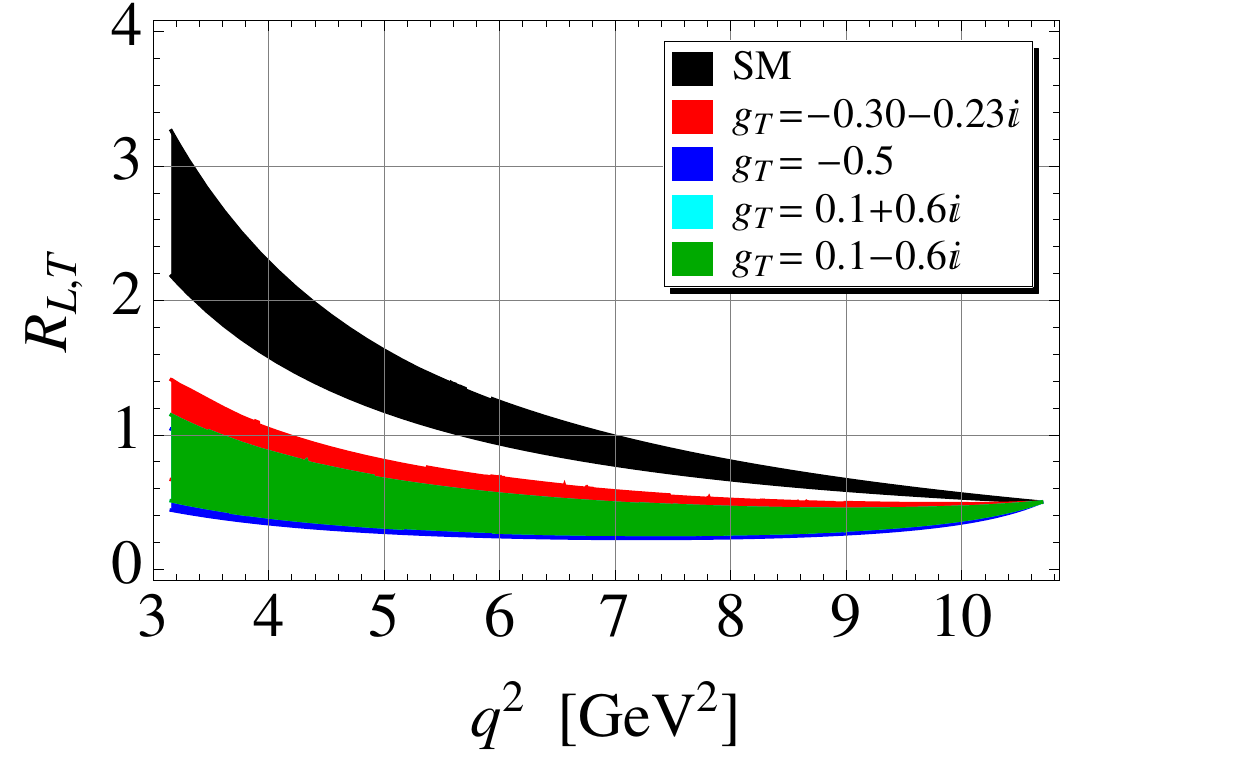}}}~{\resizebox{5.4cm}{!}{\includegraphics{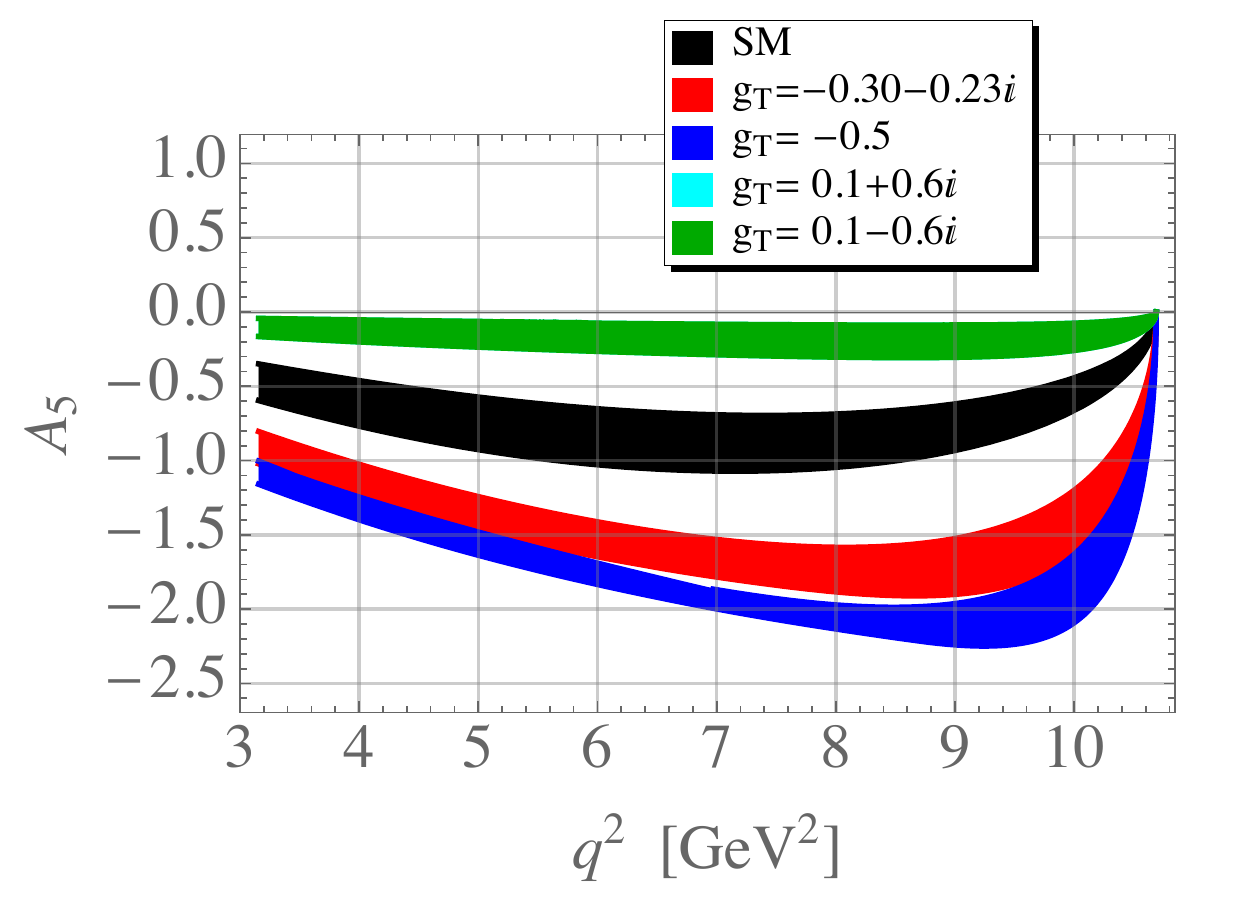}}}~{\resizebox{5.4cm}{!}{\includegraphics{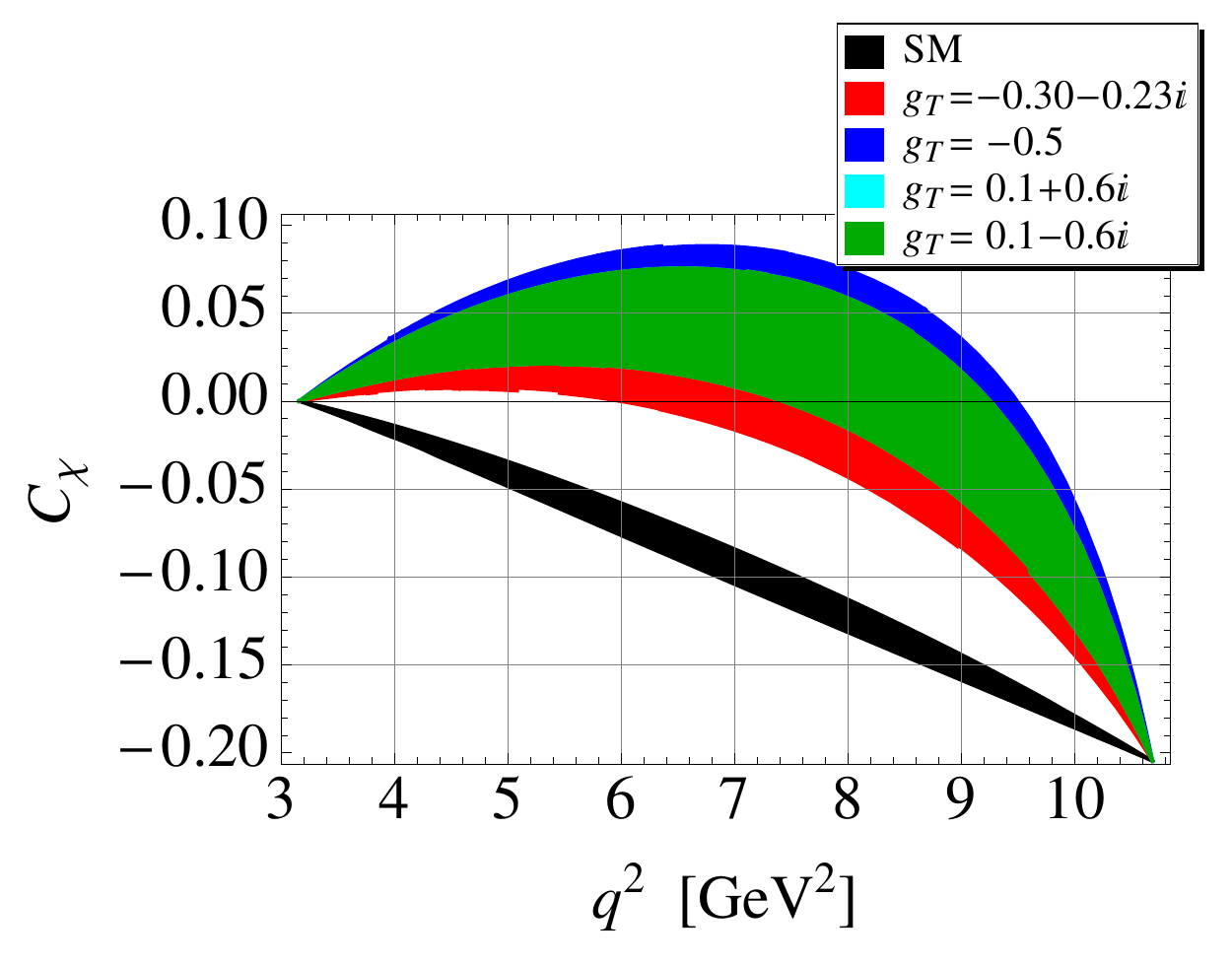}}} 
\caption{\label{fig:C}\footnotesize{\sl 
Sensitivity of the observables deduced from the angular distribution of $\bar B \to D^\ast\tau\bar \nu_\tau$ on $g_T$.}} 
\end{center}
\end{figure}
%%%%%%%%%%%%%%%%%%%%%%%%%%%%%%%%%%%%%%%%%%%%%%%%%%%%%%%%%%%
\item $S_\chi$ is very sensitive to $\Im\ g_{V,A,T}$ and is independent on $g_P$. It is a null-test of the SM because $S_\chi(q^2)\neq 0$ would represent a clear signal of NP;
\item $A_8$ is also sensitive to the imaginary part of the couplings, $\Im\ g_{V,A,P,T}$ so that a measurement of its non-zero value would be a signal of a NP phase. Notice, however, 
that the deviations with respect to the SM, are particularly pronounced in the case of $\Im\ g_T\neq 0$, as shown in Fig.~\ref{fig:D}; 
%%%%%%%%%%%%%%%%%%%%%%%%%%%%%%%%%%%%%%%%%%%%%%%%%%%%%%%%%%%
 \begin{figure}[h!!!]
\begin{center}
{\resizebox{5.4cm}{!}{\includegraphics{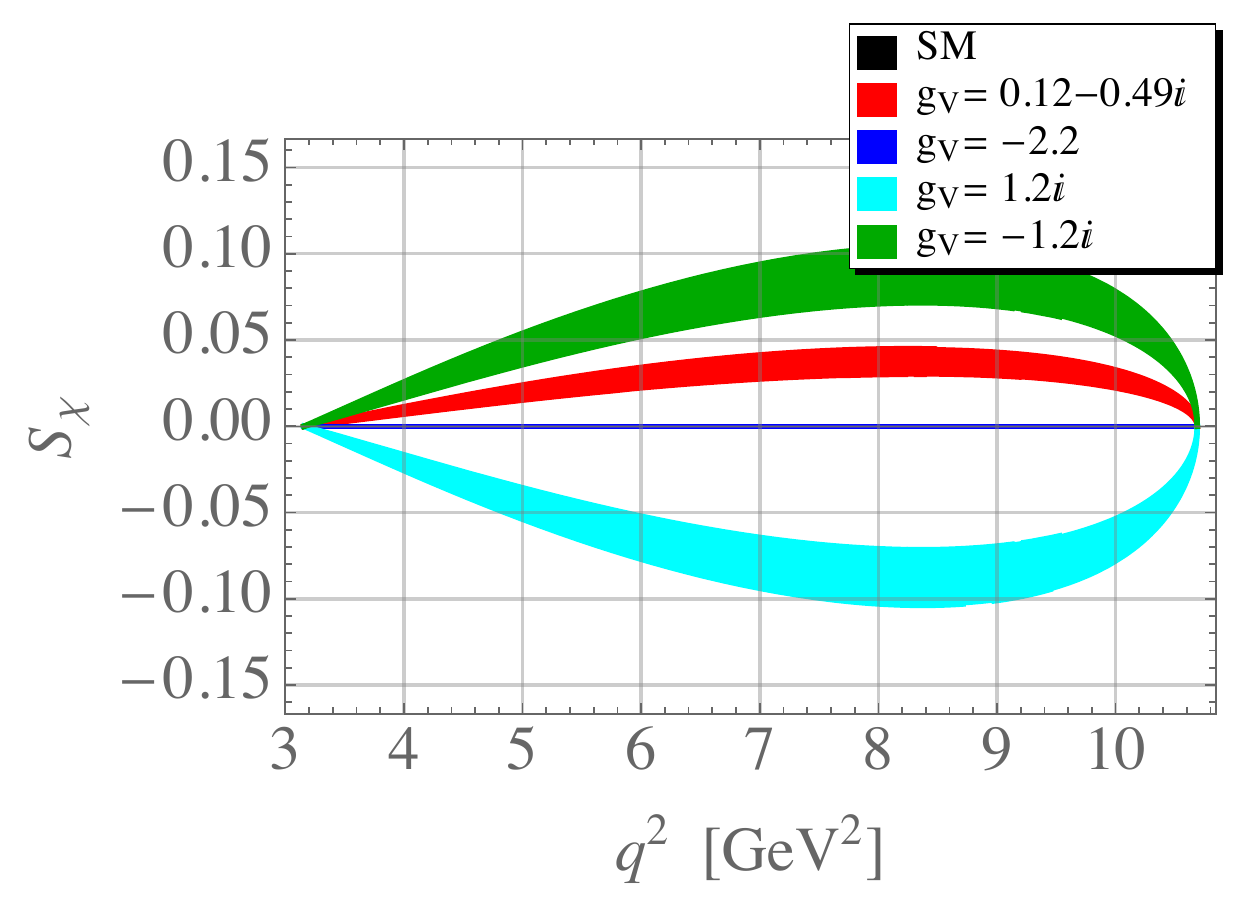}}}~{\resizebox{5.4cm}{!}{\includegraphics{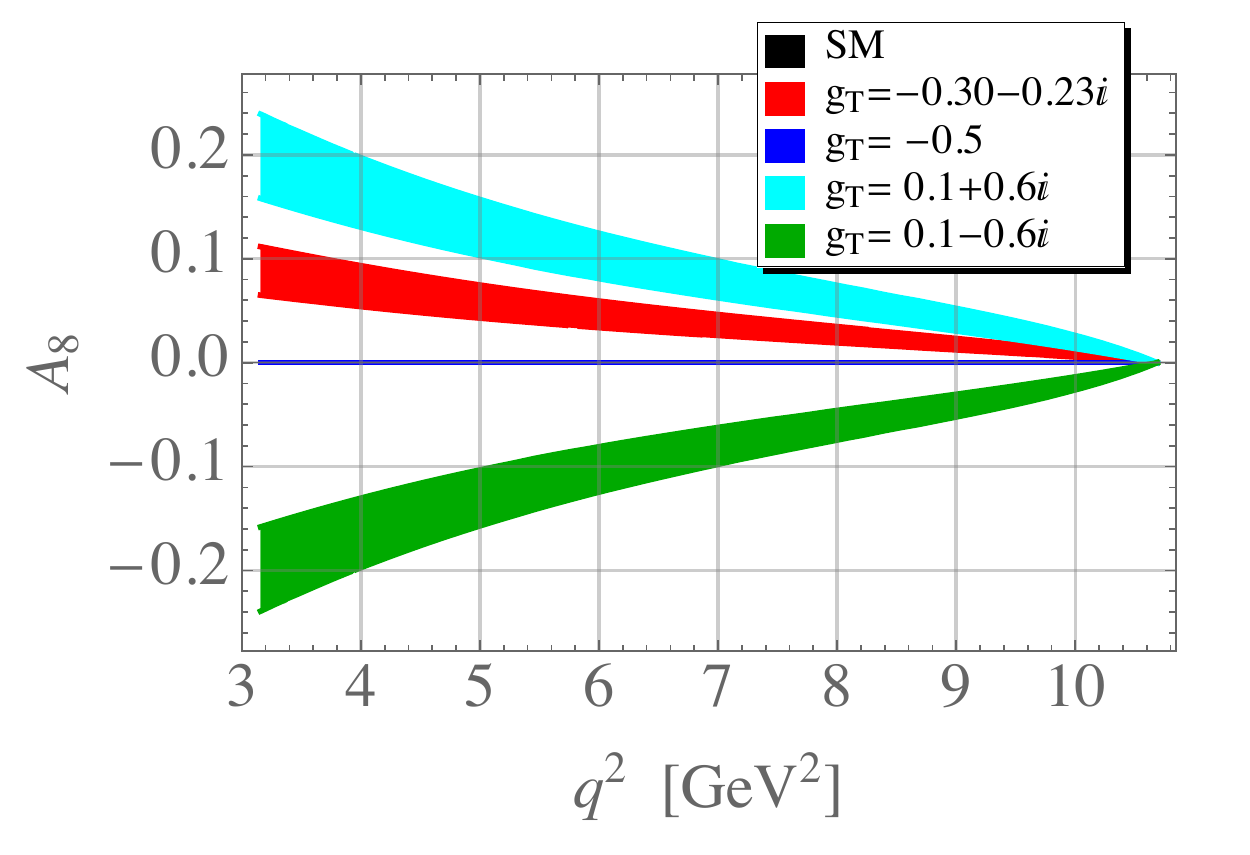}}}~{\resizebox{5.4cm}{!}{\includegraphics{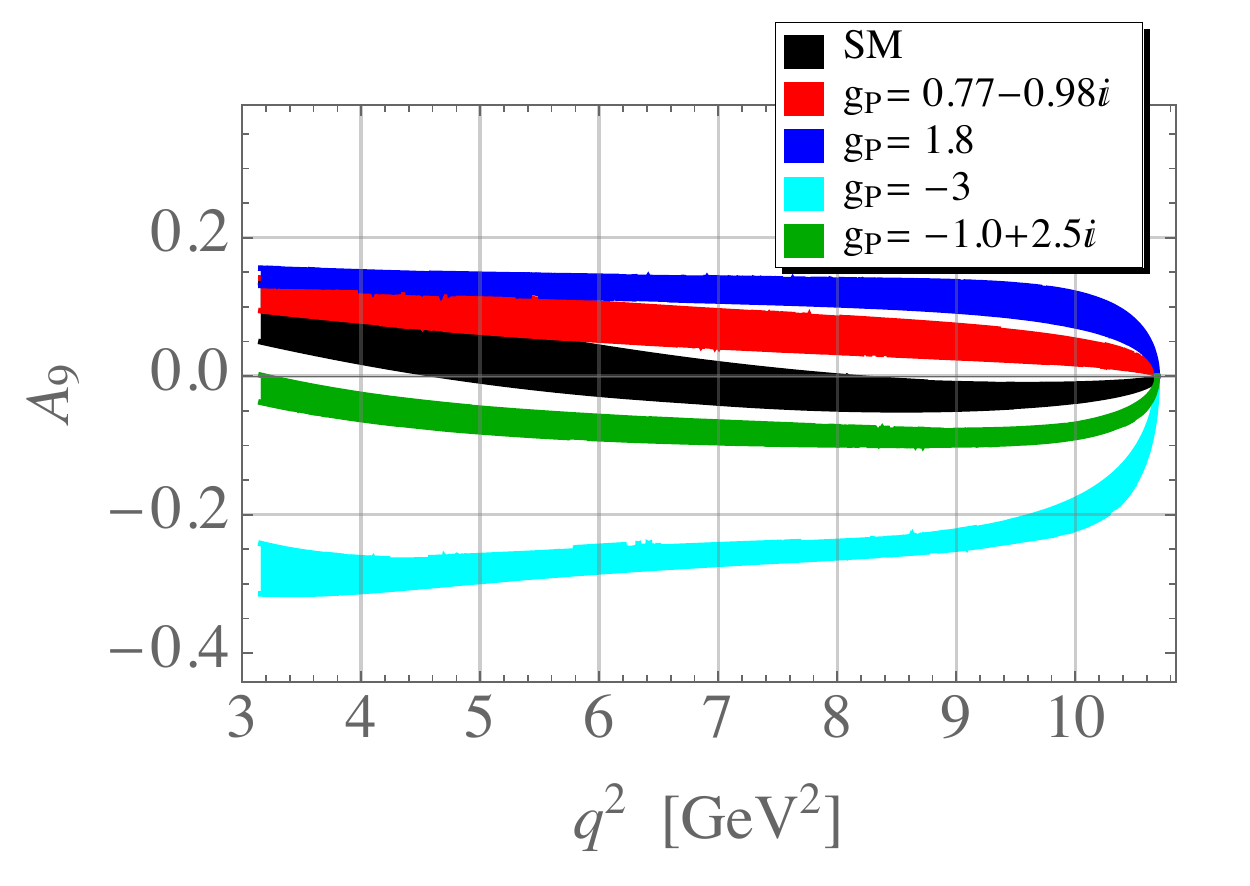}}} 
\caption{\label{fig:D}\footnotesize{\sl 
Nonzero values of $S_\chi(q^2)$, $A_8(q^2)$ are related to the nonzero $\Im\ g_i$, which would in turn represent a signal of a NP phase. Variation of $A_9(q^2)$ with respect to the change in $g_P$ is also illustrated for the case of $\bar B \to D^\ast\tau\bar \nu_\tau$.}} 
\end{center}
\end{figure}
%%%%%%%%%%%%%%%%%%%%%%%%%%%%%%%%%%%%%%%%%%%%%%%%%%%%%%%%%%%
\item $A_9$ depends on  the real part of $g_{V,P,T}$ and only mildly on $g_A$. $A_9(q^2)$ only slightly varies with $q^2$ and its value can significantly change for $\Re\ g_P<0$; 
\item $A_{10}$, just like $S_\chi$ and $A_8$, is  sensitive to the imaginary part of $g_{V,A,T}$, while it is independent on $g_P$. Note that even for large (allowed) $\Im \ g_{V,P,T}$ this asymmetry is small, i.e. never larger than $7\%$;
\item The shape of $A_{11}(q^2)$ changes for $\Re\ g_{P,T}\neq 0$, but it is insensitive to $g_{V,A}\neq 0$. Its deviation from the SM can be probed at large $q^2$'s.
\end{itemize}
The above observations are summarized in Tab.~\ref{tab:9}. Clearly, a combination of several observables would be a good testing ground of a presence of physics BSM in the low-energy semileptonic processes. The combination of the above observables would help understanding the Lorentz structure of the NP contributions (if any), their size. More specifically, three of them can be used as a check of the presence of additional NP phase(s).  

%%%%%%%%%%%%%%%%%%%%%%%%%%%%%%%%%%%%%%%%%%%%%%%
%%%%%%%%%%%%%%%%%%%%%%%%%%%%%%%%%%%%%%%%%%%%%%%
%%%%%%%%%%%%%%%%%%%%%%%%%%%%%%%%%%%%%%%%%%%%%%%
%\begin{footnotesize}
\begin{table}[ht!]
\renewcommand{\arraystretch}{1.5}
\centering{}%
%\resizebox{\textwidth}{!}{
\tabcolsep=0.09cm
\begin{tabular}{|c|ccccc|}
\hline
{\sl Quantity} & $\quad g_V\quad $  & $\quad g_A\quad $ & $\quad g_S\quad $  & $\quad g_P\quad $ & $\quad g_T\quad $ \\ \hline\hline
 $A_{ FB}^{D}$             & $\times$ & -- & $\star\star\star$  & --  & $\star$ \\ 
 $A_{\lambda_\tau}^{D}$ & $\times$ & -- & $\star\star\star$  & --  & $\star\star$ \\ \hline
$A_{ FB}^{D^\ast}$                 & $\star$ & $\star\star\star$ & --  & $\star\star\star$  & $\star$ \\ 
$A_{\lambda_\tau}^{D^\ast}$ & $\times$ & $\times$ & --  & $\star\star$  & $\star$ \\ 
$R_{L,T}$                                       & $\times$ & $\times$ & --  & $\star\star$  & $\star\star$ \\ 
$A_5$     & $\star\star$ & $\star\star$ & --  & $\star $  & $\star\star\star$ \\ 
$C_\chi$ & $\star$ & $\times$ & --  & $\star \star$  & $\star\star$  \\ 
$S_\chi$ & $\star\star\star$ & $\star\star\star$ & --  & $\times $  & $\star\star\star$  \\ 
$A_8$     & $\star\star$ & $\star\star$ & --  & $\star\star $  & $\star\star\star$  \\ 
$A_9$     & $\star$ & $\star $ & --  & $\star \star$  & $\star\star$  \\ 
$A_{10}$ & $\star\star$ & $\star\star$ & --  & $\times $  & $\star\star$  \\ 
$A_{11}$ &  $\times$ & $\times$ & --  & $\star\star $  & $\star\star$  \\  \hline
\end{tabular}
%}
\caption{{\footnotesize{}\label{tab:9} \sl Sensitivity to $g_i\neq 0$: $\times$ stands for ``{\sl not sensitive}", and $\star\star\star$ for ``maximally sensitive". }}
\end{table}
%\end{footnotesize}
%%%%%%%%%%%%%%%%%%%%%%%%%%%%%%%%%%%%%%%%%%%%%%%
%%%%%%%%%%%%%%%%%%%%%%%%%%%%%%%%%%%%%%%%%%%%%%%
%%%%%%%%%%%%%%%%%%%%%%%%%%%%%%%%%%%%%%%%%%%%%%%

\section{Summary\label{sec:4}}
In this paper we provided the general expressions for the full angular distribution of the semileptonic decays of a pseudoscalar meson to a daughter pseudoscalar or vector meson. From these formulas, apart from the differential decay widths, we were able to construct $2$ ($10$) observables when considering the decay to a pseudoscalar (vector) meson. High luminosity experimental facilities are likely to allow us to measure the detailed angular distribution of these decays and the resulting observables discussed in this paper can be used for searching the effects of physics BSM at low energies. 

We focused on the case of  $\bar B \to D^{(\ast )}\tau\bar \nu_\tau$ to illustrate the benefits of the observables discussed in this paper. In particular, three observables [$S_\chi(q^2)$, $A_8(q^2)$ and $A_{10}(q^2)$] are sensitive to the NP phase(s). Other quantities we discussed here can be used to disentangle the Lorentz structure of the NP contributions ($V$, $A$, $S$, $P$ or $T$) and perhaps to deduce its size if we had a clean QCD information about the relevant hadronic form factors at our disposal. We examined the sensitivity of each of the mentioned observables to the presence of $g_i\neq 0$, $i\in\{V,A,S,P,T\}$. 

It is known that in the case of semileptonic decay to a vector meson, a part of the decay amplitude is polluted by the $S$-wave contribution of the similar decay to a scalar meson. The resulting interference might induce important uncertainties. This is due to the fact that the scalar states are usually broad and often do not respect the BW shape. We showed that there are particular terms in the angular distribution of $\bar B\to D^\ast (\to D\pi)\ell\bar \nu_\ell$ that are non-null if there is interference with 
$\bar B\to D_0^\ast (\to D\pi)\ell\bar \nu_\ell$. If the shape of $D_0^\ast$ is close to that of BW, those interference terms should be negligibly small. Conversely, a sizable interference would suggest that either the shape of $D_0^\ast$ is not BW-like, or that there is still a part of $(D\pi)_{S-\rm wave}$ which is unaccounted for by the nearby resonances. That information could be useful for our understanding of scalar open-flavored mesons. 
\vspace*{2cm}

\section*{Acknowledgement}
S.F. acknowledges support of the Slovenian Research Agency. A.T. has been supported in part by JSPS KAKENHI Grant Number 2402804 and carried out thanks to the support of the OCEVU Labex (ANR-11-LABX-0060) and the A*MIDEX project (ANR-11-IDEX-0001-02) funded by the ``{\sl Investissements d'Avenir}" French government program managed by the ANR. I. N. is supported in part by the {\sl Bundesministerium f\"ur Bildung und Forschung} (BMBF).
\newpage
\appendix

\section{Polarization vectors\label{app:polarization}}

%%%%%%%%%%%%%%%%%%%%%%%%%%%%%%%%%%%%%%%%%%%%%%%%%%%%%%%%%%%
 \begin{figure}[t!]
\begin{center}
{\resizebox{11.4cm}{!}{\includegraphics{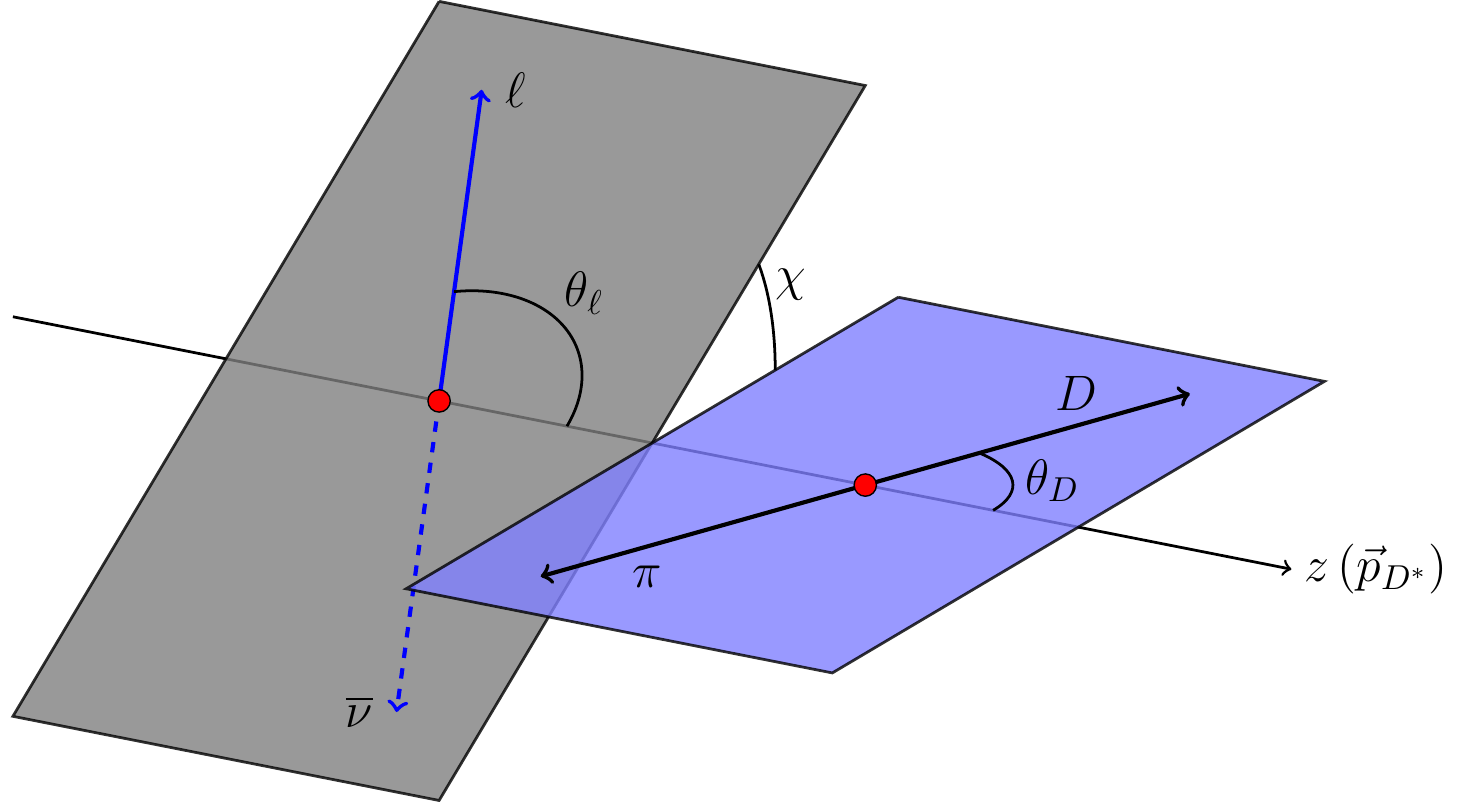}}} 
\caption{\label{fig:App}\footnotesize{\sl 
Kinematics of the $\bar B\to D^\ast (\to D\pi)\ell\bar \nu_\ell$ decay. Angles are defined as in Ref.~\protect\cite{Korner:1989qb}. }} 
\end{center}
\end{figure}
%%%%%%%%%%%%%%%%%%%%%%%%%%%%%%%%%%%%%%%%%%%%%%%%%%%%%%%%%%%
In this paper we use the convention of Ref.~\cite{Korner:1989qb} and define the angles $\thl$,$\thD$ and $\chi$ as depicted in Fig.~\ref{fig:App}. The helicity axis is chosen along the $\Dst$ momentum while the polarization vectors of $D^*$ ($\varepsilon$) and the virtual vector boson ($\widetilde\varepsilon$) are defined with lower indices as
\begin{equation}
   \varepsilon_\pm = \mp{1 \over \sqrt2}\left(
      \begin{array}{c}
         0 \\
         1 \\
         \pm i \\
         0
      \end{array}\right)\,,\quad\quad
      \varepsilon_0 = {1 \over m_\Dst}\left(
      \begin{array}{c}
         |\q| \\
         0 \\
         0 \\
         E_\Dst
      \end{array}\right)\,,
\end{equation}
and
\begin{equation}
   \widetilde\varepsilon_\pm = {1 \over \sqrt2}\left(
      \begin{array}{c}
         0 \\
         \pm 1 \\
         -i \\
         0
      \end{array}\right)\,,\quad\quad
      \widetilde\varepsilon_0 = {1 \over \sqrt{q^2}}\left(
      \begin{array}{c}
         |\q| \\
         0 \\
         0 \\
         -q_0
      \end{array}\right)\,, \quad\quad
      \widetilde\varepsilon_t = {1 \over \sqrt{q^2}}\left(
      \begin{array}{c}
         q_0 \\
         0 \\
         0 \\
         -|\q|
      \end{array}\right)\,,
\end{equation}
respectively. In the $B$-meson rest frame
\begin{equation}
   q_0={m_B^2-m_\Dst^2+q^2 \over 2m_B}\,, \quad\quad E_\Dst={m_B^2+m_\Dst^2-q^2 \over 2m_B}\,.
\end{equation}

\section{Four-body phase space}

The four-body phase space can be reduced to the product of the two-body phase spaces:
\begin{equation}
	\begin{split}
		d\Phi_4 =& (2\pi)^4\int\prod_{i=1}^4{d^3p_i \over (2\pi)^32E_i}\delta\left(P-\sum_{j=1}^4 p_j\right) \\
			=& {dm_{12}^2 \over 2\pi}{dm_{34}^2 \over 2\pi}\,d\Phi_2(m_{12},\,m_{34})\,d\Phi_2(\hat p_1,\,\hat p_2)\,d\Phi_2(\hat p_3,\,\hat p_4) \,,
	\end{split}
	\label{eq:PS4}
\end{equation}
where $m_{ij}^2=(p_i+p_j)^2$. The two-body phase space is given by standard form
\begin{equation}
	d\Phi_2(\hat p_i,\,\hat p_j)={1 \over 16\pi^2}{|\p_i| \over m_{ij}}\,d\cos\theta_i\,d\phi_i \,,
\end{equation}
with three-momentum $\p_i$ defined in the $ij$ rest frame.

Using Eq.~\eqref{eq:PS4}, one can write the phase space for the $\Bbar\to D_{(0)}^*(\to D\pi)\ell\nubar_\ell$
\begin{equation}
	d\Phi_4 = {1 \over 64(2\pi)^8}dm_{D\pi}^2dq^2\,{|\p_{D\pi}| \over m_B}\,d\cos\theta_{D\pi}\,d\phi_{D\pi}\,{|\p_D| \over m_{D\pi}}\,d\cos\thD\,d\phi_D\,{|\p_\ell| \over \sqrt{q^2}}\,d\cos\thl\,d\phi_\ell \,,
	\label{eq:PS4_2}
\end{equation}
where $\p_{D\pi}(=-\q)$, $\p_D$, $\p_\ell$ and the corresponding angles are defined in the $B$, $D\pi$ and $\ell\nubar$ rest frames respectively,
\begin{equation}
	|\p_{D\pi}| = {\sqrt{\lambda(m_B^2,m_{D\pi}^2,q^2)} \over 2m_B},\quad |\p_D| = {\sqrt{\lambda(m_{D\pi}^2,m_D^2,m_\pi^2)} \over 2m_{D\pi}},\quad |\p_\ell| = {\sqrt{\lambda(q^2,m_\ell^2,0)} \over 2\sqrt{q^2}}
\end{equation}
where $\lambda(a,b,c)=a^2+b^2+c^2-2(ab+ac+bc)$.

Integrating over the polar and azimuthal angles of the $D_{(0)}^*$ momentum ($\theta_{D\pi},\,\phi_{D\pi}$) and over the azimuthal angle of the $D$ momentum ($\phi_D$), one obtains
\begin{equation}
	d\Phi_4 = {1 \over 64(2\pi)^6}{|\q| \over m_B}{|\p_D| \over m_{D\pi}}\left(1-{m_\ell^2\over q^2}\right) dq^2dm_{D\pi}^2d\cos\thD d\cos\thl d\chi \,.
	\label{eq:PS4_3}
\end{equation}
Here we defined the angle $\phi_\ell=\chi$ with respect to the $D\pi$ rest frame.

\section{Full angular distribution in $\Bbar\to (D\pi)_{\Dst,\DOst}\ell\nubar_\ell$\label{app:2}}

The full angular distribution in $\Bbar\to D\pi\ell\nubar_\ell$ is determined by the total amplitude squared,
\begin{equation}
	|\M(\Bbar\to D\pi\ell\nubar_\ell)|^2 = |\M_\Dst|^2+|\M_\DOst|^2+2\Re[\M_\Dst \M_\DOst^*] \,,
\end{equation}
where
\begin{subequations}
	\begin{align}
		\label{eq:MDst2}
		\begin{split}
			& |\M_\Dst|^2 =  N|g_{\Dst D\pi}|^2 |\p_D|^2 |\widetilde{BW}_\Dst|^2 \times \biggl\{\biggr. \\
			& ~(|H_+|^2+|H_-|^2)\left(1+\cos^2\thl+{m_\ell^2 \over q^2}\sin^2\thl\right)\sin^2\thD + 2(|H_+|^2-|H_-|^2)\cos\thl\sin^2\thD \\
			& +4|H_0|^2\left(\sin^2\thl+{m_\ell^2 \over q^2}\cos^2\thl\right)\cos^2\thD + 4|H_t|^2{m_\ell^2 \over q^2}\cos^2\thD \\
			& -2\beta_\ell^2\left(\Re[H_+H_-^*]\cos2\chi+\Im[H_+H_-^*]\sin2\chi\right)\sin^2\thl\sin^2\thD \\
			& -\beta_\ell^2\left(\Re[H_+H_0^* + H_-H_0^*]\cos\chi+\Im[H_+H_0^* - H_-H_0^*]\sin\chi\right)\sin2\thl\sin2\thD \\
			& -2\left(\Re\left[H_+H_0^* - H_-H_0^* - {m_\ell^2 \over q^2}\left(H_+H_t^* + H_-H_t^*\right)\right]\cos\chi \right. \\
			& \quad\quad+\left.\Im\left[H_+H_0^* + H_-H_0^* - {m_\ell^2 \over q^2}\left(H_+H_t^* - H_-H_t^*\right)\right]\sin\chi\right)\sin\thl\sin2\thD \\
			& +8\Re[H_0H_t^*]{m_\ell^2 \over q^2}\cos\thl\cos^2\thD \,\biggl.\biggr\} \,,
		\end{split} \\
		& \nonumber \\
		\begin{split}
			& |\M_\DOst|^2 = 4N|g_{\DOst D\pi}|^2 |\widetilde{BW}_\DOst|^2 \times \biggl\{\biggr. \\
			& |H_0^\prime|^2\left(\sin^2\thl+{m_\ell^2 \over q^2}\cos^2\thl\right) + |H_t^\prime|^2{m_\ell^2 \over q^2} + 2\Re[H_0^\prime H_t^{\prime\,*}]{m_\ell^2 \over q^2}\cos\thl \,\biggl.\biggr\} \,,
		\end{split} \\
		& \nonumber \\
		\begin{split}
			& 2\Re[\M_\Dst\M_\DOst^*] = 2Ng_{\Dst  D\pi} g_{\DOst  D\pi}|\p_D| \times \biggl\{\biggr. \\
			& -\beta_\ell^2\left(\Re\left[(H_+H_0^{\prime\,*} + H_-H_0^{\prime\,*})\widetilde{BW}_\Dst\widetilde{BW}_\DOst^*\right]\cos\chi \right. \\
			 & \quad\quad+\left.\Im\left[(H_+H_0^{\prime\,*} - H_-H_0^{\prime\,*})\widetilde{BW}_\Dst\widetilde{BW}_\DOst^*\right]\sin\chi\right)\sin2\thl\sin\thD \\
			 & -2\left(\Re\left[\left(H_+H_0^{\prime\,*} - H_-H_0^{\prime\,*} - {m_\ell^2 \over q^2}\left(H_+H_t^{\prime\,*}+ H_-H_t^{\prime\,*}\right)\right)\widetilde{BW}_\Dst\widetilde{BW}_\DOst^*\right]\cos\chi \right. \\
			& \quad\quad+\left.\Im\left[\left(H_+H_0^{\prime\,*} + H_-H_0^{\prime\,*} - {m_\ell^2 \over q^2}\left(H_+H_t^{\prime\,*} - H_-H_t^{\prime\,*}\right)\right)\widetilde{BW}_\Dst\widetilde{BW}_\DOst^*\right]\sin\chi\right)\sin\thl\sin\thD \\
			& +4\Re[H_0H_0^{\prime\,*}\widetilde{BW}_\Dst\widetilde{BW}_\DOst^*]\left(\sin^2\thl+{m_\ell^2 \over q^2}\cos^2\thl\right)\cos\thD \\
			& +4\Re[H_t H_t^{\prime\,*}\widetilde{BW}_\Dst\widetilde{BW}_\DOst^*]{m_\ell^2 \over q^2}\cos\thD \\
			& +4\Re[(H_0H_t^{\prime\,*}+H_t H_0^{\prime\,*})\widetilde{BW}_\Dst\widetilde{BW}_\DOst^*]{m_\ell^2 \over q^2}\cos\thD \,\biggl.\biggr\} \,,
		\end{split}
	\end{align}
\end{subequations}
where
\begin{equation}
	N\equiv N(q^2) = {G_F^2 \over 2}|V_{cb}|^2 q^2\beta_\ell^2(q^2) \,,\quad \beta_\ell(q^2) = \sqrt{1-{m_\ell^2 \over q^2}} \,.
\end{equation}
We checked that using the narrow width approximation and assuming the helicity amplitudes to be real, the result of Eq.~\eqref{eq:MDst2} combined with the phase space factor reproduces the result of Ref.~\cite{Korner:1989qb}.

\bibliographystyle{utphys}

\begin{thebibliography}{99}



%\cite{Charles:2015gya}
\bibitem{CKMology}
  J.~Charles {\it et al.},
  %``Current status of the Standard Model CKM fit and constraints on $\Delta F=2$ New Physics,''
  Phys.\ Rev.\ D {\bf 91} (2015) 7,  073007
  [arXiv:1501.05013 [hep-ph]];
  %%CITATION = doi:10.1103/PhysRevD.91.073007;%%
  %34 citations counted in INSPIRE as of 16 Dec 2015
M.~Bona {\it et al.} [UTfit Collaboration],
  %``An Improved Standard Model Prediction Of BR(B ---> tau nu) And Its Implications For New Physics,''
  Phys.\ Lett.\ B {\bf 687} (2010) 61
  [arXiv:0908.3470 [hep-ph]].
  %%CITATION = doi:10.1016/j.physletb.2010.02.063;%%
  %125 citations counted in INSPIRE as of 16 Dec 2015

%\cite{Aoki:2013ldr}
\bibitem{FLAG}
  S.~Aoki {\it et al.},
  %``Review of lattice results concerning low-energy particle physics,''
  Eur.\ Phys.\ J.\ C {\bf 74} (2014) 2890
  [arXiv:1310.8555 [hep-lat]].
  %%CITATION = doi:10.1140/epjc/s10052-014-2890-7;%%
  %308 citations counted in INSPIRE as of 22 Dec 2015


%\cite{Aubert:2008yv}
\bibitem{Aubert:2008yv}
  B.~Aubert {\it et al.} [BaBar Collaboration],
  %``Measurements of the Semileptonic Decays anti-B ---> D l anti-nu and anti-B ---> D* l anti-nu Using a Global Fit to D X l anti-nu Final States,''
  Phys.\ Rev.\ D {\bf 79} (2009) 012002
  [arXiv:0809.0828 [hep-ex]].
  %%CITATION = doi:10.1103/PhysRevD.79.012002;%%
  %51 citations counted in INSPIRE as of 22 Dec 2015

%\cite{Dungel:2010uk}
\bibitem{Dungel:2010uk}
  W.~Dungel {\it et al.} [Belle Collaboration],
  %``Measurement of the form factors of the decay B0 -> D*- ell+ nu and determination of the CKM matrix element |Vcb|,''
  Phys.\ Rev.\ D {\bf 82} (2010) 112007
  [arXiv:1010.5620 [hep-ex]].
  %%CITATION = doi:10.1103/PhysRevD.82.112007;%%
  %38 citations counted in INSPIRE as of 22 Dec 2015

%\cite{Amhis:2014hma}
\bibitem{Amhis:2014hma}
  Y.~Amhis {\it et al.} [Heavy Flavor Averaging Group (HFAG) Collaboration],
  %``Averages of $b$-hadron, $c$-hadron, and $\tau$-lepton properties as of summer 2014,''
  arXiv:1412.7515 [hep-ex].
  %%CITATION = ARXIV:1412.7515;%%
  %148 citations counted in INSPIRE as of 22 Dec 2015

\bibitem{Feldmann:2015xsa}
 T.~Feldmann, B.~Müller and D.~van Dyk,
 %``Analyzing $b\to u$ transitions in semileptonic $\bar{B}_s \to K^{*+}(\to K \pi)\ell^-\bar\nu_\ell$ decays,''
 Phys.\ Rev.\ D {\bf 92} (2015) 3,  034013
% doi:10.1103/PhysRevD.92.034013
 [arXiv:1503.09063 [hep-ph]].
 %%CITATION = doi:10.1103/PhysRevD.92.034013;%%



%\cite{Lees:2013uzd}
\bibitem{RDst_exp}
  J.~P.~Lees {\it et al.} [BaBar Collaboration],
  %``Measurement of an Excess of $\bar{B} \to D^{(*)}\tau^- \bar{\nu}_\tau$ Decays and Implications for Charged Higgs Bosons,''
  Phys.\ Rev.\ D {\bf 88} (2013) 7,  072012
  [arXiv:1303.0571 [hep-ex]];
  %%CITATION = doi:10.1103/PhysRevD.88.072012;%%
  %86 citations counted in INSPIRE as of 16 Dec 2015
  M.~Huschle {\it et al.} [Belle Collaboration],
  %``Measurement of the branching ratio of $\bar{B} \to D^{(\ast)} \tau^- \bar{\nu}_\tau$ relative to $\bar{B} \to D^{(\ast)} \ell^- \bar{\nu}_\ell$ decays with hadronic tagging at Belle,''
  Phys.\ Rev.\ D {\bf 92} (2015) 7,  072014
  [arXiv:1507.03233 [hep-ex]];
  %%CITATION = doi:10.1103/PhysRevD.92.072014;%%
  %25 citations counted in INSPIRE as of 16 Dec 2015
G. Ciezarek [LHCb Collaboration], talk presented at Flavor Physics \& CP violation 2015 (Nagoya, Japan,
25-29 May 2015), [fpcp2015.hepl.phys.nagoya-u.ac.jp].

\bibitem{RDst_th}
  S.~Fajfer, J.~F.~Kamenik and I.~Nisandzic,
  %``On the $B \to D* \tau \bar{nu)_\tau$ Sensitivity to New Physics,''
  Phys.\ Rev.\ D {\bf 85} (2012) 094025
  [arXiv:1203.2654 [hep-ph]]; 
  S.~Fajfer, J.~F.~Kamenik, I.~Nisandzic and J.~Zupan,
  %``Implications of Lepton Flavor Universality Violations in B Decays,''
  Phys.\ Rev.\ Lett.\  {\bf 109} (2012) 161801
  [arXiv:1206.1872 [hep-ph]];
  %%CITATION = doi:10.1103/PhysRevLett.109.161801;%%
  %91 citations counted in INSPIRE as of 22 Dec 2015
  %%CITATION = ARXIV:1203.2654;%%
  P.~Biancofiore, P.~Colangelo and F.~De Fazio,
  %``On the anomalous enhancement observed in $B \to D^{(*)}\tau{\bar \nu}_\tau$ decays,''
  Phys.\ Rev.\ D {\bf 87} (2013) 7,  074010
  [arXiv:1302.1042 [hep-ph]];
  %%CITATION = doi:10.1103/PhysRevD.87.074010;%%
  %44 citations counted in INSPIRE as of 22 Dec 2015
A.~Celis, M.~Jung, X.~Q.~Li and A.~Pich,
  %``B ? D(*)??? decays in two-Higgs-doublet models,''
  J.\ Phys.\ Conf.\ Ser.\  {\bf 447} (2013) 012058
  [arXiv:1302.5992 [hep-ph]];
  %%CITATION = doi:10.1088/1742-6596/447/1/012058;%%
  %11 citations counted in INSPIRE as of 22 Dec 2015
A.~Crivellin, A.~Kokulu and C.~Greub,
  %``Flavor-phenomenology of two-Higgs-doublet models with generic Yukawa structure,''
  Phys.\ Rev.\ D {\bf 87} (2013) 9,  094031
  [arXiv:1303.5877 [hep-ph]];
  %%CITATION = doi:10.1103/PhysRevD.87.094031;%%
  %79 citations counted in INSPIRE as of 22 Dec 2015
M.~Tanaka and R.~Watanabe,
  %``Tau longitudinal polarization in B -> D tau nu and its role in the search for charged Higgs boson,''
  Phys.\ Rev.\ D {\bf 82} (2010) 034027
  [arXiv:1005.4306 [hep-ph]];
  %%CITATION = ARXIV:1005.4306;%%
  D.~Becirevic, N.~Kosnik and A.~Tayduganov,
  %``$\bar B\to D\tau\bar \nu_\tau$ vs. $\bar B\to D\mu\bar \nu_\mu$,''
  Phys.\ Lett.\ B {\bf 716} (2012) 208
  [arXiv:1206.4977 [hep-ph]];
  %%CITATION = ARXIV:1206.4977;%%
J.~A.~Bailey {\it et al.} [MILC Collaboration],
  %``B?D?? form factors at nonzero recoil and |V$_{cb}$| from 2+1-flavor lattice QCD,''
  Phys.\ Rev.\ D {\bf 92} (2015) 3,  034506
  [arXiv:1503.07237 [hep-lat]];
  %%CITATION = doi:10.1103/PhysRevD.92.034506;%%
  %11 citations counted in INSPIRE as of 16 Dec 2015
H.~Na {\it et al.} [HPQCD Collaboration],
  %``B?Dl? form factors at nonzero recoil and extraction of |V$_{cb}$|,''
  Phys.\ Rev.\ D {\bf 92} (2015) 5,  054510
  [arXiv:1505.03925 [hep-lat]];
  %%CITATION = doi:10.1103/PhysRevD.92.054510;%%
  %7 citations counted in INSPIRE as of 16 Dec 2015
M.~Atoui, V.~Morenas, D.~Becirevic and F.~Sanfilippo,
  %``$B_{s} \to D_{s}\ell\nu_\ell$  near zero recoil in and beyond the Standard Model,''
  Eur.\ Phys.\ J.\ C {\bf 74} (2014) 5,  2861
  [arXiv:1310.5238 [hep-lat]];
  %%CITATION = doi:10.1140/epjc/s10052-014-2861-z;%%
  %19 citations counted in INSPIRE as of 16 Dec 2015
I.~Dor\v{s}ner, S.~Fajfer, N.~Ko\v{s}nik and I.~Ni\v{s}and\v{z}i\'c,
  %``Minimally flavored colored scalar in $\bar B \to D^{(*)} \tau \bar \nu$ and the mass matrices constraints,''
  JHEP {\bf 1311} (2013) 084
  [arXiv:1306.6493 [hep-ph]];
  %%CITATION = doi:10.1007/JHEP11(2013)084;%%
  %36 citations counted in INSPIRE as of 22 Dec 2015
Y.~Sakaki, M.~Tanaka, A.~Tayduganov and R.~Watanabe,
  %``Testing leptoquark models in $\bar B \to D^{(*)} \tau \bar\nu$,''
  Phys.\ Rev.\ D {\bf 88} (2013) 9,  094012
  [arXiv:1309.0301 [hep-ph]];
  %%CITATION = doi:10.1103/PhysRevD.88.094012;%%
  %31 citations counted in INSPIRE as of 22 Dec 2015
A.~Abada, A.~M.~Teixeira, A.~Vicente and C.~Weiland,
  %``Sterile neutrinos in leptonic and semileptonic decays,''
  JHEP {\bf 1402} (2014) 091
  [arXiv:1311.2830 [hep-ph]];
  %%CITATION = doi:10.1007/JHEP02(2014)091;%%
  %38 citations counted in INSPIRE as of 22 Dec 2015
Y.~Sakaki, M.~Tanaka, A.~Tayduganov and R.~Watanabe,
  %``Probing New Physics with $q^2$ distributions in $\bar{B} \to D^{(*)} \tau \bar\nu$,''
  Phys.\ Rev.\ D {\bf 91} (2015) 11,  114028
  [arXiv:1412.3761 [hep-ph]];
  %%CITATION = doi:10.1103/PhysRevD.91.114028;%%
  %5 citations counted in INSPIRE as of 22 Dec 2015
A.~Celis, M.~Jung, X.~Q.~Li and A.~Pich,
  %``Sensitivity to charged scalars in $\boldsymbol{B\to D^{(*)}\tau\nu_\tau}$ and $\boldsymbol{B\to\tau\nu_\tau}$ decays,''
  JHEP {\bf 1301} (2013) 054
  [arXiv:1210.8443 [hep-ph]];
  %%CITATION = doi:10.1007/JHEP01(2013)054;%%
  %82 citations counted in INSPIRE as of 22 Dec 2015
 M.~A.~Ivanov, J.~G.~Körner and C.~T.~Tran,
  %``Exclusive decays $B \to \ell^-\bar\nu$ and $B \to D^{(\ast)}\ell^-\bar\nu$ in the covariant quark model,''
  Phys.\ Rev.\ D {\bf 92} (2015) 11,  114022
%  doi:10.1103/PhysRevD.92.114022
  [arXiv:1508.02678 [hep-ph]];
  %%CITATION = doi:10.1103/PhysRevD.92.114022;%%
  %1 citations counted in INSPIRE as of 10 févr. 2016
K.~Hagiwara, M.~M.~Nojiri and Y.~Sakaki,
  %``$CP$ violation in $B \to D\tau \nu_{\tau}$ using multipion tau decays,''
  Phys.\ Rev.\ D {\bf 89} (2014) 9,  094009
  [arXiv:1403.5892 [hep-ph]];
  Y.~Sakaki and H.~Tanaka,
  %``Constraints on the charged scalar effects using the forward-backward asymmetry on B¯?D(*)??¯?,''
  Phys.\ Rev.\ D {\bf 87} (2013) 5,  054002
  [arXiv:1205.4908 [hep-ph]];
  %%CITATION = doi:10.1103/PhysRevD.87.054002;%%
  %37 citations counted in INSPIRE as of 22 Dec 2015
  %%CITATION = doi:10.1103/PhysRevD.89.094009;%%
  %5 citations counted in INSPIRE as of 22 Dec 2015
S.~Faller, T.~Mannel and S.~Turczyk,
  %``Limits on New Physics from exclusive $B \to D^{(*)}\ell \bar\nu$ Decays,''
  Phys.\ Rev.\ D {\bf 84} (2011) 014022
  [arXiv:1105.3679 [hep-ph]];
  %%CITATION = ARXIV:1105.3679;%%
  R.~Feger, T.~Mannel, V.~Klose, H.~Lacker and T.~Luck,
  %``Limit on a Right-Handed Admixture to the Weak $b \to c$ Current from Semileptonic Decays,''
  Phys.\ Rev.\ D {\bf 82} (2010) 073002
  [arXiv:1003.4022 [hep-ph]];
  %%CITATION = ARXIV:1003.4022;%%
  S.~Bhattacharya, S.~Nandi and S.~K.~Patra,
  %``Optimal-observable analysis of possible new physics in $B\to D^{(\ast)}\tau\nu_{\tau}$,''
  arXiv:1509.07259 [hep-ph];
  %%CITATION = ARXIV:1509.07259;%%
  %3 citations counted in INSPIRE as of 10 févr. 2016
M.~Freytsis, Z.~Ligeti and J.~T.~Ruderman,
  %``Flavor models for $\bar{B} \to D^{(*)} \tau \bar{\nu}$,''
  Phys.\ Rev.\ D {\bf 92} (2015) 5,  054018
%  doi:10.1103/PhysRevD.92.054018
  [arXiv:1506.08896 [hep-ph]].
  %%CITATION = doi:10.1103/PhysRevD.92.054018;%%
  %21 citations counted in INSPIRE as of 03 févr. 2016


%\cite{Datta:2012qk}
\bibitem{distrBD}
  A.~Datta, M.~Duraisamy and D.~Ghosh,
  %``Diagnosing New Physics in $b \to c \, \tau \, \nu_\tau$ decays in the light of the recent BaBar result,''
  Phys.\ Rev.\ D {\bf 86} (2012) 034027
  [arXiv:1206.3760 [hep-ph]];
  %%CITATION = doi:10.1103/PhysRevD.86.034027;%%
  %68 citations counted in INSPIRE as of 22 Dec 2015
M.~Duraisamy and A.~Datta,
  %``The Full $B \to D^{*} \tau^{-} \bar{\nu_\tau}$ Angular Distribution and CP violating Triple Products,''
  JHEP {\bf 1309} (2013) 059
  [arXiv:1302.7031 [hep-ph]];
  %%CITATION = doi:10.1007/JHEP09(2013)059;%%
  %21 citations counted in INSPIRE as of 22 Dec 2015
M.~Duraisamy, P.~Sharma and A.~Datta,
  %``Azimuthal $B \to D^{*} \tau^{-} \bar{\nu_\tau}$ angular distribution with tensor operators,''
  Phys.\ Rev.\ D {\bf 90} (2014) 7,  074013
  [arXiv:1405.3719 [hep-ph]].
  %%CITATION = doi:10.1103/PhysRevD.90.074013;%%
  %11 citations counted in INSPIRE as of 22 Dec 2015



%\cite{Korner:1989qb}
\bibitem{Korner:1989qb}
  J.~G.~K\"orner and G.~A.~Schuler,
  %``Exclusive Semileptonic Heavy Meson Decays Including Lepton Mass Effects,''
  Z.\ Phys.\ C {\bf 46} (1990) 93.
  %%CITATION = ZEPYA,C46,93;%%


%\cite{Agashe:2014kda}
\bibitem{PDG}
  K.~A.~Olive {\it et al.} [Particle Data Group Collaboration],
  %``Review of Particle Physics,''
  Chin.\ Phys.\ C {\bf 38} (2014) 090001.
  %%CITATION = doi:10.1088/1674-1137/38/9/090001;%%
  %2572 citations counted in INSPIRE as of 22 Dec 2015



\bibitem{s-wave-kpi}
  D.~Becirevic and A.~Tayduganov,
  %``Impact of $B\to K^\ast_0 \ell^+\ell^-$ on the New Physics search in $B\to K^\ast \ell^+\ell^-$ decay,''
  Nucl.\ Phys.\ B {\bf 868} (2013) 368
  [arXiv:1207.4004 [hep-ph]];
  %%CITATION = doi:10.1016/j.nuclphysb.2012.11.016;%%
  %55 citations counted in INSPIRE as of 22 Dec 2015
J.~Matias,
  %``On the S-wave pollution of B-> K* l+l- observables,''
  Phys.\ Rev.\ D {\bf 86} (2012) 094024
  [arXiv:1209.1525 [hep-ph]];
  %%CITATION = doi:10.1103/PhysRevD.86.094024;%%
  %47 citations counted in INSPIRE as of 22 Dec 2015
T.~Blake, U.~Egede and A.~Shires,
  %``The effect of S-wave interference on the $B^0 \to K^{\ast 0}\ell^+\ell^-$ angular observables,''
  JHEP {\bf 1303} (2013) 027
  [arXiv:1210.5279 [hep-ph]];
  %%CITATION = doi:10.1007/JHEP03(2013)027;%%
  %38 citations counted in INSPIRE as of 22 Dec 2015
M.~D\"oring, U.~G.~Meissner and W.~Wang,
  %``Chiral Dynamics and S-wave Contributions in Semileptonic B decays,''
  JHEP {\bf 1310} (2013) 011
  [arXiv:1307.0947 [hep-ph]];
  %%CITATION = doi:10.1007/JHEP10(2013)011;%%
  %26 citations counted in INSPIRE as of 22 Dec 2015
D.~Das, G.~Hiller, M.~Jung and A.~Shires,
  %``The $ \overline{B}\to \overline{K}\pi \ell \ell $ and $ {\overline{B}}_s\ \to \overline{K}K\ell \ell $ distributions at low hadronic recoil,''
  JHEP {\bf 1409} (2014) 109
  [arXiv:1406.6681 [hep-ph]].
  %%CITATION = doi:10.1007/JHEP09(2014)109;%%
  %15 citations counted in INSPIRE as of 22 Dec 2015



%\cite{Dassinger:2008as}
\bibitem{Dassinger:2008as}
  B.~Dassinger, R.~Feger and T.~Mannel,
  %``Complete Michel Parameter Analysis of inclusive semileptonic b ---> c transition,''
  Phys.\ Rev.\ D {\bf 79} (2009) 075015
  [arXiv:0803.3561 [hep-ph]].
  %%CITATION = ARXIV:0803.3561;%%

%\cite{Cheng:2003sm}
\bibitem{cheng}
  H.~-Y.~Cheng, C.~-K.~Chua and C.~-W.~Hwang,
  %``Covariant light front approach for s wave and p wave mesons: Its application to decay constants and form-factors,''
  Phys.\ Rev.\ D {\bf 69} (2004) 074025
  [hep-ph/0310359].
  %%CITATION = HEP-PH/0310359;%%
  %190 citations counted in INSPIRE as of 25 Oct 2013



%\cite{Becirevic:2012pf}
\bibitem{notre-g}
  D.~Becirevic and F.~Sanfilippo,
  %``Theoretical estimate of the $D^* \to D\pi$ decay rate,''
  Phys.\ Lett.\ B {\bf 721} (2013) 94
  [arXiv:1210.5410 [hep-lat]].
  %%CITATION = ARXIV:1210.5410;%%
  %4 citations counted in INSPIRE as of 25 Jul 2013

%\cite{Lees:2013zna}
\bibitem{babar-g}
 J.~P.~Lees {\it et al.}  [BaBar Collaboration],
  %``Measurement of the $D*(2010)^+$ meson width and the $D*(2010)^+ - D^0$ mass difference,''
  Phys.\  Rev.\  Lett.\  111, {\bf 111801} (2013)
   [Phys.\ Rev.\ Lett.\  {\bf 111} (2013) 111801]
  [arXiv:1304.5657 [hep-ex]].
  %%CITATION = ARXIV:1304.5657;%%
  %3 citations counted in INSPIRE as of 25 Oct 2013


%\cite{Becirevic:2012zza}
\bibitem{hs}
C.~McNeile {\it et al.} [UKQCD Collaboration],
  %``Hadronic decay of a scalar B meson from the lattice,''
  Phys.\ Rev.\ D {\bf 70} (2004) 054501
  [hep-lat/0404010];
  %%CITATION = doi:10.1103/PhysRevD.70.054501;%%
  %29 citations counted in INSPIRE as of 22 Dec 2015
  B.~Blossier, N.~Garron and A.~G\'erardin,
  %``Pion couplings to the scalar B meson,''
  Eur.\ Phys.\ J.\ C {\bf 75} (2015) 103
  [arXiv:1410.3409 [hep-lat]];
  %%CITATION = doi:10.1140/epjc/s10052-015-3321-0;%%
  %1 citations counted in INSPIRE as of 22 Dec 2015
  D.~Becirevic, E.~Chang and A.~L.~Yaouanc,
  %``Pionic couplings to the lowest heavy-light mesons of positive and negative parity,''
  arXiv:1203.0167 [hep-lat].
  %%CITATION = ARXIV:1203.0167;%%
  %9 citations counted in INSPIRE as of 22 Dec 2015


\bibitem{hc}
D.~Mohler, S.~Prelovsek and R.~M.~Woloshyn,
  %``$D \pi$ scattering and $D$ meson resonances from lattice QCD,''
  Phys.\ Rev.\ D {\bf 87} (2013) 3,  034501
  [arXiv:1208.4059 [hep-lat]].
  %%CITATION = doi:10.1103/PhysRevD.87.034501;%%
  %65 citations counted in INSPIRE as of 22 Dec 2015


%\cite{Link:2002ev}
\bibitem{Link:2002ev}
  J.~M.~Link {\it et al.}  [FOCUS Collaboration],
  %``Evidence for new interference phenomena in the decay D+ ---> K- pi+ mu+ nu,''
  Phys.\ Lett.\ B {\bf 535} (2002) 43
  [hep-ex/0203031];
  %%CITATION = HEP-EX/0203031;%%
  %111 citations counted in INSPIRE as of 25 Oct 2013
R.~A.~Briere {\it et al.}  [CLEO Collaboration],
  %``Analysis of D^+ -> K^- pi^+ e^+ nu_e and D^+ -> K^- pi^+ mu^+ nu_mu Semileptonic Decays,''
  Phys.\ Rev.\ D {\bf 81} (2010) 112001
  [arXiv:1004.1954 [hep-ex]];
  %%CITATION = ARXIV:1004.1954;%%
  %6 citations counted in INSPIRE as of 25 Oct 2013
P.~del Amo Sanchez {\it et al.}  [BaBar Collaboration],
  %``Analysis of the $D^+ \to K^- \pi^+ e^+ \nu_e$ decay channel,''
  Phys.\ Rev.\ D {\bf 83} (2011) 072001
  [arXiv:1012.1810 [hep-ex]].
  %%CITATION = ARXIV:1012.1810;%%
  %13 citations counted in INSPIRE as of 25 Oct 2013


\bibitem{melikhov}
D.~Melikhov and B.~Stech,
  %``Weak form-factors for heavy meson decays: An Update,''
  Phys.\ Rev.\ D {\bf 62} (2000) 014006
  [hep-ph/0001113].
  %%CITATION = HEP-PH/0001113;%%
  %230 citations counted in INSPIRE as of 25 Oct 2013




\end{thebibliography}

\end{document}